\documentclass[paper,11pt]{JHEP}
\usepackage[centertags]{amsmath}
\usepackage{amsfonts} \usepackage{amssymb} \usepackage{amsthm}
\usepackage{graphicx}
\newcommand{\ch}{\mbox{(Ch)}}
\newcommand{\cc}{\,\mbox{c.c.}}
\newcommand{\eq}[1]{Eq. (\ref{#1})}

\newcommand{\acomm}[2]{\left\{#1,#2\right\}}

\newcommand{\ket}[1]{|{#1}\rangle}
\newcommand{\dket}[1]{|{#1}\rangle\!\rangle}

\newcommand{\ii}{\mathrm{i}}
\newcommand{\ee}{\mathrm{e}}
\newcommand{\one}{{\rm 1\kern -.9mm l}}

\newcommand{\Tr}{\mathrm{Tr}\,}
\newcommand{\tr}{\mathrm{tr}\,}

\newcommand{\re}{\mbox{Re}\,}
\newcommand{\im}{\mbox{Im}\,}

\title{Flux interactions on D-branes and instantons
}
\author{\parbox{11.5cm}{Marco Bill\`o$^1$, Livia Ferro$^{1,2}$, Marialuisa Frau$^1$, Francesco Fucito$^3$,
Alberto Lerda$^4$ and Jose F. Morales$^3$}
\\
~\\
~\\
$^1$Dipartimento di Fisica Teorica, Universit\`a di Torino\\
and I.N.F.N. - sezione di Torino \\
Via P. Giuria 1, I-10125 Torino, Italy\\
\vspace{0.3cm}
$^2$Laboratoire de Physique Th\'{e}orique\\
\'{E}cole Normale Sup\'{e}rieure\\
24, rue Lhomond, 75231 Paris Cedex 05, France\\
\vspace{0.3cm}
$^3$I.N.F.N. - sezione di Roma II
\\
Via della Ricerca Scientifica, I-00133 Roma, Italy\\
\vspace{0.3cm}
$^4$Dipartimento di Scienze e Tecnologie Avanzate, Universit\`a del Piemonte Orientale\\
and I.N.F.N. - Gruppo Collegato di Alessandria - sezione di Torino\\
Via V. Bellini 25/G, I-15100 Alessandria, Italy\\
\vspace{0.3cm}
\email{billo,ferro,frau,lerda@to.infn.it, Francesco.Fucito,Francisco.Morales@roma2.infn.it}
}
\abstract{We provide a direct world-sheet derivation of the
couplings of NS-NS and R-R fluxes to various types of D-branes
(including instantonic ones) by evaluating disk amplitudes among two
open string vertex operators at a generic brane intersection and one
closed string vertex representing the background fluxes. This
world-sheet approach is in full agreement with the derivation of the
flux couplings in the brane effective actions based on supergravity
methods, but it is applicable also to more general brane
configurations involving fields with twisted boundary conditions. As
an application, we consider an orbifold compactification of Type IIB
string theory with fractional D-branes preserving $\mathcal N=1$
supersymmetry and study the flux-induced fermionic mass terms both
on space-filling branes and on instantonic ones. Our results show
the existence of a relation between the soft supersymmetry breaking
and the lifting of some instanton fermionic zero-modes, which may
lead to new types of non-perturbative couplings in brane-world
models. }
\preprint{DFTT/14/2008\\ROM2F/2008/19\\LPTENS 08/40\\}

\begin{document}

\section{Introduction}
\label{sec:intro}

A promising scenario for phenomenological applications of string theory and realistic
model building is provided by four dimensional compactifications of
Type II string theories preserving $\mathcal N=1$ supersymmetry in the presence
of intersecting or magnetized D-branes
\cite{Blumenhagen:2005mu,Blumenhagen:2006ci,Marchesano:2007de}.
In these compactifications, gauge interactions
similar to those of the supersymmetric extensions of the Standard Model
can be engineered with space-filling D-branes that partially or totally wrap
the internal six-dimensional space. By introducing several
stacks of such D-branes one can realize adjoint gauge fields for various
groups by means of the massless excitations of open strings that start and end
on the same stack, while open strings that stretch between different
stacks provide  bi-fundamental matter fields. On the other hand, from the closed
string point of view, (wrapped) D-branes are sources for various fields
of Type II supergravity, which acquire a non-trivial profile in the bulk.
Thus the effective actions of these brane-world models describe interactions
of both open string (boundary) and closed string (bulk) degrees of freedom and have the
generic structure of $\mathcal N=1$ supergravity in four dimensions
coupled to vector and chiral multiplets.
Several important aspects of such effective actions have been intensively investigated over the years
from various points of view \cite{Blumenhagen:2005mu,Blumenhagen:2006ci,Marchesano:2007de}.

One of the main ingredients of these string compactifications is
the possibility of adding internal antisymmetric fluxes both in the Neveu-Schwarz-Neveu-Schwarz
(NS-NS) and in the Ramond-Ramond (R-R) sector of the bulk theory
\cite{Grana:2005jc,Douglas:2006es,Denef:2007pq}. These fluxes may bear important consequences
on the low-energy effective action of the brane-worlds,
such as moduli stabilization, supersymmetry breaking and, possibly, also the generation of non-perturbative superpotentials. At a perturbative level internal
3-form fluxes are encoded in a bulk superpotential \cite{Gukov:1999ya,Taylor:1999ii}
from which F-terms for the various compactification moduli can be obtained using standard
supergravity methods. These terms can also be interpreted as
the $\theta^2$ ``auxiliary'' components of the kinetic functions for the gauge theory
defined on the space-filling branes,
and thus are soft supersymmetry breaking terms for the brane-world
effective action. These soft terms have been computed in Refs.  \cite{Grana:2002nq}
\nocite{Camara:2003ku,Grana:2003ek,Camara:2004jj,Lust:2004fi,Conlon:2005ki,Conlon:2006wz}-
\cite{Berg:2007wt}
and their effects, such as flux-induced masses for the gauginos and the gravitino,
have been analyzed in various scenarios of flux compactifications
relying on the structure of the bulk supergravity Lagrangian and on
$\kappa$-symmetry considerations.

In addition to fluxes, another important issue to study is the
non-perturbative sector of the effective actions coming from string
theory compactifications \cite{Witten:1995gx,Douglas:1995bn}. Only
in the last few years, concrete computational techniques have been
developed to analyze non-perturbative effects using systems of
branes with different boundary conditions
\cite{Green:2000ke,Billo:2002hm}. These methods not only allow to
reproduce
\cite{Billo:2002hm}\nocite{Billo:2006jm,Akerblom:2006hx,Billo:2007sw}-\cite{Billo:2007py}
the known instanton calculus of (supersymmetric) field theories
\cite{Dorey:2002ik}, but can also be generalized to more exotic
configurations where field theory methods are not yet available
\cite{Blumenhagen:2006xt}\nocite{Ibanez:2006da,Florea:2006si,Bianchi:2007fx,Argurio:2007vqa,
Bianchi:2007wy,Ibanez:2007rs,Antusch:2007jd,Blumenhagen:2007zk,Aharony:2007pr,Blumenhagen:2007bn,Camara:2007dy,Ibanez:2007tu,GarciaEtxebarria:2007zv,Petersson:2007sc,Bianchi:2007rb,Blumenhagen:2008ji,Argurio:2008jm,Cvetic:2008ws,Kachru:2008wt,GarciaEtxebarria:2008pi}-\cite{Buican:2008qe}.
The study of these exotic instanton configurations has led to
interesting results in relation to moduli stabilization, (partial)
supersymmetry breaking and even fermion masses and Yukawa couplings
\cite{Blumenhagen:2006xt,Ibanez:2006da,Blumenhagen:2007zk}. A
delicate point about these stringy instantons concerns the presence
of neutral anti-chiral fermionic zero-modes which completely
decouple from all other instanton moduli, contrarily to what happens
for the usual gauge theory instantons where they act as Lagrange
multipliers for the fermionic ADHM constraints \cite{Billo:2002hm}.
In order to get non-vanishing contributions to the effective action
from such exotic instantons, it is therefore necessary to remove
these anti-chiral zero modes \cite{Argurio:2007vqa,Bianchi:2007wy}
or lift them by some mechanism \cite{Blumenhagen:2007bn}. The
presence of internal background fluxes may allow for such a lifting
and points to the existence of an intriguing interplay among soft
supersymmetry breaking, moduli stabilization, instantons and
more-generally non-perturbative effects in the low-energy theory
which may lead to interesting developments and applications. Some
preliminary results along these lines have recently appeared in Ref.
\cite{GarciaEtxebarria:2008pi}.

So far the consequences of the presence of
internal NS-NS or R-R flux backgrounds onto the world-volume
theory of space-filling or instantonic branes
have been investigated relying entirely on space-time supergravity methods \cite{Grana:2002tu}\nocite{Marolf:2003vf,Marolf:2003ye,Tripathy:2005hv,Martucci:2005rb}
-\cite{Bergshoeff:2005yp}, rather than through a string world-sheet approach
\footnote{For some recent developments using world-sheet methods see Ref.
\cite{Linch:2008rw}.}.
In this paper we fill this gap and derive the flux induced fermionic
terms of the D-brane effective actions with an explicit conformal
field theory calculation of scattering amplitudes among two open
string vertex operators describing the fermionic excitations at a
generic brane intersection and one closed string vertex operator
describing the background flux. Our world-sheet approach is quite
generic and allows to obtain the flux induced couplings in a unified
way for a large variety of different cases: space-filling or
instantonic branes, with or without magnetization, with twisted or
untwisted boundary conditions. Indeed, the scattering amplitudes we
compute are generic mixed disk amplitudes, {\it{i.e.}} mixed
open/closed string amplitudes on disks with mixed boundary
conditions, similar to the ones considered in Refs.
\cite{Billo:2004zq,Lust:2004cx,Billo:2005jw,Bertolini:2005qh,Billo:2006jm}.

Besides being interesting from a technical point of view, our approach
not only reproduces correctly all known results but can be applied also
to cases where the supergravity methods are less obvious, like
for example to study how NS-NS or R-R fluxes couple to fields with twisted
boundary conditions or how they modify the action which gives the measure of integration
on the moduli space of instantons. Finding the flux-induced soft terms on instantonic
branes of both ordinary and exotic type is a necessary step towards the investigations
of the non-perturbative aspects of flux compactifications we have mentioned above.

In this paper, after discussing the general conformal field theory
calculation of the flux couplings to boundary fermions in ten
dimensions, we select a specific compactification of Type IIB string
theory that leads to a brane-world theory with $\mathcal N=1$
supersymmetry in four dimensions. In particular we consider $\mathbb
Z_2 \times \mathbb Z_2$ orbifold of type IIB on ${\cal T}^6$ with
fractional D-branes. In this compactification scheme string theory
remains calculable and our explicit world-sheet approach is viable;
moreover the existence of inequivalent types of fractional branes
gives rise to a quiver structure allowing to engineer gauge theories
with interesting contents, such as pure super Yang-Mills theory or
SQCD. The non-perturbative side can then be explored by means of
instantonic fractional branes which can be of both ordinary or
exotic type.

More specifically, this paper is organized as follows:
in Section \ref{sec:CFT} we describe in detail the
world-sheet derivation of the flux induced fermionic terms of the D-brane effective action
from mixed open/closed string scattering amplitudes. The explicit results for
various unmagnetized or magnetized branes as well as for instantonic branes are spelled
out in Section \ref{sec:effects} in the case of untwisted open strings and in Section \ref{sec:twisted}
in some case of twisted open strings. The flux-induced fermionic couplings are further analyzed
for the $\mathbb Z_2 \times \mathbb Z_2$ orbifold compactification which
we briefly review in Section \ref{sec:N1}. Later in Section \ref{sec:n1int}
we compare our world-sheet results for the flux couplings on fractional D3-branes
with the effective supergravity approach to the soft supersymmetry breaking terms,
finding perfect agreement. In Section \ref{sec:fD-1} we exploit the
generality of our world-sheet based results to determine the soft terms
of the action on the instanton moduli space, and finally in Section \ref{sec:summary}
we summarize our results.
Our conventions on spinors, on the $\mathbb Z_2 \times \mathbb Z_2$ orbifold
and on the flux couplings for wrapped fractional D9-branes
are contained in the Appendix.

\section{Flux interactions on D-branes from string diagrams}
\label{sec:CFT}
In this section, using world-sheet methods, we study the interactions between
closed string background fluxes and massless open string excitations living on a generic D-brane intersection.
We focus on fermionic terms (like for example mass terms for gauginos),
but our conformal field theory techniques could be applied to other terms of the brane
effective action.
In order to keep the discussion as general as possible, we adopt here a ten-dimensional notation.
Later, in Sections \ref{sec:effects} and \ref{sec:twisted} we will rephrase our findings using
a four-dimensional language suitable to discuss compactifications of Type IIB string theory to $d=4$.

\begin{figure}[t]
\begin{center}
\begin{picture}(0,0)%
\includegraphics{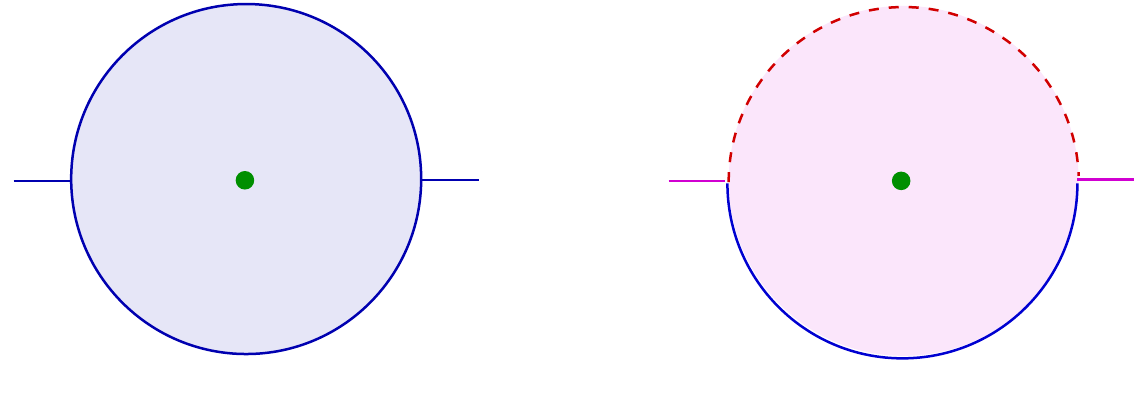}%
\end{picture}%
\setlength{\unitlength}{1579sp}%
\begingroup\makeatletter\ifx\SetFigFontNFSS\undefined%
\gdef\SetFigFontNFSS#1#2#3#4#5{%
  \reset@font\fontsize{#1}{#2pt}%
  \fontfamily{#3}\fontseries{#4}\fontshape{#5}%
  \selectfont}%
\fi\endgroup%
\begin{picture}(13638,4765)(271,-3920)
\put(286,-1825){\makebox(0,0)[lb]{\smash{{\SetFigFontNFSS{10}{12.0}{\familydefault}{\mddefault}{\updefault}$V_\Theta$}}}}
\put(5596,-1810){\makebox(0,0)[lb]{\smash{{\SetFigFontNFSS{10}{12.0}{\familydefault}{\mddefault}{\updefault}$V_{\Theta^\prime}$}}}}
\put(8146,-1823){\makebox(0,0)[lb]{\smash{{\SetFigFontNFSS{10}{12}{\familydefault}{\mddefault}{\updefault}$V_\Theta$}}}}
\put(13456,-1808){\makebox(0,0)[lb]{\smash{{\SetFigFontNFSS{10}{12.0}{\familydefault}{\mddefault}{\updefault}$V_{\Theta^\prime}$}}}}
\put(451,434){\makebox(0,0)[lb]{\smash{{\SetFigFontNFSS{10}{12.0}{\familydefault}{\mddefault}{\updefault}\emph{a)}}}}}
\put(8221,404){\makebox(0,0)[lb]{\smash{{\SetFigFontNFSS{10}{12.0}{\familydefault}{\mddefault}{\updefault}\emph{b)}}}}}
\put(1546,-3811){\makebox(0,0)[lb]{\smash{{\SetFigFontNFSS{10}{12.0}{\familydefault}{\mddefault}{\updefault}$\vec\vartheta = 0$}}}}
\put(9331,-3811){\makebox(0,0)[lb]{\smash{{\SetFigFontNFSS{10}{12.0}{\familydefault}{\mddefault}{\updefault}$\vec\vartheta \not= 0$}}}}
\put(10606,-983){\makebox(0,0)[lb]{\smash{{\SetFigFontNFSS{10}{12.0}{\familydefault}{\mddefault}{\updefault}$V_F, V_H$}}}}
\put(2731,-970){\makebox(0,0)[lb]{\smash{{\SetFigFontNFSS{10}{12.0}{\familydefault}{\mddefault}{\updefault}$V_F, V_H$}}}}
\end{picture}%
\end{center}
\caption{Quadratic coupling of untwisted, $a)$, and twisted, $b)$, open string states to closed
string fluxes. }
\label{fig:flux}
\end{figure}
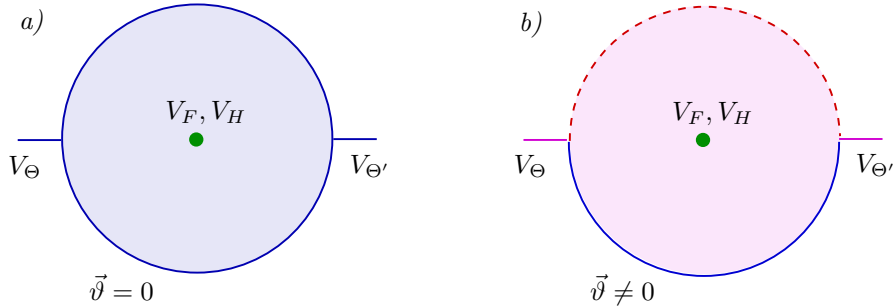

At the lowest order, the fermionic interaction terms can be derived from disk 3-point
correlators involving two vertices describing massless open-string fermions and one closed string
vertex describing the background flux, as represented in Fig. \ref{fig:flux}.
At a brane intersection massless open string modes can arise either from open strings
starting and ending on the same stack of D-branes, or from open strings connecting
two different sets of branes. In the former case the open string fields satisfy
the standard untwisted boundary conditions and the corresponding vertex operators transform in the
adjoint representation of the gauge group. In the latter case the string coordinates
satisfy twisted boundary conditions characterized by twist parameters $\vartheta$ and the
associated vertices carry Chan-Paton factors in the bi-fundamental representation of the gauge
group; by inserting twisted open string vertices, one splits the disk boundary into different portions
distinguished by their boundary conditions and Chan-Paton labels, see Fig. \ref{fig:flux}\emph{b)}. We now give some details on
these boundary conditions and later describe the physical vertex operators and their interactions with
R-R and NS-NS background fluxes.

\subsection{Boundary conditions and reflection matrices}
\label{sec:bc}
The boundary conditions for the bosonic coordinates $x^M$ ($M=0,\ldots, 9$) of the open string
are given by
\begin{equation}
\Big(\delta_{MN}\,\partial_\sigma x^N+{\rm i}({\cal F}_\sigma)_{MN}\,\partial_\tau
x^N\Big)\Big|_{\sigma=0,\pi}=0~,
\label{bc}
\end{equation}
where $\delta_{MN}$ is the flat background metric%
\footnote{Here, for convenience, we assume the space-time to have an Euclidean signature. Later, in Section \ref{sec:effects} we revert to a Minkowskian signature when appropriate.} and
\begin{equation}
({\mathcal F}_\sigma)_{MN}
=B_{MN} + 2\pi\alpha' \,({F}_\sigma)_{MN}
\label{calF}
\end{equation}
with $B_{MN}$ the anti-symmetric tensor of the NS-NS sector and $({F}_\sigma)_{MN}$
the background gauge field strength at the string end points $\sigma=0,\pi$.
Introducing the complex variable $z={\rm e}^{\tau+{\rm i}\sigma}$ and the reflection matrices
\begin{equation}
{R}_\sigma=\big(1-{\mathcal F}_\sigma\big)^{-1}\,\big(1+{\mathcal F}_\sigma\big)~,
\label{R}
\end{equation}
the conditions (\ref{bc}) become
\begin{equation}
\overline\partial x^M\Big|_{\sigma=0,\pi}=({R}_\sigma)^M_{~N}\,\partial
x^N\Big|_{\sigma=0,\pi}~.
\label{bc1}
\end{equation}
The standard Neumann boundary conditions ({\it i.e.} ${R}_\sigma=1$)
are obtained by setting ${\mathcal F}_\sigma=0$, whereas the Dirichlet case
({\it i.e.} ${R}_\sigma=-1$) is recovered in the limit ${\mathcal F}_\sigma\to \infty$.
A convenient way to solve (\ref{bc1}) is to
define multi-valued holomorphic fields $X^M(z)$ such that
\begin{equation}
X^M({\rm e}^{2\pi{\rm i}}z) =
\big({R}_\pi^{-1}\,{R}_0\big)^{M}_{~N}\,X^N(z) \equiv R^{M}_{~N}\,X^N(z)
\label{chiraly}
\end{equation}
where $R \equiv {R}_\pi^{-1} {R}_0$ is the monodromy matrix.
Then, putting the branch cut just below the negative real axis of the $z$-plane,
the conditions (\ref{bc1}) are solved by
\begin{equation}
x^M(z,\overline z) = q^M+ \frac{1}{2}\Big[X^M(z) +\big({R}_0\big)^{M}_{~N}\,X^N(\overline
z)\Big]~,
\label{sol}
\end{equation}
where $z$ is restricted to the upper half-complex plane, and $q^M$
are constant zero-modes.

For simplicity, in this paper we take the reflection matrices ${R}_0$ and ${R}_\pi$
to be commuting. Then, with a suitable $\mathrm{SO}(10)$ transformation
we can simultaneously diagonalize both matrices and write
\begin{subequations}
\begin{align}
{R}_\sigma &= \mathrm{diag}\left(\ee^{2\pi\ii\theta_\sigma^1},\ee^{-2\pi\ii\theta_\sigma^1}\,
\ldots,\ee^{2\pi\ii\theta_\sigma^5},\ee^{-2\pi\ii\theta_\sigma^5}\right)~,
\label{diagonalRs}\\
{R}   &= \mathrm{diag}\left(\ee^{2\pi\ii\vartheta^1},\ee^{-2\pi\ii\vartheta^1},\ldots,
\ee^{2\pi\ii\vartheta^5},\ee^{-2\pi\ii\vartheta^5}\right)~,
\label{diagonalR1}
\end{align}
\end{subequations}
with $\vartheta^I=\theta_0^I-\theta_\pi^I$.
In this basis the resulting (complex) coordinates, denoted by $Z^I$ and ${\overline Z}^I$
with%
\footnote{In the subsequent sections we will take the space-time to be the
product of a four-dimensional part and an internal six-dimensional part.
For notational convenience, we label the complex coordinates of the four-dimensional part by $I=4,5$ and those of the internal six-dimensional part by $I\equiv i=1,2,3$.} $I=1,\ldots,5$, satisfy
\begin{equation}
\partial{Z}^{I}({\ee}^{2\pi{\ii}}z) = {\ee}^{2\pi\ii\vartheta^I}\partial{Z}^{I}(z)
\quad\mbox{and}\quad
\partial{\overline Z}^{I}({\ee}^{2\pi\ii}z) = {\ee}^{-2\pi\ii\vartheta^I}
\partial{\overline Z}^{I}(z)~,
\label{calY1}
\end{equation}
and hence have an expansion in powers of $z^{n+\vartheta^I}$ and $z^{n-\vartheta^I}$, respectively,
with $n\in \mathbb{Z}$. The corresponding oscillators act on a twisted vacuum $|\vec{\vartheta}\rangle$
created by the twist operator $\sigma_{\vec{\vartheta}}(z)$, which is a conformal field of dimension
$h_{\sigma_{\vec{\vartheta}}}=\frac{1}{2}\sum_I\vartheta^I(1-\vartheta^I)$ satisfying the following OPE
\begin{equation}
\sigma_{\vec{\vartheta}}(z) \, \sigma_{-\vec{\vartheta}}(w)
\sim (z-w)^{\sum_I\vartheta^I(1-\vartheta^I)}~.
\label{OPE}
\end{equation}

For our purposes it is necessary to consider also the boundary conditions on the
fermionic fields $\psi^M$ of the open superstring in the RNS formalism, which are
\begin{equation}
\psi^M({\rm e}^{2\pi{\rm i}}z) = \eta\, R^{M}_{~N}\,\psi^N(z)
\label{chiralpsi}
\end{equation}
where $\eta=1$ in the NS sector and $\eta=-1$ in the R sector.
In the complex basis these boundary conditions become
\begin{equation}
\Psi^I({\ee}^{2\pi{\ii}}z) = \eta\,{\ee}^{2\pi\ii\vartheta^I}{\Psi}^{I}(z)\quad\mbox{and}\quad
{\overline \Psi}^{I}({\ee}^{2\pi\ii}z) = \eta\,{\ee}^{-2\pi\ii\vartheta^I}{\overline \Psi}^{I}(z)
~.
\label{psi}
\end{equation}
Thus, in the NS sector $\Psi^I$ and ${\overline\Psi}^{I}$ admit an
expansion in powers of $z^{n+\vartheta^I}$ and
$z^{n-\vartheta^I}$, respectively, with $n\in \mathbb{Z}$, so that
their oscillators are of the type
$\psi^I_{n+\vartheta^I + \frac 12}$. In the R sector they have a mode
expansion in powers of $z^{n+\vartheta^I}$ and
$z^{n-\vartheta^I}$, respectively, with $n\in
\mathbb{Z}+\frac{1}{2}$. Note that if $\vartheta^I\not=0$ neither
the NS nor the R sector possesses zero-modes and the corresponding
fermionic vacuum is non-degenerate. On the other hand if
$\vartheta^I=0$ there are zero-modes in the R sector, while if
$\vartheta^I=\frac{1}{2}$ there are zero-modes in the NS sector.
In these cases the corresponding fermionic vacuum is degenerate
and carries the spinor representation of the rotation group acting
on the directions in which the $\vartheta$'s vanish. For this
reason it is in general necessary to determine the boundary
reflection matrices also in the spinor representation, which we
will denote by $\mathcal{R}_\sigma$.

To find these matrices, we first note that in the vector representation $R_\sigma$
simply describes the product of five rotations with angles $2\pi\theta_\sigma^I$
in the five complex planes defined by the complex coordinates $Z^I$ and $\overline{Z}^I$, as is clear from (\ref{diagonalRs}).
Then, recalling that the infinitesimal generator of such rotations in the spinor representation is
$\frac{\ii}{2}\Gamma^{I\bar I} =\frac{\ii}{4}\left[\Gamma^I,\Gamma^{\bar I}\right]$
with $\Gamma^I$ being the $\mathrm{SO}(10)$ $\Gamma$-matrices in the
complex basis (see Appendix \ref{app:conventions} for our conventions),
we easily conclude that
\begin{equation}
\mathcal{R}_\sigma = \pm \prod_{I=1}^5\ee^{\ii\pi\theta_\sigma^I\Gamma^{I\bar I}}
= \pm \prod_{I=1}^5\frac{\left(1+\ii f_\sigma^I\Gamma^{I\bar I}\right)}{\sqrt{1+(f_\sigma^I)^2}}
\label{rspinor}
\end{equation}
where $f_\sigma^I=\tan\pi\theta_\sigma^I$ and the overall sign
depends on whether we have a D-brane or an anti D-brane. This
general formula is particularly useful to derive the explicit
expression for $\mathcal{R}_\sigma$ in the limits $f_\sigma^I\to 0$
or $f_\sigma^I\to\infty$ corresponding, respectively, to Neumann or
Dirichlet boundary conditions in the $I$-plane. For example, for an
open string starting from a D$p$-brane extending in the directions
$(01\ldots p)$ we have
\begin{equation}
\mathcal{R}_0 = \prod_{I=1}^{\frac{9-p}{2}} \big(\ii\Gamma^{I\bar I}\big)
= \Gamma^{(p+1)}\cdots\Gamma^9~.
\label{R0D3}
\end{equation}
Being particular instances of rotations, the reflection matrices in the vector and spinor representations satisfy the
following relation
\begin{equation}
\mathcal{R}_\sigma^{-1} \Gamma^M\mathcal{R}_\sigma= ({R}_\sigma)^M_{~N}\Gamma^N~.
\label{vectspin}
\end{equation}

\subsection{Open and closed string vertices}
\label{subsec:vertices}
A generic brane intersection can describe different physical situations depending
on the values of the five twists $\vartheta^I$.

When $\vartheta^I=0$ for all $I$'s, all fields are untwisted:
this is the case of the open strings
starting and ending on the same stack of D-branes which account for dynamical gauge excitations
in the adjoint representation when the branes are space-filling, or for neutral instanton moduli
when the branes are instantonic.

When $\vartheta^4=\vartheta^5=0$ but the $\vartheta^i$'s with
$i=1,2,3$ are non vanishing, only the string coordinates in the
space-time directions are untwisted and describe open strings
stretching between different stacks of D-branes. The corresponding
excitations organize in multiplets that transform in the
bi-fundamental representation of the gauge group and always contain
massless chiral fermions. When suitable relations among the
non-vanishing twists are satisfied ({\it e.g.}
$\vartheta^1+\vartheta^2+\vartheta^3=2$) also massless scalars
appear in the spectrum and they can be combined with the fermions to
form $\mathcal{N}=1$ chiral multiplets suitable to describe the
matter content of brane-world models.

Finally, when $\vartheta^4=\vartheta^5=\frac{1}{2}$, the string
coordinates have mixed Neumann-Dirichlet boundary conditions in the
last four directions and correspond to open strings connecting a
space-filling D-brane  with an instantonic brane. In this situation,
if the $\vartheta^i$'s ($i=1,2,3$) are vanishing, the instantonic
brane describes an ordinary gauge instanton configuration and the
twisted open strings account for the charged instanton moduli of the
ADHM construction
\cite{Witten:1995gx,Douglas:1995bn,Green:2000ke,Billo:2002hm}; if
instead also the $\vartheta^i$'s are non vanishing the instantonic
branes represent exotic instantons of truly stringy nature whose
role in the effective low-energy field theory has been recently the
subject of intense investigation
\cite{Blumenhagen:2006xt}\nocite{Ibanez:2006da,Bianchi:2007fx,Argurio:2007vqa,
Bianchi:2007wy,Ibanez:2007rs,Antusch:2007jd,Blumenhagen:2007zk,Aharony:2007pr,Blumenhagen:2007bn,Camara:2007dy,Ibanez:2007tu,GarciaEtxebarria:2007zv,Petersson:2007sc,Bianchi:2007rb,Blumenhagen:2008ji,Argurio:2008jm,Cvetic:2008ws,Kachru:2008wt,GarciaEtxebarria:2008pi}-\cite{Buican:2008qe}. {F}rom these
considerations it is clear that by considering open strings that are
generically twisted we can simultaneously treat all configurations
that are relevant  for the applications mentioned in the
Introduction.

\paragraph{Open String Vertices} Let us now focus on the  R sector
of the open strings at a generic brane intersection. Here the vertex operator for the
lowest fermionic excitation $\Theta_{\mathcal{A}}$ is
\begin{equation}
V_{\Theta}(z) = \mathcal{N}_{\Theta}\,\Theta_{\!\mathcal{A}}
\big[\sigma_{\vec{\vartheta}}\,s_{ \vec{\epsilon}_{\mathcal{A}}+\vec{\vartheta}}\,
\ee^{- {\frac{1}{2}\phi}} \,\ee^{\ii \,k \cdot X}\big](z)
\label{vertexferm}
\end{equation}
where we understand that the momentum $k$ is defined only in untwisted directions.
In this expression the
index $\mathcal{A}=1,\ldots,16$ labels a spinor representation of $\mathrm{SO}(10)$ with definite
chirality and runs over all possible choices of signs in the weight vector
\begin{equation}
\vec{\epsilon}_{\mathcal{A}} = \frac{1}{2}\Big(\pm,\pm,\pm,\pm,\pm\Big)
\label{epsilonA}
\end{equation}
with, say, an odd number of $+$'s, and the symbol $s_{\vec{q}}(z)$ stands for the fermionic
spin field
\begin{equation}
s_{\vec{q}}(z)=\ee^{\ii\sum_Iq^I\varphi^I(z)}
\label{spinfield}
\end{equation}
where $\varphi^I(z)$ are the fields that bosonize the world-sheet fermions according to
$\Psi^I=\ee^{\ii\varphi^I}$ (up to cocycle factors). Finally, $\phi(z)$ is the boson entering the superghost
fermionization formulas, $\sigma_{\vec{\vartheta}}(z)$ is the bosonic twist field introduced
above and $\mathcal{N}_\Theta$ is a normalization factor which will be discussed in the following
sections.

The conformal weight of the vertex operator (\ref{vertexferm}) is
\begin{equation}
h=\frac{k^2}{2}+\frac{1}{2}\sum_I\left[|\vartheta^I|(1-|\vartheta^I|)
+(\epsilon_{\mathcal{A}}^I+\vartheta^I)^2\right]+\frac{3}{8} =
\frac{k^2}{2}+1+\frac{1}{2}\sum_I\left(|\vartheta^I|+2\vartheta^I\epsilon_{\mathcal{A}}^I\right)
\label{confweight}
\end{equation}
and hence $V_{\Theta}$ describes a physical massless fermion $h=1$,
$k^2=0$, when the last term vanishes. This condition restricts the
number of the allowed polarization components of $\Theta$ as follows
\begin{equation}
\Theta_{\!\mathcal{A}}\not = 0 \quad\mbox{only if}
\quad \epsilon_{\mathcal{A}}^I = \left\{  \begin{array}{ll}
 \pm \frac{1}{2}   &  ~~\mbox{for}~~ \vartheta^I=0  \\
  -\frac{1}{2}   &  ~~\mbox{for}~~\vartheta^I > 0\\
  ~~\frac{1}{2}   &  ~~\mbox{for}~~\vartheta^I < 0\\
\end{array}
\right.
\label{theta0}
\end{equation}
For example, when all $\vartheta^I$'s are vanishing we have a chiral
spinor in ten dimensions but if only $\vartheta^4=\vartheta^5=0$
  we have a chiral spinor in the four untwisted directions
along the $(Z^4,Z^5)$ complex plane. On the other hand, in the
instantonic brane constructions mentioned above, for which
$\vartheta^4=\vartheta^5=\frac{1}{2}$, we see from the second line
in (\ref{theta0}) that the R sector describes fermions that do not
carry a spinor index under Lorentz rotations along the ND
four-dimensional plane, in perfect agreement with the ADHM
realization of the charged fermionic instanton moduli.

\paragraph{Closed String Vertices}
We now describe the closed string vertex operators corresponding to background fluxes.
In the closed string sector all fields (both bosonic and fermionic) are untwisted
due to the periodic boundary conditions%
\footnote{Even if in later sections we will consider an orbifold compactification, we will
include background fluxes from the untwisted closed string sector only. The study of the effect of background fluxes from twisted
sectors of the orbifold theory is left to future work.}. However,
in the presence of D-branes a suitable identification
between the left and the right moving components of the closed string has to be
enforced at the boundary and a non-trivial dependence on the angles $\theta_\sigma^I$ appears through
the matrices $R_\sigma$ or $\mathcal{R}_\sigma$.

Let us first consider the R-R sector of the Type IIB theory,
where the physical vertex operators for the field strengths of the
anti-symmetric tensor fields are, in the $(-\frac{1}{2},-\frac{1}{2})$ superghost picture,
\begin{equation}
V_{F}(z,\overline z)=\mathcal{N}_{F}\,F_{\mathcal{AB}}~
\ee^{-\ii\pi\alpha'k_{\mathrm L}\cdot k_{\mathrm R}} \big[s_{
\vec{\epsilon}_{\mathcal{A}}} \,\ee^{-\frac{1}{2}\phi} \,\ee^{\ii
\,k_{\mathrm L} \cdot X}\big](z)
 \times
\big[\widetilde s_{ \vec{\epsilon}_{\mathcal{B}}}
 \,\ee^{-\frac{1}{2}\widetilde\phi} \,\ee^{\ii \,k_{\mathrm R} \cdot \widetilde X}\big](\overline z)~.
\label{vertexRR}
\end{equation}
In this expression $\mathcal{N}_F$ is a normalization factor that will be discussed later,
$k_{\mathrm L}$ and $k_{\mathrm R}$ are the left and right momenta, and the
tilde sign denotes the right-moving components.
Furthermore, the factor $\ee^{-\ii\pi\alpha'k_{\mathrm L}\cdot k_{\mathrm R}}$ is a cocycle that allows
for an off-shell extension of the closed string vertex with $k_{\mathrm L}\not =k_{\mathrm R}$, as discussed in Ref. \cite{Bertolini:2005qh}. The bi-spinor polarization $F_{\mathcal{AB}}$
comprises all R-R field strengths
of the Type IIB theory according to
\begin{equation}
F_{\mathcal{AB}}=\sum_{n=1,3,5}\frac{1}{n!}\,F_{M_1\ldots M_n}
\left(\Gamma^{M_1\ldots M_n}\right)_{\mathcal{AB}}~,
\label{F135}
\end{equation}
even if in our applications only the 3-form part will play a role.
In the presence of D-branes the left and right
moving components of the vertex operator $V_F$ must be identified using the reflection rules
discussed above. In practice (see for example Ref. \cite{Bertolini:2005qh} for more details)
this amounts to set
\begin{equation}
\widetilde X^M(\overline{z}) = (R_0)^M_{~N}\,X^N(\overline{z})
\quad,\quad \widetilde s_{ \vec{\epsilon}_{\mathcal{A}}}(\overline{z})=
(\mathcal{R}_0)^{\mathcal{A}}_{~\,\mathcal B}\,
s_{ \vec{\epsilon}_{\mathcal{B}}}(\overline{z})\quad,\quad\widetilde\phi(\overline{z})=
\phi(\overline{z})
\label{leftright}
\end{equation}
and modify the cocycle factor in the vertex operator (\ref{vertexRR}) to
$\ee^{-\ii\pi\alpha'k_{\mathrm L}\cdot (k_{\mathrm R}R_0)}$.
As a consequence of the identifications (\ref{leftright}),
the R-R field-strength $F_{\mathcal{AB}}$
gets replaced by the bi-spinor polarization $(F\,\mathcal{R}_0)_{\mathcal{AB}}$ that
incorporates also the information on the type of boundary conditions enforced by the D-branes.

Let us now turn to the NS-NS sector of the closed string. Here it is possible to write
an effective BRST invariant vertex operator for the derivatives of the anti-symmetric tensor $B$
that are related to the 3-form flux $H$.
In the $(0,-1)$ superghost picture%
\footnote{This particular asymmetric picture
is chosen in view of the calculations of the disk amplitudes described in
Section \ref{subsec:RR}.}, this effective vertex is
\begin{equation}
V_H(z,\overline{z}) = {\mathcal{N}_H}\,\big(\partial_MB_{NP}\big)\,
\ee^{-\ii\pi\alpha'k_{\mathrm L}\cdot k_{\mathrm R}}
\big[\psi^M\psi^N\ee^{\ii \,k_{\mathrm L} \cdot X}\big](z) \times
\big[\widetilde \psi^P\,\ee^{-\widetilde\phi} \,\ee^{\ii
\,k_{\mathrm R} \cdot \widetilde X}\big] (\overline z)
\label{vertexNS}
\end{equation}
where again we have introduced a normalization factor and a cocycle. When we insert this vertex in
a disk diagram, we must identify the left and right moving sectors using the reflection rules
(\ref{leftright}) supplemented by
\begin{equation}
\widetilde \psi^M(\overline{z}) = (R_0)^M_{~N}\psi^N(\overline{z})~.
\label{leftright1}
\end{equation}
Consequently, in (\ref{vertexNS}) the polarization $(\partial B)$ is effectively replaced by
$(\partial BR_0)$. Notice that the NS-NS polarization combines with the boundary reflection matrix in the
vector representation $R_0$, in contrast to the R-R case
where one finds instead the reflection matrix in the spinor representation $\mathcal{R}_0$.

\subsection{The string correlator with R-R and NS-NS fluxes}
\label{subsec:RR}
We now evaluate the string correlation functions among two massless open string fermions and the
background closed string flux, as represented in Fig. \ref{fig:flux}.
It is a mixed open/closed string amplitude on a disk which, generically, has mixed boundary conditions. From the conformal field theory point of view such fermionic correlation functions are similar to the mixed amplitudes considered in
Ref.s \cite{Billo:2004zq,Lust:2004cx,Billo:2005jw,Bertolini:2005qh}.
Let us analyze first the interaction with the R-R flux.

\paragraph{R-R flux} We take two fermionic open string vertices
(\ref{vertexferm}) and one closed string R-R vertex (\ref{vertexRR}), and compute the amplitude
\begin{equation}
\mathcal{A}_{F}= \Big\langle
V_{\Theta}(x)\,V_F(z,\overline{z})\,V_{\Theta'}(y) \Big\rangle
= c_F~ {\Theta}_{\!\mathcal {A}_1}(F\mathcal{R}_0)_{\mathcal{A}_2\mathcal{A}_3}\,
{\Theta'}_{\!\!\mathcal{A}_4}\,\times\,
A^{\mathcal{A}_1\mathcal{A}_2\mathcal{A}_3\mathcal{A}_4}
\label{amplF}
\end{equation}
where the prefactor
\begin{equation}
c_F =
\mathcal{C}_{(p+1)}\,\mathcal{N}_{\Theta}\,\mathcal{N}_{\Theta'}\,\mathcal{N}_F~,
\label{cF}
\end{equation}
accounts for the normalizations of the vertex operators and the topological
normalization $\mathcal{C}_{(p+1)}$
of any disk amplitude with the boundary conditions of a D$p$-brane \cite{Billo:2002hm,DiVecchia:1996uq}, whose explicit
expression will be given in Section \ref{subsec:masses} for D3-branes and D-instantons,
see Eqs. (\ref{c4}) and (\ref{c0}).
The last factor in (\ref{amplF}) is the 4-point correlator
\begin{equation}
A^{\mathcal{A}_1\mathcal{A}_2\mathcal{A}_3\mathcal{A}_4}
=\int\frac{\prod_{i=1}^4dz_i}{dV_{\mathrm{CKG}}}~\ee^{-\ii\pi\alpha'k_2\cdot k_3}\,
\Big\langle\prod_{i=1}^4
\big[\sigma_{\vec{\vartheta}_i}\,s_{ \vec{\epsilon}_{i}+\vec{\vartheta}_i}\,
\ee^{- {\frac{1}{2}\phi}} \,\ee^{\ii \,k_i \cdot X}\big](z_i)\Big\rangle
\label{amplF1}
\end{equation}
where we have used the convenient notation
\begin{equation}
\begin{aligned}
z_1&=x~\,,~\,z_2=z~\,,~\,z_3=\overline{z}~\,,~\, z_4=y~,\\
k_1&=k~\,,~k_2=k_{\mathrm{L}}~\,,~\,k_3=k_{\mathrm{R}}R_0~\,,~k_4=k'~,\\
\vec{\vartheta}_1&=\vec{\vartheta}~\,,~\,
\vec{\vartheta}_2=0~\,,~\,\vec{\vartheta}_3=0~\,,~\,
\vec{\vartheta}_4=-\vec{\vartheta}~,
\end{aligned}
\label{zi}
\end{equation}
and we have set
$\vec{\epsilon}_i\equiv\vec{\epsilon}_{\mathcal{A}_i}$. Since the
closed string vertex is untwisted, the two open string vertices
must have opposite twists in order to have a non-vanishing
amplitude. This explains the third line above, which, according to
Eq. (\ref{theta0}), implies that when $\vartheta^I\not=0$
the polarizations $\Theta_{\!\mathcal{A}_1}$ and $\Theta'_{\!\mathcal{A}_4}$ are 
not vanishing only if $\epsilon^I_{1}=-\epsilon^I_{4}$. Therefore, if
$\vartheta^I\not=0$ for all $I$'s, the spinor weights
$\vec{\epsilon}_1$ and $\vec{\epsilon}_4$ have different GSO
parity and the amplitude (\ref{amplF}) ceases to exist. To avoid
this, from now on we will assume that
at least one of the $\vartheta^I$'s be vanishing%
\footnote{As pointed out in Ref. \cite{Billo:2007py}
when all five $\vartheta^I$'s are non vanishing, the simplest tree-level diagram involving
massless fermions of the twisted R sector requires at least three different types of boundary conditions
and thus it is not of the type of amplitudes we are discussing here, which involve only two boundary
changing operators.}. The evaluation of the correlator in (\ref{amplF1}) can be simplified by
assuming that the closed string vertex does not carry momentum in the twisted directions ({\it i.e.}
$k_2^I=k_3^I=0$ if $\vartheta^I\not=0$). This is not a restrictive choice for our purposes,
since we will be interested in the effects induced by \emph{constant} background fluxes.

In the correlator (\ref{amplF1}) the open string positions $z_1$ and $z_4$ are integrated
on the real axis while the closed string variables $z_2$ and $z_3$ are integrated in the
upper half complex plane, modulo the $\mathrm{Sl}(2;\mathbb{R})$ projective invariance
that is fixed by the conformal Killing group volume $dV_{\mathrm{CKG}}$. Using this fact we have
\begin{equation}
\frac{\prod_{i=1}^4dz_i}{dV_{\mathrm{CKG}}}= d\omega \,(1-\omega)^{-2} \,\big(z_{14}z_{23}\big)^2
\label{dz}
\end{equation}
where $\omega$ is the anharmonic ratio
\begin{equation}
\omega= \frac{z_{12}z_{34}}{z_{13}z_{24}} \quad\quad(\,|\omega|=1\,)
\label{omega}
\end{equation}
with $z_{ij}=z_i-z_j$. Due to our kinematical configuration, the contribution
of the twist fields and the bosonic exponentials to the correlator (\ref{amplF1}) can be
factorized and becomes
\begin{equation}
\Big\langle
\sigma_{\vec{\vartheta}}(z_1)\,\sigma_{-\vec{\vartheta}}(z_4)\Big\rangle
\,\Big\langle\prod_{i=1}^4 \ee^{\ii \,k_i \cdot X(z_i)}\Big\rangle
= z_{14}^{\,\sum_I\vartheta^I(1-\vartheta^I)}~\omega^{\alpha't}\,(1-\omega)^{\alpha' s}
\label{x}
\end{equation}
where we have used (\ref{OPE}), introduced the two kinematic invariants
\begin{equation}
s=(k_1+k_4)^2=(k_2+k_3)^2\quad\mbox{and}\quad t=(k_1+k_3)^2=(k_2+k_4)^2~,
\label{stu}
\end{equation}
and understood the momentum conservation.

Also the contribution of the spin fields and the
superghosts can be easily evaluated using the bosonization formulas,
that allow to write
\begin{equation}
\Big\langle \prod_{i=1}^4 s_{\vec{\epsilon}_{i}
+\vec{\vartheta}_i}(z_i)\,\ee^{-\frac{1}{2}\phi(z_i)}
\Big\rangle =
\Big\langle \prod_{i=1}^4 s_{\vec{\epsilon}_i}(z_i)
\,\ee^{-\frac{1}{2}\phi(z_i)}\Big\rangle\,\times\,
\prod_{i<j}\,z_{ij}^{\vec{\epsilon}_i
\cdot \vec{\vartheta}_j + \vec{\epsilon}_j
\cdot \vec{\vartheta}_i +\vec{\vartheta}_i\cdot \vec{\vartheta}_j}~.
\label{spinc}
\end{equation}
The first factor in the right hand side is the
four fermion correlator of the Type IIB superstring in ten dimensions which has been computed
for example in Ref. \cite{Friedan:1985ge}, namely
\begin{equation}
\Big\langle \prod_{i=1}^4 s_{\vec{\epsilon}_i}(z_i)
\,\ee^{-\frac{1}{2}\phi(z_i)}\Big\rangle
=\frac{1}{2}\prod_{i<j}z_{ij}^{-1}
\left[z_{13}z_{24}\big(\Gamma_M\big)^{\!\mathcal{A}_1\mathcal{A}_4}
\!\big(\Gamma^M\big)^{\!\mathcal{A}_2\mathcal{A}_3}
\!+\!z_{14}z_{23}\big(\Gamma_M\big)^{\!\mathcal{A}_1\mathcal{A}_3}
\!\big(\Gamma^M\big)^{\!\mathcal{A}_2\mathcal{A}_4}
\right]
\label{4spin}
\end{equation}
where we have understood the ``charge'' conservation $\sum_i \vec{\epsilon}_i=0$.
Furthermore, the $\vartheta$-dependent factor in (\ref{spinc}) can be simplified using
the relations
\begin{equation}
\vec{\epsilon}_2\cdot\vec{\vartheta}=-\vec{\epsilon}_3\cdot\vec{\vartheta}
\quad,\quad\vec{\epsilon}_1\cdot\vec{\vartheta}=
-\vec{\epsilon}_4\cdot\vec{\vartheta}=\frac{1}{2}\sum_I\vartheta^I~,
\label{rel}
\end{equation}
that follow from (\ref{theta0}) and the ``charge'' conservation of the spinor weights. Indeed, using
(\ref{rel}) we have
\begin{equation}
\prod_{i<j}\,z_{ij}^{\vec{\epsilon}_i
\cdot \vec{\vartheta}_j + \vec{\epsilon}_j
\cdot \vec{\vartheta}_i +\vec{\vartheta}_i\cdot \vec{\vartheta}_j}
=z_{14}^{\,\sum_I\vartheta^I(\vartheta^I-1)}~\omega^{-\vec{\epsilon}_3
\cdot\vec{\vartheta}}~.
\label{spin41}
\end{equation}
Collecting everything we find that the amplitude (\ref{amplF1}) can be written as
\begin{equation}
A^{\mathcal{A}_1\mathcal{A}_2\mathcal{A}_3\mathcal{A}_4}
=\big(\Gamma_M\big)^{\!\mathcal{A}_1\mathcal{A}_4}
\!\big(\Gamma^MI_1\big)^{\!\mathcal{A}_2\mathcal{A}_3}
+\big(\Gamma_MI_2\big)^{\!\mathcal{A}_1\mathcal{A}_3}
\!\big(\Gamma^M\big)^{\!\!\mathcal{A}_2\mathcal{A}_4}
\label{ampl2}
\end{equation}
where we have introduced the two $\vec\vartheta$-dependent diagonal matrices with entries
\begin{equation}
\begin{aligned}
\big(I_1\big)_{\!\mathcal{A}_3}^{~\mathcal{A}_3} &=\frac{1}{2}~\ee^{-\frac{\ii\pi\alpha's}{2}}
\int_\gamma d\omega\,(1-\omega)^{\alpha's-1}\,\omega^{\alpha't
-\vec{\vartheta}\cdot \vec{\epsilon}_3-1}~,\\
\big(I_2\big)_{\!\mathcal{A}_3}^{~\mathcal{A}_3} &=\frac{1}{2}~\ee^{-\frac{\ii\pi\alpha's}{2}}
\int_\gamma d\omega\,(1-\omega)^{\alpha's}\,\omega^{\alpha't
-\vec{\vartheta}\cdot \vec{\epsilon}_3-1}~,
\end{aligned}
\label{integrals}
\end{equation}
where $\mathcal{A}_3$ is the spinor index corresponding to the spinor weight $\vec{\epsilon}_3$. 
Here the integrals run around the clockwise oriented unit circle $\gamma: |\omega|=1$,
and can be evaluated to be \cite{Bertolini:2005qh}
\begin{equation}
\begin{aligned}
\big(I_1\big)_{\!\mathcal{A}_3}^{~\mathcal{A}_3}
&=\frac{1}{2}~\ee^{-\frac{\ii\pi\alpha's}{2}}
\Big(\ee^{-2\pi\ii\big(\alpha't-\vec{\vartheta}\cdot
\vec{\epsilon}_3\big)}-1\Big)
\,B\big(\alpha's;\alpha't-\vec{\vartheta}\cdot \vec{\epsilon}_3\big)~,\\
\big(I_2\big)_{\!\mathcal{A}_3}^{~\mathcal{A}_3}
&=\frac{1}{2}~\ee^{-\frac{\ii\pi\alpha's}{2}}
\Big(\ee^{-2\pi\ii\big(\alpha't-\vec{\vartheta}\cdot
\vec{\epsilon}_3\big)}-1\Big)
\,B\big(\alpha's+1;\alpha't-\vec{\vartheta}\cdot
\vec{\epsilon}_3\big)~,
\end{aligned}
\label{integrals1}
\end{equation}
where $B(a;b)$ is the Euler $\beta$-function.
Plugging (\ref{ampl2}) into (\ref{amplF}), with some simple manipulations we find
\begin{equation}
\mathcal{A}_{F}= -c_F\Big[\Theta'\Gamma^M\Theta\,\,\mathrm{tr}\big(
F\mathcal{R}_0I_1\Gamma_M\big)+
\Theta'\Gamma^MF\mathcal{R}_0I_2\Gamma_M\Theta\Big]
\end{equation}
where the trace is understood in the $16\times16$ block spanned by the spinor indices $\mathcal{A}_i$'s.
This expression can be further simplified by expanding the matrices $F\mathcal{R}_0I_{1}$
and $F\mathcal{R}_0I_{2}$ as
\begin{equation}
\big(F\mathcal{R}_0I_{a}\big)_{\mathcal{AB}}= \sum_{n=1,3,5}\frac{1}{n!}\,\big(F\mathcal{R}_0I_{a}\big)_{N_1\ldots N_n}
\left(\Gamma^{N_1\ldots N_n}\right)_{\mathcal{AB}}\quad\quad(a=1,2)~,
\label{FRI}
\end{equation}
and by using the $\Gamma$-matrix identities
\begin{equation}
\begin{aligned}
\tr\big(\Gamma^M\Gamma^N\big)=&16\,\delta^{MN}\quad,\quad
\tr\big(\Gamma^M\Gamma^{N_1N_2N_3}\big)= \tr\big(\Gamma^M\Gamma^{N_1\ldots N_5}\big)=0~,\\
&\Gamma_M\Gamma^{N_1\ldots N_n}\Gamma^M = (-1)^n (10-2n)\,\Gamma^{N_1\ldots N_n}~.
\end{aligned}
\label{identities}
\end{equation}
After some straightforward algebra we find
\begin{equation}
\mathcal{A}_{F}= -8c_F\Theta'\Gamma^M\Theta\,\big[F\mathcal{R}_0(2I_{1}-I_2)\big]_M
+\frac{4c_F}{3!}\Theta'\Gamma^{MNP}\Theta\,\big[F\mathcal{R}_0I_2\big]_{MNP}~.
\label{amplFfinal}
\end{equation}
This formula is one of the main results of this section. It describes the tree-level bilinear
fermionic couplings induced by R-R fluxes on a general brane intersection.

\paragraph{NS-NS flux}
Let us now turn to the fermionic couplings induced by the NS-NS 3-form flux effectively
described by the vertex operator (\ref{vertexNS}). Such couplings arise from
the following mixed disk amplitude
\begin{equation}
\mathcal{A}_{H}= \Big\langle
V_{\Theta}(x)\,V_H(z,\overline{z})\,V_{\Theta'}(y) \Big\rangle
= {c_H}~ {\Theta}_{\!\mathcal A}(\partial B{R}_0)_{{MNP}}\,
{\Theta'}_{\!\mathcal B}\,\times\,
A^{\mathcal{AB};MNP}
\label{amplH}
\end{equation}
where the normalization factor is
\begin{equation}
c_H =
\mathcal{C}_{(p+1)}\,\mathcal{N}_{\Theta}\,\mathcal{N}_{\Theta'}\,\mathcal{N}_H
\label{cH}
\end{equation}
and the 4-point correlator is
\begin{eqnarray}
A^{\mathcal{AB};MNP} &=&\int\frac{\prod_{i=1}^4dz_i}{dV_{\mathrm{CKG}}}
~\ee^{-\ii\pi\alpha'k_2\cdot k_3}\,
\Big\langle
\sigma_{\vec{\vartheta}}(z_1)\,\sigma_{-\vec{\vartheta}}(z_4)\Big\rangle
\,\Big\langle\prod_{i=1}^4 \ee^{\ii \,k_i \cdot X(z_i)}\Big\rangle
\label{amplH1}\\
&&\times
\Big\langle
s_{\vec{\epsilon}_{\mathcal A}+\vec{\vartheta}}(z_1)
\,\psi^M\!\psi^N(z_2)\,\psi^P(z_3)\,
s_{\vec{\epsilon}_{\mathcal B}-\vec{\vartheta}}(z_4)
\Big\rangle
\Big\langle
\ee^{- {\frac{1}{2}\phi(z_1)}}
\ee^{- {\phi(z_3)}}
\ee^{- {\frac{1}{2}\phi(z_4)}}\Big\rangle~.
\nonumber
\end{eqnarray}
Here we have used a notation similar to that of Eq. (\ref{zi}) for the
bosonic and twist fields, whose contribution is the
same as in Eq. (\ref{x}) because of our kinematical configuration.
Due to the Lorentz structure of the fermionic correlator, the
second line of (\ref{amplH1}) can be written as
\begin{equation}
\begin{aligned}
\Big\langle
s_{\vec{\epsilon}_{\mathcal A}+\vec{\vartheta}}(z_1)
\,\psi^M\!&\psi^N(z_2)\,\psi^P(z_3)\,
s_{\vec{\epsilon}_{\mathcal B}-\vec{\vartheta}}(z_4)
\Big\rangle
\Big\langle
\ee^{- {\frac{1}{2}\phi(z_1)}}
\ee^{- {\phi(z_3)}}
\ee^{- {\frac{1}{2}\phi(z_4)}}\Big\rangle\\
&= f(z_{ij}) \big(\Gamma^{MNP}\big)^{\!\mathcal{A}\mathcal{B}}
+g(z_{ij}) \Big[\delta^{MP}\big(\Gamma^{N}\big)^{\!\mathcal{A}\mathcal{B}}
-\delta^{NP}\big(\Gamma^{M}\big)^{\!\mathcal{A}\mathcal{B}}\Big]
\end{aligned}
\end{equation}
where the two functions $f$ and $g$ can be determined, for example,
by using the bosonization technique.
If we pick a configuration such that the field $\psi^M\!\psi^N(z_2)$
can be bosonized as $\ee^{\ii\vec{\epsilon}_2\cdot\vec{\varphi}}$
with weight vectors of the form
\begin{equation}
\vec{\epsilon}_2=\big(0,\ldots,\pm1,\ldots,\pm1,\ldots,0)~,
\label{epsilon2}
\end{equation}
corresponding to roots of $\mathrm{SO}(10)$,
we can use the same strategy we have described before in the R-R case to find
\begin{equation}
\begin{aligned}
f(z_{ij}) &= \prod_{i<j}z_{ij}^{-1}\times \big(z_{14}z_{23}\big)\times
z_{14}^{\,\sum_I\vartheta^I(\vartheta^I-1)}~\omega^{-\vec{\epsilon}_3
\cdot\vec{\vartheta}}~,\\
g(z_{ij})  &= \prod_{i<j}z_{ij}^{-1}\times\big(z_{12}z_{34}+z_{13}z_{24}\big)\times
z_{14}^{\,\sum_I\vartheta^I(\vartheta^I-1)}~\omega^{-\vec{\epsilon}_3
\cdot\vec{\vartheta}}~,
\end{aligned}
\end{equation}
where the last factors are the same as in Eq. (\ref{rel}) and
$\vec{\epsilon}_3$ is the weight vector in the vector representation associated to $\psi^P(z_3)$, of the form
\begin{equation}
\vec{\epsilon}_3=\big(0,\ldots,\pm1,\ldots,0)~.
\label{epsilon3}
\end{equation}
Collecting everything, and introducing the diagonal matrices (with vector indices) $(I_1)^P_{~P}$ and $(I_2)^P_{~P}$
defined analogously to Eq. (\ref{integrals}), after some simple manipulations we obtain
\begin{equation}
\mathcal{A}_{H}= -{4c_H}\Theta'\Gamma^N\Theta \,\delta^{MP}\,\big[\partial BR_0(2I_{1}-I_2)\big]_{[MN]P}
+{2c_H}\Theta'\Gamma^{MNP}\Theta\,\big[\partial B R_0I_2\big]_{MNP}
\label{amplHfinal}
\end{equation}
which is the NS-NS counterpart of the R-R amplitude (\ref{amplFfinal}) on a generic D brane intersection
and shares with it the same type of fermionic structures.

\section{Flux couplings with untwisted open strings ($\vec{\vartheta}=0$)}
\label{sec:effects}
We now exploit the results obtained in the previous section to analyze how constant background fluxes
couple to untwisted open strings, {\it i.e.} strings starting and ending on a single stack of D-branes.
This corresponds to set $\vec{\vartheta}=0$ in all previous formulas which drastically simplify.
Note that the condition $\vec{\vartheta}=0$ implies that $\vec{\theta}_0=\vec{\theta}_\pi$, so that the
reflection rules are the same at the two string end-points. We can
distinguish two cases, namely when these reflection rules are just signs
({\it i.e.} $\theta_\sigma^I=0$ or $1$) and when they instead depend on generic angles $\theta_\sigma^I$.
In the first case the branes are unmagnetized,  while the second corresponds to
magnetized branes.

Since we are interested in constant background fluxes, we can set the momentum of the closed string
vertices to zero; this corresponds to take the limit $s=-2t \to 0$ in the integrals (\ref{integrals1})
which yields
\begin{equation}
 2I_1=I_2= -\ii\pi~.
\label{intr0}
\end{equation}
Using this result in the R-R and NS-NS amplitudes (\ref{amplFfinal}) and (\ref{amplHfinal}),
we see that the fermionic couplings with a single $\Gamma$ matrix vanish and only the terms with three
$\Gamma$'s survive, so that the total flux amplitude is
\begin{equation}
\mathcal A \equiv \mathcal{A}_F + \mathcal{A}_H =-2\pi\ii\,
\Theta\Gamma^{MNP}\Theta\,\Big[\frac{c_F}{3}\big(F\mathcal{R}_0\big)_{MNP}
+c_H\big(\partial B R_0\big)_{MNP}\Big]~. \label{ampltot}
\end{equation}
Here we used the fact that the untwisted fermions $\Theta$ and
$\Theta'$ in (\ref{amplFfinal}) and (\ref{amplHfinal}) actually
describe the same field and only differ because they carry opposite
momentum. For this reason we multiplied the above amplitudes by a
symmetry factor of 1/2 and dropped the $'$ without introducing
ambiguities.

It is clear from Eq. (\ref{ampltot}) that once the flux configuration is given, the structure of the
fermionic couplings for different types of D-branes depends crucially
on the boundary reflection matrices $R_0$ and $\mathcal R_0$.
Notice that the R-R piece of the amplitude (\ref{ampltot}) is generically non zero for 1-form, 3-form
and 5-form fluxes. However, from now on we will restrict our analysis only to the 3-form  and hence the
bi-spinor to be used is simply
\begin{equation}
F_{\mathcal{AB}} = \frac{1}{3!} F_{MNP}
\left(\Gamma^{MNP}\right)_{\mathcal{AB}}~.
\label{F3}
\end{equation}
We can now specify better
the normalization factors $c_F$ and $c_H$. In fact the vertex (\ref{vertexRR}) for a R-R 3-form and the NS-NS vertex (\ref{vertexNS})
should account for the following quadratic terms of the bulk theory
in the ten-dimensional Einstein frame:
\begin{equation}
\frac{1}{2\kappa_{10}^2}\int d^{10}x\,\sqrt{g_{(E)}}\,\Big(\frac{1}{3!}\,\ee^{\varphi}F^2
+ \frac{1}{3!}\,\ee^{-\varphi}dB^2\Big)~,
\label{bulk10}
\end{equation}
where $\varphi$ is the dilaton and $\kappa_{10}$ is the gravitational Newton constant in ten dimensions.
In order to reproduce the above dilaton dependence, the normalization factors $\mathcal{N}_F$
and $\mathcal{N}_H$ of the R-R and NS-NS vertex operators must scale with the string coupling $g_s
=\ee^\varphi$ as
\begin{equation}
\mathcal{N}_F \sim g_s^{1/2}\quad\mbox{and}\quad\mathcal{N}_H\sim g_s^{-1/2}~,
\label{NfNh}
\end{equation}
so that from Eqs. (\ref{cF}) and (\ref{cH}) we obtain
\begin{equation}
c_H=c_F/g_s~.
\label{cfch}
\end{equation}
Taking all this into account, we can rewrite the total amplitude (\ref{ampltot}) as
\begin{equation}
\mathcal A \equiv \mathcal{A}_F + \mathcal{A}_H =-\frac{2\pi\ii}{3!}\,c_F\,
\Theta\Gamma^{MNP}\Theta\,
T_{MNP}
\label{ampltot1}
\end{equation}
where
\begin{equation}
T_{MNP} = \big(F\mathcal{R}_0\big)_{MNP}+\frac{3}{g_s}\,
\big( \partial B R_0\big)_{[MNP]} ~.
\label{tmnl}
\end{equation}

Up to now we have used a ten-dimensional notation. However, since we are interested
in studying the flux induced couplings for gauge theories and instantons in four dimensions, it becomes
necessary to split the indices $M,N,\ldots=0,1,\ldots,9$ appearing in the above equations into four-dimensional space-time indices $\mu,\nu,\ldots=0,1,2,3$, and six-dimensional indices
$m,n,\ldots=4,5,\ldots,9$ labeling the directions of the internal space
(which we will later take to be compact, for example a 6-torus $\mathcal T_6$ or an orbifold thereof).
Clearly background fluxes carrying indices along the space-time break the four-dimensional
Lorentz invariance and generically give rise to deformed gauge theories. Effects of this
kind have already been studied using world-sheet techniques in
Refs. \cite{Billo:2004zq,Billo:2005jw} where
a non vanishing R-R 5-form background of the type $F_{\mu\nu mnp}$ was shown to originate the
${\mathcal N}\!=\!1/2$ gauge theory, and in Ref. \cite{Billo:2006jm} where the so-called $\Omega$
deformation of the $\mathcal N =2$ gauge theory was shown to derive from a R-R 3-form flux of
the type ${F}_{\mu\nu m}$. In the following, however, we will consider only internal fluxes,
like $F_{mnp}$ or $(\partial B)_{mnp}$, which preserve the four-dimensional Lorentz invariance, and
the fermionic amplitudes we will compute are of the form
\begin{equation}
\mathcal A =-\frac{2\pi\ii}{3!}\,c_F\,\Theta\Gamma^{mnp}\Theta\,T_{mnp}
\label{ampltot2}
\end{equation}
We now analyze the structure of these couplings beginning with the simplest case of space-filling
unmagnetized D-branes; later we examine
unmagnetized Euclidean branes and finally branes with a non-trivial world-volume magnetic field.

\subsection{Unmagnetized D-branes}
\label{subsec:unmagnetizedD}

Even if the fermionic couplings (\ref{ampltot1}) have been derived in Section \ref{sec:CFT} assuming a
Euclidean signature, when we discuss space-filling D-branes with $\vec{\vartheta}=0$, the
rotation to a Minkowskian signature poses no problems. In this case,
$\Theta$ becomes a Majorana-Weyl spinor in ten dimensions which in particular satisfies
\begin{equation}
\Theta\Gamma^{mnp}\Theta =-\,\big(
\Theta\Gamma^{mnp}\Theta\big)^*~.
\label{cc}
\end{equation}
Furthermore for an unmagnetized D$p$-brane that fills the four-dimensional Minkowski space
and possibly extends also in some internal directions, the reflection matrices $R_0$
and $\mathcal R_0$ are very simple: indeed in the vector representation
\begin{equation}
R_0 = \mathrm{diag}(\pm {1},\pm 1, \ldots)~,
\label{r0vec}
\end{equation}
where the entries specify whether a direction is longitudinal ($+$) or transverse ($-$), while in the spinor representation
\begin{equation}
\mathcal R_0 =  \Gamma^{p+1}\cdots\Gamma^{9}~.
\label{r0spin}
 \end{equation}
Using these matrices we easily see that $T_{mnp}$ is a {\it real} tensor, so that
in view of Eq. (\ref{cc}) also the total fermionic amplitude (\ref{ampltot2})
is real, as it should be.

The explicit expression of $T_{mnp}$ is particularly simple in the case of brane configurations
which respect the $4+6$ structure of the space-time, {\it i.e.} D3- and D9- branes. For space-filling
D3-branes all internal indices are transverse, so that $\left.R_0\right|_{\mathrm{int}}=-1$
and $\mathcal R_0=\Gamma^4\cdots\Gamma^9$. From Eq. (\ref{tmnl}) it follows then
\begin{equation}
T_{mnp}= (*_6 F)_{mnp} -\frac{1}{g_s}\,H_{mnp}
\label{td3}
\end{equation}
where $*_6$ denotes the Poincar\`{e} dual in the six-dimensional internal space and
$H=dB$
\footnote{In our conventions $(*_6 F)_{mnp}= \frac{1}{3!}\epsilon_{mnprst}\,F^{rst}$ and $H_{mnp}=
3\partial_{[m} B_{np]}= \big(\partial_m B_{np}+\partial_n B_{pm}+\partial_p B_{mn}\big)$.}.

For D9-branes, instead, all internal indices are longitudinal, and to emphasize this fact  we denote them
as $\hat m, \hat n, \ldots$ In this case we simply have $R_0=1$ and $\mathcal R_0=1$
so that
\begin{equation}
T_{\hat m\hat n \hat p}= F_{\hat m\hat n \hat p}+\frac{1}{g_s}\,H_{\hat m \hat n \hat p}~.
\label{td9}
\end{equation}
Note however that D9-branes must always be accompanied by orientifold 9-planes (O9)
for tadpole cancellation and that the corresponding orientifold projection kills the
NS-NS flux $H_{\hat m \hat n \hat p}$. If we take this fact into account,
the coupling tensor for D9-branes reduces to
\begin{equation}
T_{\hat m\hat n \hat p}= F_{\hat m\hat n \hat p}~.
\label{td91}
\end{equation}

The case of space-filling D7- and D5-branes is slightly more involved since for these branes
the internal directions are partially longitudinal and partially transverse. In particular,
for D7-branes the longitudinal internal indices $\hat m, \hat n\ldots$ take four values while the transverse indices $p,q,\ldots$ take two values. Eq.~(\ref{tmnl}) implies then that the only non vanishing
components of the $T$ tensor for D7-branes are
\begin{equation}
T_{\hat m \hat n \hat p}=\frac{1}{g_s}\,H_{\hat m \hat n \hat p}\quad,\quad
T_{\hat m \hat n p} = F_{\hat m \hat n}^{\phantom{\hat m \hat n}\!q}\,\epsilon_{qp}
+\frac{1}{g_s}\,H_{\hat m \hat n p}
\quad\mbox{and}\quad T_{\hat mnp} = -\frac{1}{g_s}\,H_{\hat mnp}~.
\label{tD7}
\end{equation}
If one introduces O7-planes to cancel the tadpoles produced by the D7-branes, one can see that
the corresponding orientifold projection\footnote{The extra $(-1)^{F_L}$ appearing in the case of
O3/O7-planes ensure that the corresponding orientifold actions square to one, i.e.
$(\Omega I_{4n+2} (-1)^{F_L})^2=I^2_{4n+2} (-1)^{F_L+F_R}=1$. } $\Omega I_2 (-1)^{F_L}$
removes all $F$ and $H$ components with an even number of
transverse indices so that the only surviving couplings are
\begin{equation}
T_{\hat m \hat n p} = F_{\hat m \hat n}^{\phantom{\hat m \hat n}\!q}\,\epsilon_{qp}
+\frac{1}{g_s}\,H_{\hat m \hat n p}~.
\label{tD7a}
\end{equation}

For D5-branes the situation is somehow complementary, since
the longitudinal internal indices take two values while the transverse ones run over four values.
In this case one can show that
the non vanishing components of the $T$ tensor are
\begin{equation}
T_{\hat m \hat n p}=\frac{1}{g_s}\,H_{\hat m \hat n p}\quad,\quad
T_{\hat m np}=-\frac{1}{2}\,F_{\hat m}^{\phantom{\hat m}\!qr}\epsilon_{qrnp}
-\frac{1}{g_s}\,H_{\hat m np}
\quad\mbox{and}\quad T_{mnp} = -\frac{1}{g_s}\,H_{mnp}~.
\label{tD5}
\end{equation}
Again the O5-planes required for tadpole cancellation enforce an orientifold projection $\Omega I_4$
which removes the components of $H$($F$) with an even(odd) number of transverse indices.
Thus, the coupling $T_{\hat m np}$ reduces to
\begin{equation}
T_{\hat m np}=-\frac{1}{2}\,F_{\hat m}^{\phantom{\hat m}\!qr}\epsilon_{qrnp}~.
\label{tD5a}
\end{equation}
The fermionic couplings for the various D-branes we have discussed,
taking into account the appropriate orientifold projections, are summarized in Tab.~\ref{Dbranes}.
\begin{table}[ht]
\centering
\begin{tabular}{cccccccccccc}
\hline\hline
  & \phantom{\vdots}\!0\!&\!1\!&\!2\!&\!3\!&\!4\!&\!5\!&\!6\!&\!7\!&\!8\!&\!9\! & $T_{mnp}$ \\ [1ex]
 \hline
D3     & \phantom{\vdots}\!$-$\!&\!$-$\!&\!$-$\!&\!$-$\!
&\!$\times$\!&\!$\times$\!&\!$\times$\!&\!$\times$\!&\!$\times$\!&\!$\times$\! &
$(*_6F)_{mnp} - \frac{1}{g_s}H_{mnp} $
\\ [1ex]
D5    & \phantom{\vdots}\!$-$\!&\!$-$\!&\!$-$\!&\!$-$\!
&\!$-$\!&\!$-$\!&\!$\times$\!&\!$\times$\!&\!$\times$\!&\!$\times$\! &
$\frac{1}{g_s}\,H_{\hat m \hat n p}~\,;~
-\frac{1}{2}\,F_{\hat m}^{\phantom{\hat m}\!qr}\epsilon_{qrnp}~\,;~
-\frac{1}{g_s}\,H_{mnp}
$
\\[1ex]
D7    & \phantom{\vdots}\!$-$\!&\!$-$\!&\!$-$\!&\!$-$\!
&\!$-$\!&\!$-$\!&\!$-$\!&\!$-$\!&\!$\times$\!&\!$\times$\! &
$F_{\hat m \hat n}^{\phantom{\hat m \hat n}\!q}\,\epsilon_{qp}
+\frac{1}{g_s}\,H_{\hat m \hat n p}
$
\\[1ex]
D9    & \phantom{\vdots}\!$-$\!&\!$-$\!&\!$-$\!&\!$-$\!
&\!$-$\!&\!$-$\!&\!$-$\!&\!$-$\!&\!$-$\!&\!$-$\! &
$F_{\hat m\hat n \hat p}
$
\\[1ex]
\hline\hline
\end{tabular}
\caption{Structure of the fermionic couplings $T$ induced by background fluxes on D3, D5, D7 and
D9-branes after taking into account the appropriate orientifold projections;
longitudinal internal directions are labeled by $\hat m, \hat n, \ldots$ and
internal transverse ones by $m, n, \ldots$}
 \label{Dbranes}
 \end{table}

These results clearly exhibit the fact that the R-R and NS-NS 3-form fluxes do
not appear on equal footing in the effective couplings $T$.  This is due to the
different ${\cal R}_0$ and $R_0$ reflection matrices
entering in the definition of the R-R and NS-NS vertex operators as discussed in Section \ref{sec:CFT}.
It is interesting to observe in Tab. \ref{Dbranes} that,
while for D9- and D5- branes the fermionic couplings depend either on $F$ or on $H$,
for D3- and D7-branes they depend on a combination of the R-R and NS-NS fluxes. This
follows from the fact that O3- and O7-planes act on the same way on R-R and NS-NS 3-forms.
By introducing the complex 3-form\footnote{Self-duality of type IIB can be used to
promote this expression to its $SL(2,\mathbb{Z})$-covariant version $G=F-\tau H$ with
$\tau=C_0-\ii \ee^{-\varphi}$. A direct evaluation of the $C_0$-dependent term however requires
a string amplitude involving two closed and two open string insertions in the disk. }
\begin{equation}
G=F-\frac{\ii}{g_s}\,H~,
\label{G}
\end{equation}
it is possible to rewrite the D3 brane coupling (\ref{td3}) as
\begin{equation}
T_{mnp}=(*_6 F)_{mnp} - \frac{1}{g_s}\,H_{mnp} =  \re\big( \!*_6 \!G-\ii G\big)_{mnp}~.
\label{tD31}
\end{equation}
Thus our explicit conformal field theory calculation confirms
that an imaginary self-dual (ISD) 3-form flux $G$ does not couple to
unmagnetized D3-branes, a well-known result that has been previously obtained using
purely supergravity methods
\cite{Grana:2002tu,Marolf:2003ye,Camara:2003ku,Camara:2004jj,Martucci:2005rb}.

Also the fermionic couplings (\ref{tD7a}) for the D7 branes can be written in terms of the
3-form flux $G$.
Indeed, introducing a complex notation and denoting as $i$ and $\overline i$
the complex directions of the plane transverse to the D7-branes
(sometimes in the literature also called D$7_i$-branes), we have
\begin{equation}
T_{\hat m \hat n i}= \ii \,G_{\hat m \hat n i}\quad\mbox{and}\quad
T_{\hat m \hat n \overline{i}}= -\ii \,G^*_{\hat m \hat n \overline{i}}~,
\label{tD71}
\end{equation}
in agreement with the structure of soft fermionic mass terms found
in Ref. \cite{Camara:2004jj}.

\subsection{Unmagnetized Euclidean branes}
\label{subsec:unmagnetizedE}

Euclidean branes that are transverse to the four-dimensional space-time and extend partially
or totally in the internal directions are relevant to discuss non-perturbative instanton effects
in the framework of branes models. In this case, to treat consistently the flux induced couplings
it is necessary to work in a space with Euclidean signature as we have done in
Section \ref{sec:CFT}. Then, the massless fermions $\Theta$ cannot satisfy a Majorana condition,
and relations like (\ref{cc}) do not hold any more. On the other hand, in Euclidean space
there is no issue about the reality of a fermionic amplitude and, as we will see, also the
coupling tensor $T$ is in general complex.

Let us begin by considering the D-instantons (or D$(-1)$-branes)
for which all ten directions are transverse. In this case we have
\begin{equation}
R_0=-1\quad\mbox{and}\quad {\mathcal R}_0=\Gamma^0\Gamma^1\cdots\Gamma^9 \equiv
\ii \,\Gamma_{(11)}^{\mathrm E}
\label{rd-1}
\end{equation}
where $\Gamma_{(11)}^{\mathrm E}$ is the chirality matrix in ten Euclidean dimensions. Thus,
recalling our chirality choice for the spinors $\Theta$, we easily see that for D-instantons
the $T$ tensor (\ref{tmnl}) is simply
\begin{equation}
T_{mnp}= -\ii\, F_{mnp} - \frac{1}{g_s}\,H_{mnp} = -\ii\,G_{mnp}~.
\label{tD-1}
\end{equation}

Let us now turn to Euclidean instantonic 5-branes (or E5-branes)
extending in the six internal directions.
In this case the reflection matrix  in the vector representation entering in
the fermionic coupling $T$ is $R_0|_{\rm int}=1$ along the internal
directions while the matrix in the spinor representation is
\begin{equation}
\mathcal R_0 = \Gamma^0\Gamma^1\Gamma^2\Gamma^3
= -\ii \,\Gamma^4\cdots\Gamma^9\Gamma_{(11)}^{\mathrm E}~.
\label{re5}
\end{equation}
Therefore, for unmagnetized E5-branes we obtain from Eq. (\ref{tmnl})
\begin{equation}
T_{\hat m\hat n\hat p}= \ii\, (*_6 F)_{\hat m\hat n\hat p} + \frac{1}{g_s}\,H_{\hat m\hat n\hat p}
\label{tE5}
\end{equation}
where we have used the same index notation introduced in the previous subsection.
The above coupling simply reduces to
\begin{equation}
T_{\hat m\hat n\hat p}= \ii\, (*_6 F)_{\hat m\hat n\hat p}
\label{tE5a}
\end{equation}
in an orientifold model with O9-planes.

In the literature some attention has been devoted also to
Euclidean 3-branes (or E3-branes) extending along four of the six
internal directions \cite{Tripathy:2005hv,Bergshoeff:2005yp}.
These branes have some similarity with the D7-branes considered in
the previous subsection, and thus our discussion can follow the
same path. Using again the convention of splitting the internal
indices into longitudinal (hatted) and transverse (unhatted) ones,
we can show that the flux-induced fermionic couplings on E3-branes
are
\begin{equation}
T_{\hat m \hat n \hat p} = \frac{1}{g_s}\,H_{\hat m \hat n \hat p}\quad,\quad
T_{\hat m \hat n p} = -\frac{\ii}{2}\,\epsilon_{\hat m \hat n \hat r \hat s}\,
F^{\hat r \hat s}_{\phantom{\hat r \hat s}p}
+\frac{1}{g_s}\,H_{\hat m \hat n p}
\quad\mbox{and}\quad T_{\hat m np} = -\frac{1}{g_s}\,H_{\hat m np}~.
\label{tE3}
\end{equation}
If we consider the appropriate orientifold projections, which in this case remove both
$H_{\hat m \hat n \hat p}$ and $H_{\hat m np}$, we see that the only non-vanishing coupling is
\begin{equation}
T_{\hat m \hat n p} = -\frac{\ii}{2}\,\epsilon_{\hat m \hat n \hat r \hat s}\,
F^{\hat r \hat s}_{\phantom{\hat r \hat s}p}
+\frac{1}{g_s}\,H_{\hat m \hat n p}~.
\label{tE3a}
\end{equation}
This is in perfect agreement with the result of Refs. \cite{Tripathy:2005hv,Bergshoeff:2005yp}
that has been derived with pure supergravity methods.
To make the comparison easier, we observe that the E3-fermionic terms can be rewritten as
\begin{equation}
\Theta \Gamma^{\hat m \hat n p}\Theta\,T_{\hat m\hat n p} =
\Theta \Gamma^{\hat m \hat n p}\,{\widetilde G}_{\hat m\hat n p}\,\Theta
\label{aE3}
\end{equation}
where
\begin{equation}
{\widetilde G} = \frac{1}{g_s}\, H + \ii\,F \,\gamma_{(5)}
\label{e33}
\end{equation}
is the flux combination that is usually introduced in this case, with
$\gamma_{(5)}$ being the chirality matrix for the
four-dimensional brane world-volume. We further remark
that our general formula (\ref{ampltot1}) accounts for all flux-induced fermionic terms of the E3-brane
effective action discussed in Refs. \cite{Tripathy:2005hv,Bergshoeff:2005yp} including those
which break the Lorentz invariance in the first four directions.

For completeness we also mention that the fermionic couplings for the Euclidean 1-branes (or E1-branes)
are given by
\begin{equation}
T_{\hat m \hat n p}=\frac{1}{g_s}\,H_{\hat m \hat n p}\quad,\quad
T_{\hat m  n p} = -\ii\,\epsilon_{\hat m \hat q}\,F^{\hat q}_{\phantom{\hat q}np}
-\frac{1}{g_s}\,H_{\hat m n p}
\quad\mbox{and}\quad T_{mnp} = -\frac{1}{g_s}\,H_{mnp}~;
\label{tE1}
\end{equation}
note that $H_{\hat m n p}$ is removed by the orientifold projection
when the E1-branes are considered together with D5/D9-branes and the corresponding orientifold planes.
The structure of the various fermionic couplings for the instantonic branes discussed above is
summarized in Tab.~\ref{Ebranes}.
\begin{table}[ht]
\centering
\begin{tabular}{cccccccccccc}
\hline\hline
  & \phantom{\vdots}\!0\!&\!1\!&\!2\!&\!3\!&\!4\!&\!5\!&\!6\!&\!7\!&\!8\!&\!9\! & $T_{mnp}$ \\ [1ex]
 \hline
D$(-1)$     & \phantom{\vdots}\!$\times$\!&\!$\times$\!&\!$\times$\!&\!$\times$\!
&\!$\times$\!&\!$\times$\!&\!$\times$\!&\!$\times$\!&\!$\times$\!&\!$\times$\! &
$-\ii F_{mnp} - \frac{1}{g_s}H_{mnp} $
\\ [1ex]
E1    & \phantom{\vdots}\!$\times$\!&\!$\times$\!&\!$\times$\!&\!$\times$\!
&\!$-$\!&\!$-$\!&\!$\times$\!&\!$\times$\!&\!$\times$\!&\!$\times$\! &
$\frac{1}{g_s}\,H_{\hat m \hat n p}~\,;~
-\ii\,\epsilon_{\hat m \hat q}\,F^{\hat q}_{\phantom{\hat q}np}~\,;~
-\frac{1}{g_s}\,H_{mnp}
$
\\ [1ex]
E3    & \phantom{\vdots}\!$\times$\!&\!$\times$\!&\!$\times$\!&\!$\times$\!
&\!$-$\!&\!$-$\!&\!$-$\!&\!$-$\!&\!$\times$\!&\!$\times$\! &
$-\frac{\ii}{2}\,\epsilon_{\hat m \hat n \hat r \hat s}\,
F^{\hat r \hat s}_{\phantom{\hat r \hat s}p}
+\frac{1}{g_s}\,H_{\hat m \hat n p}$
\\[1ex]
E5    & \phantom{\vdots}\!$\times$\!&\!$\times$\!&\!$\times$\!&\!$\times$\!
&\!$-$\!&\!$-$\!&\!$-$\!&\!$-$\!&\!$-$\!&\!$-$\! &
$\ii\,(*_6F)_{\hat m \hat n \hat p}$
\\[1ex]
\hline\hline
\end{tabular}
\caption{Structure of the fermionic couplings $T$ induced by background fluxes on
D$(-1)$, E1, E3 and E5 instantonic branes
after taking into account the appropriate orientifold projections;
the longitudinal internal directions are labeled by $\hat m, \hat n, \ldots$, the
internal transverse ones by $m, n, \ldots$ .}
\label{Ebranes}
\end{table}

We conclude our analysis by observing that in presence of E-branes,
the spacetime filling D$p$-branes live in the Euclidean
ten-dimensional space. Still, the couplings of such D$p$-branes are
again given by the same linear combinations of $F$ and $H$ like in
the Minkowskian case considered in last section, since $R_0$ and
${\cal R}_0$ are trivial along the would be time direction.

\subsection{Magnetized branes}
\label{subsec:magnetized}

The results of the previous subsections can be generalized in a rather straightforward way
to branes with a non-trivial magnetization on their world-volume for which
the longitudinal coordinates satisfy non-diagonal boundary conditions.
Indeed we can start from the same brane configurations we have analyzed before, introduce a
world-volume gauge field $A$ that couples to the open string end-points and obtain a
magnetization ${\mathcal F}_0={\mathcal F}_\pi= 2\pi\alpha'(dA)$. In this way we can use the
same R-R and NS-NS background fluxes of the previous subsections and simply study the new
couplings induced by the world-volume magnetization through the reflection matrices
$R_0$ and $\mathcal R_0$ given in Eqs. (\ref{R}) and (\ref{rspinor}).

As an example we briefly discuss the case of the magnetized E5 branes which play an important
role in the instanton calculus of the gauge theory engineered with wrapped D9-branes and O9-planes \cite{Billo:2007sw,Billo:2007py}. Adopting the same index notation as before, one can
easily realize that the spinor reflection matrix (\ref{rspinor}) for a magnetized E5 brane
can be written in the real basis as
\begin{equation}
\mathcal R_0= \Gamma^0\cdots\Gamma^3\,\mathcal U_0 = -\ii
\,\Gamma^4\cdots\Gamma^9\Gamma_{(11)}^{\mathrm E}\,\mathcal U_0
\label{rrm}
\end{equation}
where
\begin{equation}
\mathcal U_0 = \frac{1}{\sqrt{\det(1-{\mathcal F}_0)}}
~;\,\ee^{\frac{1}{2}({\mathcal F}_0)_{\hat m \hat n}
\Gamma^{\hat m\hat n}}\,;
\label{uu}
\end{equation}
in which the symbol $;\cdots ;$ means antisymmetrization on the vector indices of the $\Gamma$'s,
so that only a finite number of terms appear in the expansion of the exponential.
In our case we explicitly have
\begin{equation}
\begin{aligned}
;\,\ee^{\frac{1}{2}({\mathcal F}_0)_{\hat m \hat n}
\Gamma^{\hat m\hat n}}\,;~
= \,\,&1 \,+\,\frac{1}{2}\, ({\mathcal F}_0)_{\hat m \hat n}
\Gamma^{\hat m\hat n}
\,+\,\frac{{\ii}}{16} \,
({\mathcal F}_0)^{\hat m \hat n}({\mathcal F}_0)^{\hat p \hat q}
\epsilon_{\hat m \hat n \hat p \hat q \hat r \hat s}\,
\Gamma^{\hat r \hat s}\Gamma_{(7)}
\\
&
- \frac{{\rm i}}{3! \cdot 8} ({\mathcal F}_0)^{\hat m \hat n}
({\mathcal F}_0)^{\hat p \hat q}({\mathcal F}_0)^{\hat r \hat s}
\epsilon_{\hat m \hat n \hat p \hat q \hat r \hat s} \,\Gamma_{(7)}
\end{aligned}
\label{uu5}
\end{equation}
where $\Gamma_{(7)}=\ii\Gamma^4\ldots\Gamma^9$ is the chirality matrix of the E5-brane world volume.
Using this expression in Eq. (\ref{tmnl}) and focusing for
simplicity only on R-R fluxes since the NS-NS fluxes are anyhow removed by the orientifold
projection, after simple manipulations we find that the fermionic couplings of a magnetized E5-brane
are described by the tensor
\begin{equation}
 \begin{aligned}
T_{\hat m \hat n \hat p} =& \frac{1}{\sqrt{\det(1-{\mathcal F}_0)}}\,
\Big[\,\ii\,(*_6F)_{\hat m \hat n \hat p}
+ 3\ii\,(*_6F)_{\hat m \hat n}^{\phantom{mn} \hat q}\,({\mathcal F_0})_{\hat q \hat p}
\\
&+\frac{3\ii}{8}\,F_{\hat m \hat n}^{\phantom{mn}\hat q}
\,({\mathcal F_0})^{\hat r \hat s}({\mathcal F_0})^{\hat t \hat u}\,
\epsilon_{\hat q \hat r \hat s \hat t \hat u \hat p}
 - \frac{\ii}{3! \,8}\, F_{\hat m \hat n \hat p}\,
({\mathcal F_0})^{\hat q \hat r}({\mathcal F_0})^{\hat s \hat t}
({\mathcal F_0})^{\hat u \hat v}\,\epsilon_{\hat q \hat r \hat s \hat t \hat u \hat v}\Big]~.
\end{aligned}
\label{te5m}
\end{equation}
In the same way, and always starting from the general formula (\ref{tmnl}) one can discuss all other
types of magnetized branes.

\subsection{Flux-induced fermionic mass and lifting of instanton zero-modes}
\label{subsec:masses}

To complete the previous analysis we write the fermion bilinear $\Theta\Gamma^{mnp}\Theta$
using a four-dimensional spinor notation; in this way the structure of the flux-induced
fermionic masses will be more clearly exposed.  According to our $4+6$ splitting, the anti-chiral
ten dimensional spinor $\Theta_{\mathcal A}$ decomposes as
\begin{equation}
\Theta_{\mathcal A} ~\to~\big(\Theta^{\alpha A},\Theta_{\dot\alpha A}\big)
\label{theta}
\end{equation}
where $\alpha$ ($\dot\alpha$) are chiral (anti-chiral) indices in four dimensions,
and the lower (upper) indices $A$ are chiral (anti-chiral) spinor indices of
the internal six dimensional space.
Furthermore, by decomposing the $\Gamma$ matrices according to
\begin{equation}
\Gamma^\mu= \gamma^\mu \otimes 1\quad,\quad\Gamma^m=\gamma_{(5)}\otimes\gamma^m~,
\label{gamma}
\end{equation}
one can show that
\begin{equation}
\Theta \Gamma^{mnp}\Theta = -\ii\,\Theta^{\alpha A} \Theta_{\alpha}^{{\phantom\alpha} B}
\big(\overline\Sigma^{mnp}\big)_{AB}
-\ii\,\Theta_{\dot\alpha A}
\Theta^{\dot\alpha}_{{\phantom\alpha} B}\big(\Sigma^{mnp}\big)^{AB}
\label{decomp}
\end{equation}
where $\Sigma^{mnp}$ and $\overline\Sigma^{mnp}$ are respectively the chiral and anti-chiral blocks
of $\gamma^{mnp}$ (see Appendix \ref{app:conventions} for details).
It is important to notice that
\begin{equation}
*_6\Sigma^{mnp} = -\ii\,\Sigma^{mnp}\quad,\quad
*_6\overline\Sigma^{mnp} = +\ii\,\overline\Sigma^{mnp}
\label{sigmaSD}
\end{equation}
so that $\Sigma^{mnp}$ only couples to an imaginary self-dual (ISD) tensor, while
$\overline{\Sigma}^{mnp}$ only couples to an imaginary anti-self dual (IASD)
tensor.
More explicitly, we have
\begin{equation}
\begin{aligned}
\Theta \Gamma^{mnp}\Theta\,T_{mnp} &= -\ii\,\Theta^{\alpha A}
\Theta_{\alpha}^{{\phantom\alpha} B}
\big(\overline\Sigma^{mnp}\big)_{AB}\,T_{mnp}^{\mathrm{IASD}}
-\ii\,\Theta_{\dot\alpha A} \Theta^{\dot\alpha}_{{\phantom\alpha}
B}\big(\Sigma^{mnp}\big)^{AB} \,T_{mnp}^{\mathrm{ISD}}\\
&= -\ii\,\Theta^{\alpha A} \Theta_{\alpha}^{{\phantom\alpha} B}
T_{AB} -\ii\,\Theta_{\dot\alpha A}
\Theta^{\dot\alpha}_{{\phantom\alpha} B} T^{AB}
\end{aligned}
\label{decomp1}
\end{equation}
where
\begin{equation}
T_{mnp}^{\mathrm{ISD}} =\frac{1}{2}\big(T-\ii*_6\!T\big)_{mnp}\quad,\quad
T_{mnp}^{\mathrm{IASD}} =\frac{1}{2}\big(T+\ii*_6\!T\big)_{mnp}~.
\label{isdiasd}
\end{equation}
In the second line of Eq. (\ref{decomp1}) we have adopted a $\mathrm{SU}(4)\sim \mathrm{SO}(6)$
notation and defined the IASD and ISD parts of the $T$-tensor as the following
$4\times 4$ symmetric matrices
\begin{equation}
T_{AB}=(\overline\Sigma^{mnp}\big)_{AB}\,T_{mnp}^{\mathrm{IASD}}\quad,\quad
T^{AB}=\big(\Sigma^{mnp}\big)^{AB} \,T_{mnp}^{\mathrm{ISD}}~,
\label{isdiasd1}
\end{equation}
with upper (lower) indices $A,B$ running
over the 
${\bf 4}$ (${\bf\bar 4}$) representations of $\mathrm{SU}(4)$.
Fixing a complex structure, the 3-form tensors
$T^{\mathrm{ISD}},T^{\mathrm{IASD}}$  can be decomposed into their
(3,0),(2,1),(1,2) and (0,3) parts as indicated in Tab. \ref{TIASD}.
The $(2,1)$ components are distinguished into six primitive ones (P),
satisfying $g^{j\bar k}T_{ij\bar k}=0$,
and three non-primitive ones (NP), satisfying $T_i=g^{j\bar k}T_{ij\bar k}$.
A similar decomposition holds for the $(1,2)$ part.
The various components transform in irreducible representations of
the $\mathrm{SU}(3)\in \mathrm{SU}(4)$ holonomy group under which the internal coordinates $Z^i,\bar
Z^i$ transform as ${\bf 3}$ and ${\bf \bar 3}$
respectively and spinors like ${\bf 4}={\bf 1}+{\bf 3}$ and
${ \bf \bar4}= { \bf \bar1}+{ \bf\bar 3}$.
The $\mathrm{SU}(3)$ content of the $T$-tensor is displayed in the last column in
Tab. \ref{TIASD}.
\begin{table}[ht]
\centering
\begin{tabular}{|ccc|}
\hline $\phantom{\vdots}T^{\mathrm{ISD}}$ & $\to$ & $T_{(0,3)}
\oplus T_{(1,2)_{\mathrm{NP}}}\oplus T_{(2,1)_{\mathrm{P}}}={\bf
\bar 1}
\oplus {\bf \bar 3} \oplus {\bf \bar 6}$\\
[1ex]
 \hline
$\phantom{\vdots}T^{\mathrm{IASD}}$ & $\to$  & $T_{(3,0)} \oplus
T_{(2,1)_{\mathrm{NP}}}\oplus T_{(1,2)_{\mathrm{P}}}={\bf 1} \oplus
{\bf 3} \oplus {\bf 6} $
\\[1ex]
\hline
\end{tabular}
 \caption{ Decomposition of the ISD and IASD parts of the 3-form $T$.
The $(2,1)$ and $(1,2)$ components are distinguished into primitive (P)
and non-primitive (NP) parts. The last column displays the $\mathrm{SU}(3)$
content of the various pieces.}
 \label{TIASD}
 \end{table}

Let us now use this information to rewrite the fermionic terms we have discussed in
the previous subsections, focusing in particular on D3-branes and D-instantons
on flat space.
In the case of D3-branes, we can use a Minkowski
signature and the Majorana-Weyl fermion $\Theta$ decomposes as in (\ref{decomp})
where the four-dimensional chiral and anti-chiral
components are related by charge conjugation and assembled into four Majorana spinors.
These are the four gauginos leaving on the world-volume of the D3-brane, and for future
notational convenience we will denote their chiral and anti-chiral parts
as $\Lambda^{\alpha A}$ and $\bar\Lambda_{\dot\alpha A}$ (instead of $\Theta^{\alpha A}$
and $\Theta_{\dot\alpha A}$).
Then, using Eqs. (\ref{tD31}) and (\ref{decomp}) in the general
expression (\ref{ampltot2}), we obtain
\begin{equation}
\mathcal A_{\mathrm D3} =
\frac{2\pi\ii}{3!}\,c_F\,{\mathrm{Tr}}\Big[\, \Lambda^{\alpha A}
\Lambda_{\alpha}^{{\phantom \alpha}B}
\big(\overline\Sigma^{mnp}\big)_{AB}\,G_{mnp}^{\mathrm{IASD}} -
\bar\Lambda_{\dot\alpha A}\bar\Lambda^{\dot\alpha}_{{\phantom
\alpha}B}
\big(\Sigma^{mnp}\big)^{AB}\,\big(G_{mnp}^{\mathrm{IASD}}\big)^*
\,\Big] \label{massD3}
\end{equation}
where we have made explicit the colour trace
generators%
\footnote{We use the following normalization:
${\mathrm{Tr}}\left(T^aT^b\right)=\frac{1}{2}\,\delta^{ab}$.}.

Recalling that the topological normalization of any disk
amplitude with D3-strings is \cite{Billo:2002hm}
\begin{equation}
\mathcal C_{(4)} = \frac{1}{\pi^2\,{\alpha'}^2\,g_{\mathrm{YM}}^2}~,
\label{c4}
\end{equation}
one can show that in order to obtain gauginos with canonical dimension of
(length)$^{-3/2}$ and standard kinetic term of the form
\begin{equation}
\frac{1}{g_{\mathrm{YM}}^2}\int d^4x \,{\mathrm{Tr}}\big(\!-2\ii\,
\bar\Lambda_{\dot\alpha A}\bar D\!\!\!\!/^{\,\dot\alpha \beta}
\Lambda_\beta^{\,A}\big)~,
\label{kinetic}
\end{equation}
one has to normalize the gaugino vertices with
\begin{equation}
\mathcal N_\Lambda= (2\pi\alpha')^{\frac{3}{4}}~.
\label{nlambda}
\end{equation}
Then, using these ingredients the prefactor appearing in \eq{massD3} becomes
\begin{equation}
c_F=\frac{4}{
g_{\mathrm{YM}}^2}\,(2\pi\alpha')^{-\frac{1}{2}}\,{\mathcal N}_F~.
\label{cfD3}
\end{equation}
{F}rom the explicit expression of the amplitude (\ref{massD3})
we see that an IASD $G$-flux configuration
induces a Majorana mass%
\footnote{Notice that this is the mass term for gauginos which are
not canonically normalized, as we do not rescale away the overall
factor of $1/g_{\mathrm{YM}}^2$ appearing in \eq{kinetic}.} for the
gauginos leading to supersymmetry breaking on the gauge theory
\cite{Grana:2002tu,Camara:2003ku,Grana:2003ek,Camara:2004jj}. Notice
that the mass term for the two different chiralities are complex
conjugate of each other: $T^{\mathrm{IASD}}=-\ii\,
G^{\mathrm{IASD}}$ and $T^{\mathrm{ISD}}=\ii(G^{\mathrm{IASD}})^*$. This
is a consequence of the Majorana condition that the four-dimensional
spinors inherit from the Majorana-Weyl condition of the fermions in
the original ten-dimensional theory.

If we decompose $G^{\mathrm{IASD}}$ as indicated in Tab.
\ref{TIASD}, we see  that a $G$-flux of type $(1,2)_{\mathrm P}$
gives mass to the three gauginos transforming non-trivially under
$\mathrm{SU}(3)$ but keeps the $\mathrm{SU}(3)$-singlet gaugino massless, thus
preserving ${\mathcal N}=1$ supersymmetry. On the other hand, a $G$-flux of type $(3,0)$,
or $(2,1)_{\mathrm{NP}}$ gives mass also to the $\mathrm{SU}(3)$-singlet gaugino.

Things are rather different instead on D-instantons whose fermionic
coupling is given by Eq. (\ref{tD-1}). Indeed, by inserting such
coupling in Eq. (\ref{ampltot2}) and using again Eq. (\ref{decomp})
we obtain
\begin{equation}
\mathcal A_{{\mathrm D}(-1)} =
\frac{2\pi\ii}{3!}\,c_F(\Theta)\,\Big[\, \Theta^{\alpha A}
\Theta_{\alpha}^{{\phantom\alpha} B}
\big(\overline\Sigma^{mnp}\big)_{AB}\,G_{mnp}^{\mathrm{IASD}} +\bar
\Theta_{\dot\alpha A}\bar \Theta^{\dot\alpha}_{{\phantom \alpha}B}
\big(\Sigma^{mnp}\big)^{AB}\,G_{mnp}^{\mathrm{ISD}} \,\Big]
\label{massD-1}
\end{equation}
where now the prefactor $c_F(\Theta)$ contains the topological
normalization of the D$(-1)$ disks (the value of the gauge instanton
action ), namely \cite{Billo:2002hm}
\begin{equation}
\mathcal C_{(0)} = \frac{8\pi^2}{g_{\mathrm{YM}}^2} \quad
\Rightarrow \quad c_F(\Theta)=\frac{8\pi^2}{g_{\mathrm{YM}}^2}{\cal
N}_{\Theta}^2\,{\mathcal N}_F \label{c0}
\end{equation}
{F}rom the amplitude (\ref{massD-1}) we explicitly see that
both the IASD and the ISD components of the $G$-flux couple to the D-instanton fermions;
however the couplings are different and independent for the two chiralities since
they are not related by complex conjugation, as always in Euclidean spaces.
In particular, comparing Eqs. (\ref{massD3}) and
(\ref{massD-1}), we see that an ISD $G$-flux does not give a mass to any gauginos
but instead induces a ``mass'' term for the anti-chiral instanton zero-modes which are therefore
lifted. This effect may play a crucial role in discussing the non-perturbative contributions
of the so-called ``exotic'' D-instantons for which the neutral anti-chiral zero modes $\bar \Theta_{\dot\alpha A}$ must be removed \cite{Argurio:2007vqa,Bianchi:2007wy} 
or lifted by some mechanism \cite{Blumenhagen:2007bn,Petersson:2007sc}.
Introducing an ISD $G$-flux is one of such mechanisms as we
will discuss in more detail in Section \ref{sec:fD-1}.

\section{Flux couplings with twisted open strings ($\vec{\vartheta}\not =0$)}
\label{sec:twisted}

As we have emphasized, the general world-sheet calculation presented
in Section \ref{sec:CFT} allows to obtain the couplings between
closed string fluxes and open string fermions at a generic D-brane
intersection, even for non-vanishing twist parameters
$\vec{\vartheta}$. A systematic study of the amplitudes
(\ref{amplFfinal}) and (\ref{amplHfinal}) when $\vec{\vartheta}\not
=0$ will be presented elsewhere; here we just analyze a simple case
of such twisted amplitudes which will be relevant for the
applications discussed in Section \ref{sec:fD-1}.

The case we discuss is that of the 3-form flux couplings with the
twisted fermions stretching between a D3-brane and a D-instanton
which represent the charged (or flavored) fermionic moduli of the
${\cal N}=4$ ADHM construction of instantons (see for example Refs.
\cite{Dorey:2002ik,Billo:2002hm}) and are usually denoted as $\mu^A$
and $\bar\mu^A$ depending on the orientation. In the notation of
Section \ref{sec:CFT} the D3/D$(-1)$ and D$(-1)$/D3 strings are
characterized by twist vectors of the form
\begin{equation}
\vec{\vartheta} = \Big(0,0,0,+\frac{1}{2},+\frac{1}{2}\Big)
\quad\mbox{and}\quad
\vec{\vartheta}' = \Big(0,0,0,-\frac{1}{2},-\frac{1}{2}\Big)
\label{vartheta3-1}
\end{equation}
respectively,
and thus, according to \eq{theta0}, the open string fermions in these
sectors have weight vectors
\begin{equation}
\vec{\epsilon}_1 =\Big(\vec{\epsilon}_{A},-\frac{1}{2},-\frac{1}{2}\Big)
\quad\mbox{and}\quad
\vec{\epsilon}_4 =
\Big(\vec{\epsilon}_{A},+\frac{1}{2},+\frac{1}{2}\Big)~.
\label{epsilonmu}
\end{equation}
The notation $\vec{\epsilon}_{A}$ ($\vec{\epsilon}^{\,A}$) denotes
an anti-chiral (chiral) spinor weight of the internal
$\mathrm{SO}(6)$ rotation group. The vertex operators corresponding
to (\ref{epsilonmu}) (see \eq{vertexferm}) are then
\begin{equation}
V_\mu(z) =\mathcal{N}_{\mu}\,\mu^{A}
\big[\sigma_{\vec{\vartheta}}\,s_{ \vec{\epsilon}_A}\, \ee^{-
{\frac{1}{2}\phi}}\big](z) \quad\mbox{and}\quad
V_{\bar\mu}(z) = \mathcal{N}_{\bar\mu}\,{\bar\mu}^{A}
\big[\sigma_{\vec{\vartheta}'}\,s_{ \vec{\epsilon}_A}\, \ee^{-
{\frac{1}{2}\phi}}\big](z)
 \label{vertexmu}
\end{equation}
where $\mathcal{N}_{\mu},\mathcal{N}_{\bar\mu}$ are suitable
normalizations. Notice that since the last two components of
$(\vec{\epsilon}_1+\vec{\vartheta})$ and
$(\vec{\epsilon}_4+\vec{\vartheta}')$ are zero, only a spin-field
$s_{ \vec{\epsilon}_A}=S_A$ of the internal $\mathrm{SO}(6)$ appears
in the vertices (\ref{vertexmu}); furthermore there is no momentum
in any direction since $\mu^{A},\bar\mu^A$ are moduli rather than
dynamical fields. Note also that both $\mu^A$ and $\bar\mu^A$ carry the
same $\mathrm{SO}(6)$ chirality.

On the other hand, the vertex operator for a R-R field strength contains two
parts: one with left and right weights of the type
\begin{equation}
\vec{\epsilon}_2=\big(\vec{\epsilon}^{\,A},
\vec{\epsilon}^{\,\dot\alpha} \big) \quad\mbox{and}\quad
\vec{\epsilon}_3=\big(\vec{\epsilon}^{\,B},
\vec{\epsilon}^{\,\dot\beta} \big) ~,
\label{epsilon22}
\end{equation}
and one with weights of the type
\begin{equation}
\vec{\epsilon}_2=\big(\vec{\epsilon}_{A},
\vec{\epsilon}_\alpha \big) \quad\mbox{and}\quad
\vec{\epsilon}_3=\big(\vec{\epsilon}_{B}, \vec{\epsilon}_\beta
\big)~,
\label{epsilon21}
\end{equation}
where $\vec{\epsilon}_\alpha$ ($\vec{\epsilon}^{\,\dot\alpha}$) are
the chiral (anti-chiral) spinor weights%
\footnote{In our conventions, as explained in Appendix \ref{app:conventions},
$\alpha \in \{\frac12(++),\frac12(--)\}$
and  $\dot\alpha\in\{\frac12(-+),\frac12(+-)\}$.} 
of $\mathrm{SO}(4)$.
We now show that when the R-R field is an internal 3-form $F_{mnp}$ only the part in
(\ref{epsilon22}) couples to the $\mu$ and $\bar\mu$'s.

Let us consider the general R-R amplitude (\ref{amplFfinal}) 
and take, for example, the $\sigma=0$ boundary on the D$(-1)$-brane,
{\it i.e.} $R_0=-1$ and $\mathcal R_0=\ii\Gamma_{(11)}^{\mathrm E}$.
Let us then observe that the spinor reflection matrix can be effectively replaced by
$\mathcal R_0=-\ii$, since our GSO projection selects the anti-chiral sector, and that
the two $\vec{\vartheta}\cdot \vec{\epsilon}_3$-dependent integrals $I_1$ and $I_2$,
defined in (\ref{integrals}), are scalars along the six internal directions 
because the internal components of $\vec\vartheta$ are vanishing (see Eq. (\ref{vartheta3-1})).
All this implies that the
term with a single $\Gamma$ in (\ref{amplFfinal}) vanishes, so that the only
non-trivial contribution comes from the term with three $\Gamma$'s.

To proceed we need to evaluate the integral $I_2$. In the limit $s=-2t\to 0$, from
Eq. (\ref{integrals1}) we easily find
\begin{equation}
\big(I_2\big)_{\!\mathcal{A}_3}^{~\mathcal{A}_3} =\frac{1}{2
\vec{\vartheta}\cdot \vec{\epsilon}_3}~ \Big(1-\ee^{2\pi\ii
\vec{\vartheta}\cdot \vec{\epsilon}_3 }\Big)~;
\label{integralsi2}
\end{equation}
recall that $\mathcal{A}_3$ is the spinor index corresponding to the weight $\vec{\epsilon}_3$.
There are two distinct cases, corresponding to the two possibilities
(\ref{epsilon22}) and (\ref{epsilon21}) respectively. In the first case we have 
\begin{equation}
\vec\vartheta\cdot\vec\epsilon_3=0 
\quad\quad  \Rightarrow  \quad\quad \big(I_2\big)_{\!\mathcal{A}_3}^{~\mathcal{A}_3}=-\pi \ii~,
\label{case1}
\end{equation}
while in the second case we have
\begin{equation}
 \begin{aligned}
\vec\vartheta\cdot\vec\epsilon_3&=+\frac12\quad\mbox{if}\quad
 \vec{\epsilon}_\beta=\frac12\big(++\big)
\quad\quad  \Rightarrow  \quad\quad \big(I_2\big)_{\!\mathcal{A}_3}^{~\mathcal{A}_3}=+2~,\\
\vec\vartheta\cdot\vec\epsilon_3&=-\frac12\quad\mbox{if}\quad
 \vec{\epsilon}_\beta=\frac12\big(--\big)
\quad\quad  \Rightarrow  \quad\quad \big(I_2\big)_{\!\mathcal{A}_3}^{~\mathcal{A}_3}=-2~.
 \end{aligned}
\label{case2}
\end{equation}
Using the explicit expression of the $\Gamma$ matrices given in Appendix \ref{app:conventions},
it is not difficult to realize that the above results can be summarized by writing
\begin{equation}
 I_2=-\pi\ii\,\left(\frac{1+\Gamma_{(7)}}{2}\right)-2\ii\,\Gamma^{01}
\left(\frac{1-\Gamma_{(7)}}{2}\right)
\label{I2}
\end{equation}
where $\Gamma_{(7)}$ is the chirality matrix in the six-dimensional internal space.
Indeed, restricting to the anti-chiral block, one can check that the first
term in (\ref{I2}) accounts for the matrix elements (\ref{case1}), while the second
term for the matrix elements (\ref{case2}). At this point it is clear that 
with an internal R-R 3-form flux $F_{mnp}$, the coefficient $\big(F{\mathcal R}_0I_2\big)_{mnp}$
of the amplitude (\ref{amplFfinal}) can only receive contribution from the first
term in (\ref{I2}), which yields
\begin{equation}
 {\mathcal A}_F \sim 
{\bar\mu}^{A}\mu^{B}
\,\big(\overline\Sigma^{mnp}\big)_{AB}\,F_{mnp}^{\mathrm{IASD}}~.
\label{mumubF}
\end{equation}

The evaluation of coupling with an internal NS-NS 3-form flux $H_{mnp}$ is much simpler.
In fact the left and right weights appearing in the NS-NS vertex operator are
\begin{equation}
\vec{\epsilon}_2=\big(\pm \vec{e}_{m}\pm
\vec{e}_{n}, \vec 0 \big) \quad\mbox{and}\quad
\vec{\epsilon}_3=\big(\pm\vec{e}_{p}, \vec 0 \big)~,
\label{epsilon21nsns}
\end{equation}
with $\vec{e}_{m,n,p}$ unit vectors in the $SO(6)$ weight space
specifying the $H_{mnp}$-hyperplane. Thus, we always have $\vec\theta\cdot
\vec{\epsilon}_3=0$ which implies that the entries of the two diagonal
matrices $I_1$ and $I_2$ (with vector indices) are
\begin{equation}
 2\,\big(I_1\big)^P_{\,P}= \big(I_2\big)^P_{\,P} = -\pi\ii~.
\label{i1i2}
\end{equation}
Thus, from Eq. (\ref{amplHfinal}) we see that the term with a single $\Gamma$ vanishes, while the
term with three $\Gamma$'s yields
\begin{equation}
 {\mathcal A}_H \sim 
{\bar\mu}^{A}\mu^{B}
\,\big(\overline\Sigma^{mnp}\big)_{AB}\,H_{mnp}^{\mathrm{IASD}}~.
\label{mumubH}
\end{equation}

Collecting the two contributions (\ref{mumubF}) and (\ref{mumubH}) and reinstating the
appropriate normalizations, we finally obtain
\begin{equation}
{\mathcal A}_{\mathrm{D3/D(-1)}} \equiv {\mathcal A}_F +{\mathcal
A}_H =\frac{4\pi\ii}{3!}\,c_F(\mu)\,{\bar\mu}^{A}\mu^{B}
\,\big(\overline\Sigma^{mnp}\big)_{AB}\,G_{mnp}^{\mathrm{IASD}}
\label{mumubtot}
\end{equation}
where
$c_F(\mu)=\mathcal{C}_{(0)}\,\mathcal{N}_{\mu}\,\mathcal{N}_{\bar\mu}\,\mathcal{N}_F$
with $\mathcal C_{(0)}$ given in \eq{c0}. Notice that no symmetry factors has to
be included in this amplitude, since $\mu$ and $\bar\mu$ are really
distinct and independent quantities. This amplitude together with the one
in \eq{massD-1} accounts for the flux induced fermionic couplings on
the D-instanton effective action, and their meaning will be
discussed in Section \ref{sec:fD-1}.

\section{Flux couplings in an ${\mathcal N}=1$ orbifold set-up}
\label{sec:N1}
The results of the previous sections clearly show that internal
NS-NS and R-R fluxes bear important consequences on the brane
effective action and may be relevant in phenomenological
applications. Therefore it is particularly interesting to study
such flux interactions in models with $\mathcal{N}=1$
supersymmetry. To do so we adopt a toroidal orbifold
compactification scheme where string theory remains calculable and
the flux couplings are basically those described in Section
\ref{sec:effects}. We consider type IIB theory compactified on a
Calabi-Yau 3-fold with the $\mathcal{N}=2$ bulk supersymmetry
further broken down to $\mathcal{N}=1$ by the introduction of
D-branes and O-planes. To be specific we consider the orbifold
$\mathcal{T}_6/(\mathbb{Z}_2\times\mathbb{Z}_2)$ with $\mathcal{T}_6$ completely
factorized as a product of three 2-torii. The action of
$\mathbb{Z}_2\times\mathbb{Z}_2$ on the orthonormal complex coordinates
$Z^i$ ($i=1,2,3$) of the torus is in Tab. \ref{tablez2z2}.
\begin{table}[ht]
\centering
\begin{tabular}{c|cccc}
\hline \hline
  coordinates & $h_0$ & $h_1$ & $h_2$ & $h_3$  \\
   \hline
$Z^1$   &  $+$ & $+$ & $-$  & $-$\\
$Z^2$ &  $+$ & $-$ & $+$ & $-$  \\
$Z^3$ & $+$ & $-$ & $-$  &$+$\\
  \hline \hline
\end{tabular}
\caption{Orbifold group action on the complex
coordinates $Z^i$ of $\mathcal{T}^6$.} 
\label{tablez2z2}
\end{table}
In order to be self-contained we now briefly recall the structure
of the closed string multiplets  and the pattern of fractional
D-branes that can be introduced in this orbifold.

\subsection{Closed and open string sectors in the $\mathbb Z_2\times\mathbb{Z}_2$ orbifold}

\paragraph{Closed string states} Let us start by considering the oriented closed string states
before the introduction of O-planes. The massless closed string
states in the   orbifold organize into a gravity multiplet,
$h_{2,1}$ vector multiplets and $h_{1,1}+1$ hypermultiplets of the
${\cal N}=2$ supersymmetry. For strings defined on the quotient
space, the orbifold projection onto
$\mathbb{Z}_2\times\mathbb{Z}_2$ invariant states has to be
enforced and as usual we distinguish between untwisted and twisted
sectors.

The untwisted sector follows from that on $\mathcal{T}^6$ after
restricting to $\mathbb{Z}_2\times\mathbb{Z}_2$-invariant
components. It contains: the gravity multiplet; the universal
hypermultiplet having as bosonic components the dilaton $\varphi$,
the axion $C_0$ and the dualized NS-NS and R-R 2-forms $B_2$ and
$C_2$ with four dimensional indices;
$h^{\mathrm{untw}}_{21}=3$ vector multiplets with bosonic components $(V^i,u^i)$
where the scalars $u^i$ parametrize the complex structure
deformations; $h^{\mathrm{untw}}_{11}=3$ hypermultiplets containing
the scalars $(v_i,b_i,c_i,\tilde c^i)$ with $v_i$ representing the
K\"ahler parameter of the $i$-th torus, $b_i$ and $c_i$ the components of the
NS-NS and R-R 2-forms along the $i$-th torus, and $\tilde c^i$
the R-R 4-form components along the dual 4-cycle.

Closed strings on the $\mathbb{Z}_2\times\mathbb{Z}_2$ orbifold
have also twisted sectors associated to each of the three non
trivial elements $h_i$ and localized at the 16 possible fixed loci
of their action; the total number of twisted sectors is therefore
$3\times 16 = 48$. To fully specify the
orbifold model, we must declare the action of the group elements
$h_i$ also on the twist fields. There are two consistent
possibilities, which correspond to the singular limits of two
different CY manifolds with $(h_{11},h_{21})=(51,3)$ and
$(h_{11},h_{21})=(3,51)$. Here we restrict ourselves to the first
choice%
\footnote{The second choice corresponds to declare that the
twist field $\Delta_{ij}$ in the four-dimensional plane (ij)
transforms in the representation $|\epsilon_{ijk}|R_k$.}, 
corresponding to take all twisted fields invariant under $h_i$. With
this choice, the twisted sectors contribute to the massless
spectrum with $h^{\mathrm{tw}}_{11}=48$ hypermultiplets containing the
scalars $(v_{\hat i},b_{\hat i}, c_{\hat i},\tilde c^{\hat i})$,
$\hat i=1,...48$, which describe, respectively, the deformations of the blow-up modes
of the vanishing 2-cycles and the exceptional components of the NS-NS 2-form 
and of the R-R 2- and 4-forms. It is important to recall that
the orbifold limit is attained with a non-zero background value of
the NS-NS $B$-field on the vanishing cycles \cite{Aspinwall:1996mn},
so that the scalar fields $b_{\hat i}$ mentioned above represent fluctuations
around this value.

Finally the introduction of O-planes projects the spectrum onto the
subset of $\Omega I$-invariant states with $\Omega$ the worldsheet
parity and $I$ some involution of the CY threefold. The resulting
spectrum falls into vectors and chiral multiplets of the unbroken
${\cal N}=1$ supersymmetry. The details depends on the choice of
the O-planes. For example for a vacuum built out of O3/O7-planes,
$B_2$ and $C_2$ are odd while for O5/O9 planes, $B_2$ and $C_4$
are odd. As a consequence, in the twisted sector either $b_{\hat i}$ and
$c_{\hat i}$, or $b_{\hat i}$ and $\tilde c^{\hat i}$ would be
projected out for the O3/O7 and O5/O9-choices respectively.

\paragraph{Open string states and fractional D-branes}
The fundamental types of D-branes which can be placed transversely
to an orbifold space are called fractional branes
\cite{Douglas:1996sw}. Such branes must be localized at one of the
fixed points of the orbifold group (which in our case are 64). For
simplicity we focus on fractional D3 branes sitting at a specific
fixed point (say, the origin) and work around this configuration.
Locally our system is undistinguishable from the theory living on
the non-compact orbifold $\mathbb{C}^3/(\mathbb{Z}_2\times\mathbb{Z}_2)$.

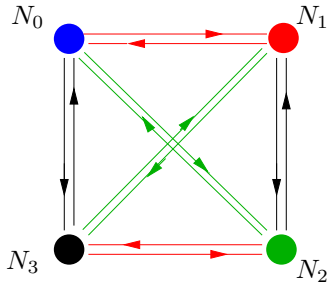
\begin{figure}[hbt]
\begin{center}
\begin{picture}(0,0)%
\includegraphics{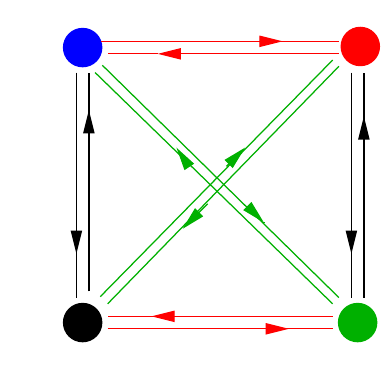}%
\end{picture}%
\setlength{\unitlength}{1579sp}%
\begingroup\makeatletter\ifx\SetFigFontNFSS\undefined%
\gdef\SetFigFontNFSS#1#2#3#4#5{%
  \reset@font\fontsize{#1}{#2pt}%
  \fontfamily{#3}\fontseries{#4}\fontshape{#5}%
  \selectfont}%
\fi\endgroup%
\begin{picture}(4559,4450)(211,-4247)
\put(4726,-136){\makebox(0,0)[lb]{\smash{{\SetFigFontNFSS{10}{12.0}{\familydefault}{\mddefault}{\updefault}$N_1$}}}}
\put(4726,-4111){\makebox(0,0)[lb]{\smash{{\SetFigFontNFSS{10}{12.0}{\familydefault}{\mddefault}{\updefault}$N_2$}}}}
\put(301,-136){\makebox(0,0)[lb]{\smash{{\SetFigFontNFSS{10}{12.0}{\familydefault}{\mddefault}{\updefault}$N_0$}}}}
\put(226,-3961){\makebox(0,0)[lb]{\smash{{\SetFigFontNFSS{10}{12.0}{\familydefault}{\mddefault}{\updefault}$N_3$}}}}
\end{picture}%
\end{center}
\caption{The quiver diagram encoding the field content and the
charges for fractional D-branes of the local orbifold
$\mathbb{C}^3/(\mathbb{Z}_2\times\mathbb{Z}_2)$. The dots
represent the branes associated with the irrep $R_A$ of the
orbifold group. A stack of $N_A$ such branes supports a
$\mathrm{U}(N_A)$ gauge theory. An oriented link from the $A$-th
to the $B$-th dot corresponds to a chiral multiplet $\phi^{AB}$
transforming in the $(N_A,\overline N_B)$ representation of the
gauge group and in the $R_A\,R_B^{-1}$ representation of the
orbifold group.} \label{fig:quiver}
\end{figure}

The fractional branes are in correspondence with the irreducible
representations $R_A$ of the orbifold group: in fact the
Chan-Paton indices of an open string connecting two fractional
branes of type $A$ and $B$ transform in the representation $R_A\otimes
R_B$. In addition the orbifold group acts on the $\mathrm{SO}(6)$ internal
indices of the open string fields as indicated in Tab. \ref{tablez2z2lp}.
\begin{table}[ht]
\centering
\begin{tabular}{c|ccc}
\hline\hline
irrep $R_A$& & ~~~~~~~~~fields &\\
\hline
$R_0\phantom{\vdots}$ & $A_\mu$ & $\Lambda^0\equiv\Lambda^{---}$ &$ \bar\Lambda_{0}\equiv\bar\Lambda_{+++}$
\\
$R_1\phantom{\vdots}$ & $\phi^1$ & $\Lambda^1\equiv\Lambda^{-++}$ &$ \bar\Lambda_{1}\equiv\bar\Lambda_{+--}$
\\
$R_2\phantom{\vdots}$ & $\phi^2$ & $\Lambda^2\equiv\Lambda^{+-+}$ &$ \bar\Lambda_{2}\equiv\bar\Lambda_{-+-}$
\\
$R_3\phantom{\vdots}$ & $\phi^3$ & $\Lambda^3\equiv\Lambda^{++-}$ &$ \bar\Lambda_{3}\equiv\bar\Lambda_{--+}$
\\
\hline \hline
\end{tabular}
\caption{Representation of the orbifold group on the ${\mathcal N}=4$ open string
fields}
\label{tablez2z2lp}
\end{table}
This action should be compensated by that on the
Chan-Paton indices in such a way that the whole field is
$\mathbb{Z}_2\times\mathbb{Z}_2$ invariant. For example the
vector $A_\mu$ and the gaugino $\Lambda^0$ are invariant under the
orbifold group and therefore they should carry indices $(N_A,\overline N_A)$ 
in the adjoint of the $\prod_{A=0}^3 \mathrm{U}(N_A)$. The remaining
fields fall into chiral multiplets transforming in the
bifundamental representations $(N_A,\overline N_B)$. 
The resulting quiver diagram is displayed in Figure \ref{fig:quiver}.

To discuss the couplings of fractional branes to closed strings,
it may be convenient to describe the branes by means of boundary
states \cite{DiVecchia:1999rh,DiVecchia:1999fx}, which we indicate
schematically as $\ket A$ for a brane of type $A$. It turns out 
(see for example Ref. \cite{Billo:2000yb}) that these boundary states 
$\ket A$ are suitable combinations of boundary
states $\dket{I}$ associated to the $h_I$-twisted sector, namely
\begin{equation}
 \label{ctoi}
\ket{A} = \frac{1}{4} \sum_I \,\ch^I_{A} \dket{I}~.
\end{equation}
with $\ch^I_{A}={\mathrm{tr}}_{R_A}(h_I)$. In our orbifold, using the
character table (\ref{frac3}) given in Appendix \ref{subapp:T6orb}, these sums explicitly read
\cite{Bertolini:2001gg}
\begin{equation}
 \label{fbbs}
\begin{aligned}
\ket{0} & = \frac 14\Big( \dket{0} + \dket{1} + \dket{2} +
\dket{3}\Big)\quad,
\quad \ket{1}  = \frac 14 \Big(\dket{0} + \dket{1} - \dket{2} - \dket{3}\Big)~,\\
\ket{2} & = \frac 14 \Big(\dket{0} - \dket{1} + \dket{2} -
\dket{3}\Big)\quad, \quad \ket{3} = \frac 14 \Big(\dket{0} -
\dket{1} - \dket{2} + \dket{3}\Big)~.
\end{aligned}
\end{equation}
These boundary states show that the fractional D3-branes
couple not only to twisted closed string fields but to
\emph{untwisted} ones as well, with a fractional tension and
a fractional charge given by $1/4$ of the ones of the regular branes.

The fractional branes corresponding to $R_A$ ($A\not= 0$) can also
be interpreted geometrically as D5-branes
suitably wrapped on exceptional%
\footnote{Since we focus on branes localized at the origin, for each
$A$ we consider only one of the 16 possible exceptional cycles
$e^{\hat A}$.} 2-cycles $e^{\hat A}$ in the blown-up space. To this
extent, the background value of the NS-NS 2-form $B_2$ in the orbifold limit plays a
crucial role in accounting for the untwisted couplings of the
branes. We will take advantage of this description in the remaining
of this section when we interpret the flux couplings computed in
Section \ref{sec:effects} in the effective low-energy supergravity
theory.

\subsection{Gauge kinetic functions and soft supersymmetry breaking on D3-branes}
\label{sec:n1int}

In presence of D-branes the $\mathcal{N}=2$ bulk supersymmetry of
the chosen compactification is reduced to a specific $\mathcal{N}=1$
slice depending on the boundary conditions imposed by the branes on
the spin fields, which are encoded  in the spinor reflection matrix
$\mathcal{R}_0$ of \eq{rspinor}. The supersymmetry left unbroken by
D-branes should be aligned to that preserved by O-planes and tadpole
conditions should be enforced.
As a consequence, the field content of the bulk theory is reorganized into $\mathcal N=1$
multiplets; in particular the compactification moduli, as well as
the dilaton and axion fields, are assembled into complex scalars
within suitable chiral superfields, which couple to the $\mathcal
N=1$ vector and chiral multiplets living on the D-branes.

The tree-level effective action on the D-branes can be obtained in the field theory
limit $\alpha'\to 0$ from disk diagrams and takes the standard
form of an $\mathcal{N}=1$ supersymmetric action
in which the couplings are actually functions of the moduli due to the
possible interactions with closed string fields. In particular, the
gauge Lagrangian depends on the bulk moduli $M$ via its ``gauge kinetic function''
$f(M)$ which encodes the information on the Yang-Mills coupling $g_{\mathrm{YM}}$ and
the $\theta$-angle $\theta_{\mathrm{YM}}$ according to
\begin{equation}
f(M)=\frac{\theta_{\mathrm{YM}}}{2\pi}\,+\,\ii\,\frac{4\pi}{g_{\mathrm{YM}}^2}~,
\label{gym}
\end{equation}
so that the quadratic part in the gauge field strengths reads
\begin{equation}
 \label{biwzf}
-\frac{1}{8\pi}\int d^4x\,\, \im f(M) \, \Tr \big(F_{\mu\nu} F^{\mu\nu}\big)
+ \frac{1}{8\pi}\int d^4x\,\,\re f(M)\,  \Tr \big(
F_{\mu\nu} {}^*\!F^{\mu\nu}\big)~.
\end{equation}
Actually, the residual supersymmetry implies that the gauge Lagrangian
takes the form
\begin{equation}
 \label{gkf}
-\frac{\ii}{8\pi}\int d^2\theta \,\, f\big(M(\theta)\big)\, \Tr \big(W^\alpha(\theta)
W_\alpha(\theta)\big) + \cc
\end{equation}
where $W^\alpha(\theta)$ is the $\mathcal{N}=1$ gauge superfield whose
lowest component is the gaugino $\Lambda^\alpha$, while
the moduli $M$ in the gauge kinetic function $f$ get promoted to chiral superfields $M(\theta)$.

This is very interesting in two respects. First, the determination
of the gauge kinetic functions for different types of branes
preserving the same $\mathcal{N}=1$ supersymmetry suggests a way
to assemble the bulk moduli and their superpartners into
$\mathcal{N}=1$ chiral multiplets. Second, the Lagrangian
(\ref{gkf}) contains a gaugino mass term, which arises whenever
the $\theta^2$ component of $f\big(M(\theta)\big)$ assumes a
non-zero vacuum expectation value. As we will see such mass terms
can be related to the flux-induced fermionic couplings computed in
Section \ref{sec:effects} (see in particular \eq{massD3}). To
establish the precise relation we need to determine both the gauge
kinetic functions for the D-branes used to engineer the gauge
theory, and the appropriate complex combinations of the
compactification moduli $M$ that can be promoted to chiral
superfields. This is what we do in the following.

\paragraph{Gauge kinetic function}
Let us take a fractional D3-brane, say of type $A$. To deduce its gauge
kinetic function $f_A$ we have several possibilities. We can derive the
quadratic terms in the gauge fields of \eq{biwzf} from disk diagrams, with the
boundary attached to the brane and with two open string vertices for the gauge field
inserted on the boundary and closed string scalar vertices in the interior.
Alternatively, we can compute the coupling among closed strings and the boundary
state $\ket{A}_F$ representing the fractional D3-brane with a
constant magnetic field $F$ turned on in the world-volume and infer from it the gauge
kinetic function $f_A$ (see {\it e.g.} Ref. \cite{DiVecchia:2005vm}).
Finally, we can simply read off the coupling from the
Dirac-Born-Infeld (DBI) and Wess-Zumino (WZ) actions of the fractional D3-brane.

The last option is viable if one regards the fractional D3-branes of type $A$
as D5-branes wrapped%
\footnote{This extends to the case of the fractional brane of type
$0$ by interpreting it as a D5-wrapped on $\sum_{A=1}^3 {\hat
e}^A$ with negative orientation and with a suitable magnetic flux
turned on.} on the twisted 2-cycle $\hat e^A$, as recalled in
Section \ref{sec:N1}. In this case the DBI action with an
additional Wess-Zumino term for the D5-brane  (in the string
frame) is
\begin{equation}
 \label{bid5t}
S_{D3,A}=-T_5\int_{D3}\int_{{\hat e}^A} \ee^{-\varphi} \sqrt{-\det\left(G +
\mathcal{F}\right)}+T_5\int_{D3}\int_{{\hat e}^A} \sum_{n=0}^3
C_{2n}\, e^{ \mathcal F   } ~,
\end{equation}
where $\mathcal F= B_2+2\pi\alpha'F$, and $T_5 =
T_3/(4\pi^2\alpha')$ with $T_3 =(2\pi)^{-1} (2\pi\alpha')^{-2}$
being the D3-brane tension.
Expanding to quadratic order in F and
using the non-zero background value of $B_2$ along
the vanishing cycles \cite{Aspinwall:1996mn}
\begin{equation}
\int_{{\hat e}^A} B_2=\frac 14(4\pi^2\alpha')
\end{equation}
one finds
\begin{equation}
 \label{ymtermt}
S_{D3,A}=-\frac{1}{16\pi}\int_{D3} \,\ee^{-\varphi}
\, F_{\mu\nu} F^{\mu\nu}+ \frac{1}{16\pi}\int_{D3} \,C_0\,F_{\mu\nu} {}^*\!F^{\mu\nu}+\ldots  ~.
\end{equation}
Promoting these expressions to the non-abelian case, which results in an extra factor of
1/2 due to the normalization of the colour trace, and comparing
with \eq{biwzf}, we can read off that the gauge kinetic function for the fractional D3-brane
of type $A$ is
\begin{equation}
 \label{gkfA}
f_A(M) = \frac{\tau}4  ~,
\end{equation}
with
\begin{equation}
 \label{gkf3}
\tau \equiv C_0 + \ii\ee^{-\varphi}
\end{equation}
the axion-dilaton field.
Combining \eq{gkfA} with Eqs. (\ref{gym}) and (\ref{gkf3}) leads to
\begin{equation}
g_{\mathrm{YM}}^2=16\pi\ee^{\varphi}\quad\mbox{and}\quad\theta_{\mathrm{YM}}=\frac{\pi C_0}{2}~.
\label{gym1}
\end{equation}
The way to arrange the remaining untwisted and twisted scalars as
the complex bosons of suitable chiral multiplets is suggested, as
remarked above, by the gauge kinetic functions of other D-branes
maintaining the same $\mathcal{N}=1$ supersymmetry selected by the
fractional D3-branes. For the untwisted scalars, we can just
consider ``regular'' branes, such as D7-ones wrapped on one of the
untwisted 4-cycles. Starting from the wrapped D7-brane DBI-WZ
action, in the end one finds (see for instance Ref.
\cite{Blumenhagen:2006ci}) that the gauge kinetic function for
these branes is
\begin{equation}
 \label{gkf7}
f_i(M) = t^i\quad\mbox{with}\quad
t^i \equiv \tilde c^i + \frac{\ii}{2} |\epsilon^{ijk}| v_j v_k
\end{equation}
where $v_i$ and $\tilde c^i$ have been defined at the beginning of this section.
The complex fields $t^i$ represent
the correct (untwisted) K\"ahler coordinates to be used for the $\mathcal{N}=1$
supergravity associated to CY compactifications with D3/D7-branes and O3/O7-planes,
together with the $\tau$ variable defined in \eq{gkf3}.
Notice also that the imaginary parts
of the coordinates (\ref{gkf7}) are related to the volume $\mathcal{V}$ of
the $\mathcal{T}_6/\mathbb{Z}_2\times\mathbb{Z}_2$ orbifold, measured in the Einstein frame; in fact
\begin{equation}
 \label{vol}
\Big(\im t^1\, \im t^2\, \im t^3\Big)^\frac{1}{2} = v_1 v_2 v_3 =
\mathcal{V}~.
\end{equation}

Let us now return to the gauge theory defined on the fractional D3-branes and on its gauge kinetic function $f_A=\tau/4$. The modulus $\tau$ is connected by
the residual supercharges to other closed string states and it
can be promoted to a chiral superfield $\tau(\theta)$. 
The complete Lagrangian of the fractional D3-branes, given in \eq{gkf},
contains then also the coupling of the gaugino $\Lambda^\alpha$ to the
auxiliary component $F^\tau$ of $\tau(\theta)$, namely
\begin{equation}
\label{gaugaux}
-\,\frac{\ii}{8\pi}\,\frac{F^\tau}{4}\,\Tr \big(\Lambda^\alpha \Lambda_\alpha\big) + \cc ~,
\end{equation}
which corresponds to the following mass
\begin{equation}
\label{cangaugmass}
m_\Lambda = \frac{1}{2 \im f_A}\, \frac{F^\tau}4
= \frac{\ee^\varphi\,F^\tau}{2}
\end{equation}
for canonically normalized gaugino fields.

The bilinear term (\ref{gaugaux}) must coincide
with the flux-induced coupling we have computed in Section \ref{subsec:masses}. In fact,
in presence of a $G$-flux the gauginos acquire mass terms given
by \eq{massD3} which must be adapted to our $\mathcal N=1$ orbifold model. This is
easily done by taking only the invariant gaugino $\Lambda^0\equiv\Lambda$.
Using Appendix \ref{app:conventions} (and in particular \eq{sigma00}) we find that
the only component of the $G^{\mathrm{IASD}}_{mnp}$ tensor
which contributes to \eq{massD3} when $A=B=0$, is its $(3,0)$ part; thus, after combining
Eqs. (\ref{cfD3}) and (\ref{gym1}), we find that the flux-induced
gaugino mass term for fractional D3-branes reads
\begin{equation}
 \label{gmt}
-\frac{\ii}{2}\,\ee^{-\varphi}\,(2\pi\alpha')^{-\frac{1}{2}}\mathcal{N}_F\,
G_{(3,0)} \, \Tr \big(\Lambda^\alpha
\Lambda_\alpha\big) + \cc ~.
\end{equation}
Comparing with \eq{gaugaux}, we finally deduce that
\begin{equation}
 \label{Ftaubi}
F^\tau =  16\pi\, \ee^{-\varphi}\,(2\pi\alpha')^{-\frac{1}{2}}\mathcal{N}_F\,
G_{(3,0)} ~.
\end{equation}
Later in this section, we will fix the normalization $\mathcal{N}_F$ of the
flux vertex by requiring that this expression for $F^\tau$
matches the one obtained by constructing the bulk low energy Lagrangian.

\paragraph{Comparison with the bulk theory}
It is well known (see for example Refs.
\cite{Grana:2003ek,Grimm:2004uq}) that the bulk theory for our
toroidal compactification yields, after a dimensional reduction to four
dimensions and a Weyl rescaling to the $d=4$ Einstein frame, a
$\mathcal{N}=1$ supergravity theory coupled to vector and matter
multiplets in the standard form. This effective theory is
therefore specified, besides the gauge kinetic function for the
bulk vector multiplets, by the K\"ahler potential $K$ and by the
holomorphic superpotential $W$ for the chiral multiplets. To
simplify the treatment, in the following we consider as dynamical
only a subset of the compactification moduli; in particular we
keep the dependence on the universal chiral multiplet $\tau$ of
\eq{gkf3}, but restrict to a slice of the K\"ahler moduli space in
which an overall scale
\begin{equation}
 \label{kms}
t \equiv t^1 = t^2 = t^3
\end{equation}
is considered. Such a scale is related to the compactification
volume by $\mathcal{V} = (\im t)^{3/2}$ as it follows from \eq{vol}.
We also freeze the complex structure moduli $u^i$ to their
``trivial'' value corresponding to $\mathcal{T}_6$ being the product
of three upright tori, {\it {i.e.}} we set $u^1= u^2 = u^3 = \ii$;
furthermore we neglect the dependence on all the remaining twisted
and untwisted moduli.

With these assumptions, the K\"ahler potential for the bulk theory is
\begin{equation}
\label{KP}
 K =-\log(\im \tau) - 3 \log(\im t)~.
\end{equation}
When internal 3-form fluxes are turned on, a non-trivial bulk superpotential appears \cite{Gukov:1999ya,Taylor:1999ii} and its expression is
\begin{equation}
 \label{WGO}
W = \frac{1}{\kappa_{10}^2} \int G\wedge \Omega =
\frac{4}{\kappa_4^2} G_{(0,3)}
\end{equation}
where $\Omega$ is the holomorphic 3-form of the internal space, and $\kappa_{10}$ and $\kappa_4$
are, respectively, the gravitational constants in ten and four dimensions%
\footnote{In our case the holomorphic 3-form is simply given by
$\Omega= dZ^1\wedge dZ^2\wedge dZ^3$. In our conventions, we have
$\int \bar\Omega\wedge \Omega = (2\pi\sqrt{\alpha'})^6$, while $\kappa_{10}^2/\kappa_4^2 =
(2\pi\sqrt{\alpha'})^6/4$, where the factor of $1/4$ represents the order of the orbifold group.}.
In \eq{WGO} the 3-form flux is
\begin{equation}
G= F -\tau \,H
\label{G3}
\end{equation}
which is the natural extension of \eq{G} when $g_s$ is promoted to $\ee^\varphi$ and the
presence of a non-vanishing axion $C_0$ is taken into account.
Note that $W$ has the correct dimensions of (length)$^{-3}$, since $\kappa_4$ is a length and
the flux is a mass, and that only the ISD component $G_{(0,3)}$ of $G$ is responsible
for a non-vanishing $W$.

In presence of a superpotential $W$, the auxiliary fields in the chiral multiplets are given
by the standard supergravity expressions which in our case become
\begin{equation}
 \label{FW}
\begin{aligned}
{\overline F}^{\,\bar \tau} & = -\ii \,\kappa_4^2\,\ee^{K/2} \,K^{\bar\tau \tau}\, D_{\tau} W
\,=\, 8 \,\frac{\ee^{-\varphi/2}}{\mathcal{V}} \,\overline{G}_{(0,3)}\quad \Rightarrow \quad
{F}^{\tau} \,=\, 8 \,\frac{\ee^{-\varphi/2}}{\mathcal{V}} \,G_{(3,0)}~,\\
{\overline F}^{\,\bar t} & =- \ii\, \kappa_4^2\,\ee^{K/2}\, K^{\bar t t} \,D_{t} W
\,=\, 8\, \frac{\ee^{\varphi/2}}{\mathcal{V}^{\frac 13}}\, {G}_{(0,3)}
\quad \Rightarrow \quad
{F}^{t} \,=\,   8\, \frac{\ee^{\varphi/2}}{\mathcal{V}^{\frac 13}}\, \overline{G}_{(3,0)}    ~,
\end{aligned}
\end{equation}
where $\overline{G}$ is the complex conjugate of $G$,
$K^{\bar\tau \tau}$ and $K^{\bar t t}$ are the inverse K\"ahler metrics for $\tau$ and
$t$ respectively, and the K\"ahler covariant derivatives of the superpotential
are defined as $D_iW=\partial_iW+\big(\partial_iK\big)W$.
Thus, by comparing the expression of $F^\tau$ derived from the
flux-induced gaugino mass and given in \eq{Ftaubi} with \eq{FW}, we find perfect
agreement in the structure and can fix the normalization
of the closed string vertex for the flux to be
\begin{equation}
 \label{NFvalue}
\mathcal{N}_F = \frac{\ee^{\varphi/2}}{2\pi\mathcal{V}} \,(2\pi\alpha')^{\frac{1}{2}}~.
\end{equation}
{F}rom \eq{NfNh} we also infer that (promoting $g_s$ to $\ee^{\varphi}$)
\begin{equation}
 \label{NHvalue}
\mathcal{N}_H = \frac{\ee^{-\varphi/2}}{2\pi\mathcal{V}}\,(2\pi\alpha')^{\frac{1}{2}}~.
\end{equation}
With these normalizations, the closed string vertices (\ref{vertexRR})
and (\ref{vertexNS}) can be used to derive directly from string amplitudes  the
terms in the four dimensional effective Lagrangian in the Einstein frame, and the
resulting expressions do indeed have the correct normalization that follows from
the dimensional reduction of the original Type IIB action in ten dimensions.
In this perspective, we point out that
the scalar potential due to the chiral multiplets, which has the form
\begin{equation}
 \label{scalpot}
 \begin{aligned}
 V_F  &= \kappa_4^2\,\,\ee^K \left(K^{\tau\bar\tau} D_\tau W D_{\bar\tau}\bar
 W + K^{t\bar t} D_t W D_{\bar t}\overline W
 - 3\, |W|^2\right) \\
 &= \frac{16}{\kappa_4^2}\,\, \frac{\ee^\varphi}{\mathcal{V}^2} \,
 {G}_{(3,0)}\,\overline{G}_{(0,3)} = \frac{1}{\kappa_4^2}\,\,
 \Big|4\,\frac{\ee^{\varphi/2}}{\mathcal{V}} \,{G}_{(3,0)}\Big|^2 ~,
 \end{aligned}
\end{equation}
coincides with the kinetic terms for the R-R
and NS-NS 3-forms in the ten dimensional Einstein frame, given in  \eq{bulk10},
after dimensional reduction to $d=4$ and rescaling
to the four dimensional Einstein frame,
if only the (3,0) and (0,3) components of the fluxes
are turned on.

Let us also recall that the last term of $V_F$ in the first line of \eq{scalpot} is a purely ``gravitational'' contribution, related to the gravitino mass
\begin{equation}
 \label{gravmass}
\big|m_{3/2}\big| = \kappa_4^2\, \ee^{K/2} \,\big|W\big| =
\Big|4\,\frac{\ee^{\varphi/2}}{\mathcal{V}} \,G_{(0,3)}\Big|~.
\end{equation}
{F}rom these equations, we see clearly the very different r\^ole of the ISD flux $G_{(0,3)}$,
which induces a gravitino mass, with respect to the
IASD flux $G_{(3,0)}$, which is instead responsible for the gaugino mass term. The latter
is described by Eq.s (\ref{gaugaux}) and (\ref{FW}) which correspond, according to \eq{cangaugmass},
to a canonical gaugino mass
\begin{equation}
 \label{gmG}
\big| m_\Lambda \big| = \Big|4\,\frac{\ee^{\varphi/2}}{\mathcal{V}}\,G_{(3,0)}\Big|~.
\end{equation}
These very well-known results \cite{Camara:2003ku,Grana:2003ek,Camara:2004jj} will
be generalized and extended to instantonic branes in the following section, and the effects
on the instanton moduli space of a flux-induced mass term for the gaugino or the gravitino
will be determined. We conclude this section by mentioning that the same analysis we have
described for fractional D3 branes can be performed without any difficulty
in the case of fractional D9 branes.
Some details on this are provided in Appendix \ref{subsec:fD9}.

\section{The r\^ole of fluxes on fractional D-instantons}
\label{sec:fD-1} 
In Sections \ref{sec:effects} and \ref{sec:twisted}
we have computed the fermionic bilinear couplings of the NS-NS and
R-R bulk fluxes to open strings with at least one end-point on the
D-instanton. The results (\ref{massD-1}) and (\ref{mumubtot})
describe deformations of the instanton moduli space of the
${\cal N}=4$ gauge theory living on the D3-brane. We now discuss the
meaning of these interaction terms in a simple example within the
context of the $\mathcal N=1$ orbifold compactification introduced
in the last section.

Consider a specific node $A$ of the quiver diagram represented in
Fig. \ref{fig:quiver} and put on it $N$ fractional D3 branes and one
fractional D-instanton. The latter describes the $k=1$ gauge
instanton for the $\mathcal N=1$ $\mathrm U(N)$ Yang-Mills theory
defined on the world-volume of the space-filling D3-branes. The open
strings with at least one end point on it account for the instanton
zero-modes in the ADHM construction
\cite{Witten:1995gx}\nocite{Douglas:1995bn,Green:2000ke}-\cite{Billo:2002hm}.
Specifically, for the D$(-1)$/D$(-1)$ strings,  we have the four bosonic
zero-modes $x^\mu$ and three auxiliary zero-modes $D_c$ from the NS
sector plus two chiral zero-modes $\theta^\alpha$ and two
anti-chiral zero-modes $\lambda_{\dot\alpha}$ from the R sector.
Besides these neutral modes, there are also charged zero-modes from
the D3/D$(-1)$ and D$(-1)$/D3 strings comprising the bosons
$w_{\dot\alpha}^{\,\,u}$ and $\bar w_{\dot \alpha u}$ from the NS
sector, and the scalar fermions $\mu^u$ and $\bar\mu_u$ from the R
sector, where the upper (or lower) index $u$ belongs to the
fundamental (or anti-fundamental) representation of $\mathrm U(N)$.

The action of the $\mathcal N=1$ fractional D-instanton zero-modes turns out to be
(see {\it e.g.} Ref. \cite{Dorey:2002ik})
\begin{equation}
S_{\mathrm{inst}}=2\pi\ii\,f_A +\ii\,\lambda_{\dot\alpha}\big(
\bar\mu_u w^{\dot\alpha u} + \bar w^{\dot\alpha}_{\,\,u} \mu^u\big)
-\ii D_c\,\bar w_{\dot\alpha u}
(\tau^c)^{\dot\alpha}_{\,\,\dot\beta}w^{\dot\beta u} \label{sd-1}
\end{equation}
where $f_A=\tau/4$ is the gauge kinetic function (\ref{gkfA}) and
$\tau^c$ are the three Pauli matrices. Note that neither $x^\mu$ nor
$\theta^\alpha$ appear in $S_{\mathrm D(-1)}$; in fact they are the
Goldstone modes of the supertranslation symmetries broken by the
instanton and as such can be identified with the superspace
coordinates of the $\mathcal N=1$ theory. On the other hand
$\lambda_{\dot\alpha}$ and $D_c$ appear only linearly in
$S_{\mathrm D(-1)}$: they are Lagrange multipliers
enforcing the so-called super ADHM constraints. The action
(\ref{sd-1}) can be easily derived by computing (mixed) disk
amplitudes with insertions of vertex operators representing the
various zero-modes \cite{Billo:2002hm}.

Let us focus in particular on the fermionic moduli. The neutral zero-modes
$\theta^\alpha$ and $\lambda_{\dot\alpha}$ are clearly described by the chiral and anti-chiral
components of the D$(-1)$/D$(-1)$ fermion that is invariant under the orbifold
action ({\it
i.e.} with an internal spinor index 0), namely
\begin{equation}
 \Theta^{\alpha 0}\, \sim\, g_0\, \theta^{\alpha}\,\quad\mbox{and}\quad
\Theta_{\dot\alpha 0} \,\sim\, \lambda_{\dot\alpha}\,~.
\label{thetalambda}
\end{equation}
The extra power of the D$(-1)$ gauge coupling
$g_0=\frac{1}{\sqrt\pi}(2\pi\alpha')^{-1}\ee^{\varphi/2}$
accounts for the correct scaling dimensions that allow to interpret $\theta^\alpha$ as
the fermionic superspace coordinate with dimensions of (length)$^{1/2}$. On the
other hand, as mentioned above, $\lambda_{\dot\alpha}$ is the Lagrange multiplier
for the fermionic ADHM constraint and carries dimensions of (length)$^{-{3}/{2}}$, so that
no rescaling is needed.

In the charged sector the fermionic moduli $\mu^u$ and $\bar\mu_u$
correspond to the $\mathbb{Z}_2\times\mathbb{Z}_2$ invariant fermions
of the strings stretching between the D3-branes and the D-instanton so that, using the
notation of Section \ref{sec:twisted}, for each colour we have
\begin{equation}
\mu^0\,\sim\,g_0\,\mu \quad\mbox{and}\quad\bar \mu^0\,\sim\,g_0\,\bar \mu~.
\label{mumubar}
\end{equation}
As before, an extra power of $g_0$ is included to account for the correct
(length)$^{{1}/{2}} $ dimensions of the charged moduli $\mu,\bar \mu$.
The normalizations%
\footnote{The extra factor of
$4\sqrt\pi\,\ee^{-\varphi/2}$ in the definition of $\mathcal{N}_\theta$ with
respect to \cite{Billo:2002hm} is needed, as we will see, in order
to identify $\theta$ with the superspace coordinates. } of the fermionic string vertices can then be
written as \cite{Billo:2002hm} 
\begin{equation}
{\mathcal N}_{\lambda} = {(2\pi\alpha')^{\frac{3}{4}}}\quad,\quad
{\mathcal N}_\theta =  4\sqrt\pi\,\ee^{-\varphi/2}\,\frac{g_0}{\sqrt{2}} \,(2\pi\alpha')^{\frac{3}{4}}\quad\,\quad
 {\mathcal N}_\mu ={\mathcal N}_{\bar\mu} =
\frac{g_0}{\sqrt{2}}{(2\pi\alpha')^{\frac{3}{4}}} ~.
\label{nmu}
\end{equation}

We are now ready to study how the bulk R-R and NS-NS fluxes modify the moduli
action. Actually in Section \ref{sec:effects} we have already computed the flux interactions
with the untwisted fermions of a D$(-1)$-brane (see Eq. (\ref{massD-1})) while in
Section \ref{sec:twisted} we computed the flux couplings to the twisted fermions
of the D3/D$(-1)$ system (see Eq. \ref{mumubtot}). So what we have to do now is simply
to insert in these equations the appropriate normalizations discussed above and
take into account the identifications of the fluxes with the bulk chiral multiplets explained
in the previous section. The flux induced terms in the instanton moduli action are thus%
\footnote{Recall that in Euclidean spaces there is a minus sign in going from an amplitude to
an action.}
\begin{equation}
S_{\mathrm{inst}}^{\mathrm{flux}} = - \mathcal A^{\rm flux}_{{\mathrm D}(-1)} - \mathcal A^{\rm flux}_{\mathrm{D3/D(-1)}}
\label{sflux}
\end{equation}
where $\mathcal A^{\rm flux}_{{\mathrm D}(-1)}$ and $\mathcal A^{\rm flux}_{\mathrm{D3/D(-1)}} $
are the $A=B=0$ parts of the amplitudes (\ref{massD-1})
and (\ref{mumubtot}), {\it i.e.}
\begin{equation}
\begin{aligned}
\mathcal A^{\mathrm{flux}}_{{\mathrm D}(-1)}{\phantom{\vdots}} &= -2\pi\ii\,c_F(\theta)\,\theta^\alpha\theta_\alpha\,{G}_{(3,0)}
+ 2\pi\ii\,c_F(\lambda)\,\lambda_{\dot\alpha}\lambda^{\dot\alpha}\,{G}_{(0,3)}~,\\
\mathcal A^{\mathrm{flux}}_{\mathrm{D3/D(-1)}}{\phantom{\vdots}}
&=-4\pi\ii\,c_F(\mu)\,\bar\mu_u\,\mu^u\,{G}_{(3,0)}~.
\end{aligned}
\label{sflux1}
\end{equation}
In these expressions we have distinguished the $c_F$ coefficients
for the various terms to account for the appropriate normalizations
of the moduli as discussed before.
Recalling that the normalization
$\mathcal C_{(0)}$ of the disk amplitudes is given by Eq. (\ref{c0})
with $g_{\mathrm{YM}}^2$ defined in (\ref{gym1}), and that the $G$
fluxes are normalized as indicated in Eq. (\ref{NFvalue}), we find
\begin{equation}
\begin{aligned}
c_F(\theta)&= \mathcal C_{(0)}\,\mathcal N_F\,\mathcal N_\theta^2=
2\,\frac{\ee^{-\varphi/2}}{\mathcal{V}}~,\\
c_F(\lambda)&=  \mathcal C_{(0)}\,\mathcal
N_F\,\mathcal N_{\lambda}^2
= (2\pi\alpha')^{2}\,\frac{\ee^{-\varphi/2}}{4\mathcal{V}}~,\\
c_F(\mu)&= \mathcal C_{(0)}\,\mathcal N_F\,\mathcal N_\mu^2
= \frac{\ee^{\varphi/2}}{8\pi\mathcal{V}}~.
\end{aligned}
\label{cfs}
\end{equation}
Notice that all factors of $\alpha'$ cancel in $c_F(\theta)$ and
$c_F(\mu)$, but they survive in $c_F(\lambda)$ whose scaling
dimension of (length)$^4$ is the correct one for the $\lambda^2$
term of $\mathcal A^{\mathrm{flux}}_{{\mathrm D}(-1)}$ in (\ref{sflux1}) to be dimensionless.
Using these coefficients in (\ref{sflux1}) and exploiting the results of the previous section
(in particular Eqs. (\ref{gmG}) and (\ref{gravmass})), we can
rewrite the flux-induced moduli action as follows
\begin{equation}
\begin{aligned}
S_{\mathrm{inst}}^{\mathrm{flux}}\,& =\,4\pi\ii\frac{\ee^{-\varphi/2} G_{(3,0)}}{\mathcal V}\,\theta^\alpha\theta_\alpha
\,-\,\ii\pi\,(2\pi{\alpha'})^2\,\frac{\ee^{-\varphi/2} G_{(0,3)}}{2{\mathcal V}} \,\lambda_{\dot\alpha}\lambda^{\dot\alpha} \,+\,\ii\,
\frac{ G_{(3,0)}}{2{\mathcal V}} \,\bar\mu_u \mu^u\\
& =\,\frac{\ii\pi}{2}\,F^\tau\,\theta^\alpha\theta_\alpha
\,-\,\frac{\ii\pi}{8}\,\kappa_4^2\,(2\pi{\alpha'})^2\,{\ee^{-\varphi}\,\ee^{K/2}} \,\lambda_{\dot\alpha}\lambda^{\dot\alpha} \,+\,
\frac{\ii\,\ee^{\varphi}}{16} F^\tau \,\bar\mu_u \mu^u~.
\end{aligned}
\label{sflux2}
\end{equation}
The $\theta^2$ term represents the auxiliary component of the gauge
kinetic function $f_A=\tau/4$, which is therefore promoted to the
full chiral superfield $f_A(\theta)=\tau(\theta)/4$ in complete (and
expected) analogy with what happened on the D3-branes. The other two
terms are less obvious: they represent the explicit effects of a
background $G$ flux on the instanton moduli space, and are the
strict analogue for the instanton action of the soft supersymmetry
breaking terms of the gauge theory. In particular the $\bar\mu\mu$
term is related to the IASD flux component $G_{(3,0)}$ which is
responsible for the gaugino mass $m_\Lambda$, while the $\lambda^2$
term represents a truly stringy effect on the instanton moduli space
and is related to the ISD flux component $G_{(0,3)}$ which gives
rise to the gravitino mass $m_{3/2}$.

The study of these terms, of their consequences for the instanton
calculus and of the non-perturbative effects that they may induce in
the gauge theory will be presented in a companion paper \cite{Billo:2008pg}. Here we
simply mention that the above analysis can be easily generalized to
SQCD models with flavored matter in the fundamental
(anti-fundamental) representation and also to configurations in
which the fractional D-instanton occupies a node of the quiver
diagram which is {\em not} occupied by the colored or flavored
space-filling branes. For these ``exotic'' instanton configurations
there are no bosonic moduli $w_{\dot\alpha}$ and $\bar
w_{\dot\alpha}$ and the action (\ref{sd-1}) simply reduces to first
term involving the gauge kinetic function. Since the neutral
anti-chiral fermionic moduli $\lambda_{\dot\alpha}$ do not couple
to anything, to avoid a trivial vanishing result upon integration
over the moduli space, it is necessary to remove them
or to lift them. As we have explicitly seen, by coupling
the fractional D-instanton to an ISD $G$-flux of type (0,3) it is
possible to achieve this goal exploiting the $\lambda^2$ term
proportional to the gravitino mass.

Finally, we observe that using the explicit expression (\ref{tE3a}) of the
fermionic coupling, the flux-induced moduli amplitude on a E3 instanton
$\mathcal A^{\mathrm{flux}}_{{\mathrm E}3}$ contains terms of the form
\begin{equation}
\theta^\alpha\theta_\alpha\,{\bar G}_{(3,0)} \quad\mbox{and}\quad
\lambda_{\dot\alpha}\lambda^{\dot\alpha}\,{\bar G}_{(0,3)}
\label{AE3}
\end{equation}
where ${\bar G}_{(3,0)}$ and ${\bar G}_{(0,3)}$ are, respectively, the $(3,0)$ and
$(0,3)$ components of $\bar G= F +\frac{\ii}{g_s}\,H$. Thus, on a E3 instanton
a $G$-flux of type (2,1) or (0,3) cannot lift the $\lambda_{\dot\alpha}$'s since its
conjugate flux $\bar G$ does not contain a $(0,3)$-component, in full agreement with the
findings of Ref. \cite{Blumenhagen:2007bn}. However, as is clear from (\ref{AE3}),
a $G$-flux of type (3,0) can lift the anti-chiral zero-modes $\lambda_{\dot\alpha}$.

\section{Summary of results}
\label{sec:summary}

In this paper we have computed the couplings of NS-NS and R-R fluxes to
fermionic bilinears living on general brane intersections (including
instantonic ones). The couplings have been extracted from disk amplitudes
among two open string vertex operators and one closed string vertex
representing the background fluxes. The results for the R-R and
NS-NS amplitudes are given in Eqs. (\ref{amplFfinal}) and (\ref{amplHfinal}).
At leading order in $\alpha'$ they describe fermionic mass terms
induced at linear order in the R-R and NS-NS fluxes for open string
modes with boundary conditions encodes in the magnetized reflection
matrices $R_0,{\cal R}_0$ and the open string twists
$\vec\vartheta$.

The case $\vec\vartheta=0$ corresponds to open strings starting and
ending on two parallel D-branes. The result in this case can be
written in the simple form
\begin{equation}
\mathcal A  =-\frac{2\pi\ii}{3!}\,c_F\, \Theta\Gamma^{MNP}\Theta\,
T_{MNP} 
\label{ampltot1s}
\end{equation}
where $c_F$ is a normalization factor and
\begin{equation}
T_{MNP} = \big(F\mathcal{R}_0\big)_{MNP}+\frac{3}{g_s}\,\big(
\partial B R_0\big)_{[MNP]} ~. 
\label{tmnls}
\end{equation}
This formula shows that different branes couple to different
combinations of the R-R and NS-NS fields.
For compactifications to $d=4$ in presence of 3-form internal fluxes
the explicit form of the $T$ tensors are displayed in
Tab.~\ref{Dbranes} for spacetime filling D-branes and in
Tab.~\ref{Ebranes} for instantonic branes. For spacetime filling
branes, the $T$-tensor describes the structure of soft fermionic
mass terms for a general D-brane intersection.  For Euclidean branes, they
accounts for fermionic mass terms in the instanton moduli space
action modifying the fermionic zero mode structure of the instanton.
Our results are in perfect agreement with
those of Refs.
\cite{Grana:2002tu,Marolf:2003ye,Camara:2003ku,Camara:2004jj,Martucci:2005rb,Tripathy:2005hv,Bergshoeff:2005yp}
that have been derived with pure supergravity methods and generalize
them to generic (instantonic or not) D-brane intersections. The
effects of open string magnetic fluxes can be easily incorporated
into these formulas via the reflection matrices  $R_0({\mathcal F})$ and
${\mathcal R}_0({\mathcal F})$. 
As an example, the explicit form of the $T$-tensor for Euclidean magnetized E5-branes 
has been given in \eq{te5m}.

The cases of open strings ending on D3-branes and D-instantons
have been studied in detail.  For D3-branes in flat space one obtains
\begin{equation}
{\mathcal A}_{\mathrm D3} = \frac{2\pi \ii}{3!}\,c_F(\Lambda)\,{\mathrm{Tr}}
\Big[\, \Lambda^{\alpha A}
\Lambda_{\alpha}^{{\phantom \alpha}B}
\big(\overline\Sigma^{mnp}\big)_{AB}\,G_{mnp}^{\mathrm{IASD}} -
\bar\Lambda_{\dot\alpha A}\bar\Lambda^{\dot\alpha}_{{\phantom\alpha}B}
\big(\Sigma^{mnp}\big)^{AB}\,\big(G_{mnp}^{\mathrm{IASD}}\big)^*
\,\Big]
\label{massD3s}
\end{equation}
with $G=F-\tau H$, $G^{\mathrm{IASD}}$ its imaginary
anti-self-dual part and $c_F(\Lambda)$ a normalization factor.
This formula encodes the structure of soft
symmetry breaking terms in ${\cal N}=4$ gauge theory induced by
NS-NS and R-R fluxes.

The coupling of fluxes to D-instantons is given instead by
\begin{equation}
{\mathcal A}_{{\mathrm D}(-1)} = \frac{2\pi\ii}{3!}\, \Big[\,c_F(\theta)\,
\theta^{\alpha A} \theta_{\alpha}^{{\phantom\alpha} B}
\big(\overline\Sigma^{mnp}\big)_{AB}\,G_{mnp}^{\mathrm{IASD}}
+c_F(\lambda)\,\lambda_{\dot\alpha A}\lambda^{\dot\alpha}_{{\phantom
\alpha}B} \big(\Sigma^{mnp}\big)^{AB}\,G_{mnp}^{\mathrm{ISD}}
\,\Big]~.
\label{massD-1s}
\end{equation}

The case  $\vec\vartheta\neq 0$ describes the couplings of open strings stretching between
non-parallel stacks of D-branes. For spacetime filling D-branes the corresponding
open string excitations describe chiral matter transforming in bi-fundamental
representations of the gauge group and always contain massless chiral fermions.
The case, where open strings are
twisted by $\vartheta=\frac{1}{2}$ along the spacetime directions,
describes the charged moduli of gauge or exotic instantons. For gauge
instantons in ${\cal N}=4$ gauge theory one finds the flux induced action
\begin{equation}
{\mathcal A}_{\mathrm{D3/D(-1)}} =\frac{4\pi\ii
}{3!}\,c_F(\mu)\,{\bar\mu}^{A}\mu^{B}
\,\big(\overline\Sigma^{mnp}\big)_{AB}\,G_{mnp}^{\mathrm{IASD}}~.
\label{mumubtots}
\end{equation}
The results obtained here extend straightforwardly to less
supersymmetric theories and to exotic instantons. In particular for pure ${\cal N}=1$ SYM,
the flux couplings for both gauge and exotic instantons follow from
(\ref{massD3s},\ref{massD-1s},\ref{mumubtots}) by restricting the
spinor components to $A=B=0$. The only contributions to fermionic
mass terms come in this case from the components $G_{(3,0)}$ and
$G_{(0,3)}$ related to the soft symmmetry breaking gaugino and
gravitino masses. Explicitly for
$\mathcal{T}_6/(\mathbb{Z}_2\times\mathbb{Z}_2)$ we have
\begin{equation}
 \label{gravmasscon}
\begin{aligned}
&\big|m_\Lambda\big| = \Big| \frac{\kappa_4^2}{2} \,\ee^\varphi\,\ee^{K/2}\,D^\tau W\Big|
= \Big|4 \,\frac{\ee^{\varphi/2}}{\mathcal{V}} \,G_{(3,0)}\Big| ~,\\
&\big|m_{3/2}\big| = \Big| \kappa_4^2\, \ee^{K/2}\,W \Big| 
= \Big|4\,\frac{\ee^{\varphi/2}}{\mathcal{V}} \,G_{(0,3)}\Big|~,
\end{aligned}
\end{equation}
and the fermionic flux couplings can be written as
\begin{equation}
\begin{aligned}
&{\mathcal A}_{\mathrm D3} = -\frac{\ii}{16\pi}\,m_\Lambda\,\ee^{-\varphi}\,
 {\mathrm{Tr}}\big[\, \Lambda^\alpha \Lambda_\alpha
\big]+{\rm c.c.}~,\\
&{\mathcal A}_{{\mathrm D}(-1)} =
-\pi\ii\,m_\Lambda\,\ee^{-\varphi}\,\theta^\alpha\theta_\alpha\,
+\frac{\pi\ii}{8}\,(2\pi{\alpha'})^2\,m_{3/2}\,\ee^{-\varphi}\, \lambda_{\dot\alpha}\lambda^{\dot\alpha}~,\\
&{\mathcal A}_{\mathrm{D3/D(-1)}} =-\frac{\ii}{8}\,m_\Lambda
\,\bar\mu_u \mu^u~.
\end{aligned}
\label{massfs}
\end{equation}
These flux couplings modify the zero mode structure of the instanton and
allow for new low energy coupling in the D3-brane action. In
particular, we observe that the presence of $\lambda$-fermionic zero modes
typically signals an obstruction to the generation of
non-perturbative superpotentials via exotic instantons. This
difficulty can be overcome by the introduction of an O-plane
leading to O(1)-instantons with no $\lambda$-modes. The presence
of the $\lambda^2$-term in (\ref{massfs}) suggests that
R-R and NS-NS fluxes can provide a valid alternative mechanism in the
case of oriented gauge theories.
 A precise study of the low energy couplings generated by instantons in presence
 of such fluxes will be presented in the companion paper \cite{Billo:2008pg}.

\vskip 1cm 
\noindent {\large {\bf Acknowledgments}} 
\vskip 0.2cm
\noindent We thank C. Bachas, P. Di Vecchia, D. Du\`o, R. Marotta, I. Pesando, R. Russo
and M. Serone for many useful discussions. This work is partially
supported by the European Commission FP6 Programme under contracts
MRTN-CT-2004-005104 (in which A.L. is associated to the
University of Torino), MRTN-CT-2004-512194 and MRTNCT-2004-503369 and by the NATO grant
PST.CLG.978785.
L.F. would like to thank C. Bachas for kind support and 
LPTENS for wonderful hospitality. 

\vskip 1cm
\appendix
\section{Appendix}
\subsection{Notations and conventions}
\label{app:conventions}

We use the following notations for space-time indices in the real basis:
\begin{itemize}
\item
$d=10$ vector indices: $M,N, ... \in \{0, ..., 9\}$;
\item
$d=4$ vector indices: $\mu,\nu, ... \in \{0, ..., 3\}$;
\item
$d=6$ vector indices: $m,n, ... \in \{4, ..., 9\}$.
\end{itemize}
The corresponding complex indices are denoted by  $I, J=1, ..., 5$ with $I=i=1,2,3$ referring
to the coordinates of the six-dimensional space and $I=4,5$ referring to the four-dimensional
space-time directions. Even with Euclidean signatures we use the real index 0.

\paragraph{$\Gamma$-matrices in ten dimensions}
In a $d=10$ Euclidean space the $\Gamma$ matrices which satisfy
$\acomm{\Gamma^M}{\Gamma^N} = 2\delta^{MN}$,
can be given the following explicit representation in terms of the Pauli matrices $\tau^c$:
\begin{equation}
\label{gm10}
\begin{aligned}
\Gamma^0 & = \tau^1 \otimes ~\mathbf{1} \otimes ~\mathbf{1} \otimes ~\mathbf{1} \otimes ~\mathbf{1}
\\
\Gamma^1 & = \tau^2 \otimes ~\mathbf{1} \otimes ~\mathbf{1} ~\otimes \mathbf{1}~ \otimes \mathbf{1}
\\
\Gamma^2 & = \tau^3 \otimes \tau^1 \otimes ~\mathbf{1} \otimes ~\mathbf{1} \otimes ~\mathbf{1}
\\
\Gamma^3 & = \tau^3 \otimes \tau^2  \otimes ~\mathbf{1} \otimes ~\mathbf{1} \otimes ~\mathbf{1}
\\
& ~~~~~~~~~~~~\vdots
\\
\Gamma^{8} & =  \tau^3 \otimes \tau^3 \otimes \tau^3 \otimes \tau^3  \otimes \tau^1
\\
\Gamma^{9} & = \tau^3 \otimes \tau^3 \otimes \tau^3 \otimes \tau^3  \otimes \tau^2
\end{aligned}
\end{equation}
The charge conjugation matrix $C$ satisfies
\begin{equation}
\label{Cdef}
C  \Gamma^M C^{-1} = - (\Gamma^M)^t\quad\mbox{with}\quad
C^t = - C~,
\end{equation}
and in the above representation is
\begin{equation}
\label{ccm2n}
C = \tau^2 \otimes \tau^1 \otimes \tau^2 \otimes \tau^1 \otimes \tau^2 ~.
\end{equation}
The charge conjugation matrix is used to raise and lower the 32-dimensional spinor indices
(${\mathcal{\widehat A}}, {\mathcal {\widehat B}},\ldots$) of the
$\Gamma$-matrices according to
\begin{equation}
(\Gamma^{M})^{{\mathcal{\widehat A}}{\mathcal {\widehat B}}} \,\equiv\,
(\Gamma^{M})^{\mathcal{\widehat A}}_{~\,\,\mathcal{\widehat C}} \,\,
(C^{-1})^{{\mathcal {\widehat C}}{\mathcal {\widehat B}}}\quad\mbox{and}
\quad
(\Gamma^{M})_{{\mathcal {\widehat A}}{\mathcal {\widehat B}}} \,\equiv\,
(C)_{{\mathcal {\widehat A}}{\mathcal {\widehat C}}} \,\, (\Gamma^{M})^{\mathcal{
\widehat C}}_{~\,\,\mathcal {\widehat B}} ~~.
\end{equation}
The chirality matrix is defined by
\begin{equation}
\label{chir2n}
 \Gamma_{(11)}^{\mathrm E}= -\ii \, \Gamma^0 \Gamma^1 \ldots \Gamma^{9}
=\tau^3 \otimes \tau^3 \otimes \tau^3 \otimes \tau^3 \otimes \tau^3 ~.
\end{equation}

The above expressions are useful to obtain the factorization of the $d=10$ matrices
when the ten-dimensional space is split into $4+6$. In fact, by writing
\begin{equation}
\Gamma^\mu =\gamma^\mu \otimes \mathbf{1}~~,~~
\Gamma^m =\gamma_{(5)} \otimes \ \gamma^m~~,~~
\Gamma_{(11)}^{\mathrm E} =\gamma_{(5)} \otimes \gamma_{(7)}~~,~~
C =C_4\otimes C_6~,
\label{split}
\end{equation}
we can read off the explicit representation of the Dirac matrices $\gamma^\mu$ and
$\gamma^m$ for $d=4$ and $d=6$, respectively, of the corresponding chirality matrices
$\gamma_{(5)}$ and $\gamma_{(7)}$, and of the charge conjugation matrices
$C_4$ and $C_6$.

\paragraph{$\Gamma$-matrices in four dimensions} The $d=4$ matrices
$\gamma^\mu$ which can be read from Eqs. (\ref{gm10}) and (\ref{split}), are
\begin{equation}
\label{gm4}
\gamma^0 = \tau^1 \otimes \mathbf{1}~~,~~
\gamma^1 = \tau^2 \otimes \mathbf{1}~~,~~
\gamma^2 = \tau^3 \otimes \tau^1~~,~~
\gamma^3 = \tau^3 \otimes \tau^2~,
\end{equation}
while the chirality and charge conjugation matrices are
\begin{equation}
\label{gamma5}
\gamma_{(5)} = -\, \gamma^0 \gamma^1 \gamma^2 \gamma^3= \tau^3\otimes\tau^3
\quad\mbox{and}\quad
C_4=\tau^2\otimes\tau^1~.
\end{equation}
In this tensor product basis the four spinor indices are ordered as
\begin{center}
\begin{tabular}{c|c}
$2 \,\vec{\epsilon}$ & chirality \\
\hline
$(++)$ & $+$ \\
$(-+)$ & $-$ \\
$(+-)$ & $-$ \\
$(--)$ & $+$ \\
\end{tabular}
\end{center}
However, it is often useful to rearrange them in order to have first the two chiral indices
$\alpha \in \{(++),(--)\}$ and then the two anti-chiral ones $\dot\alpha\in\{(-+),(+-)\}$, in
such a way that the chirality matrix takes the more conventional form
$\gamma_{(5)} =\mathbf{1}\otimes \tau^3$.
With such a rearrangement the above Euclidean Dirac matrices $\gamma^\mu$ become
\begin{equation}
\label{gamma4def}
\gamma^\mu =
\begin{pmatrix}
0 & \sigma^\mu \\
\overline \sigma^\mu & 0
\end{pmatrix}
\end{equation}
with $\sigma^\mu = \big( \mathbf{1},\, -\ii \tau^3,\, \ii \tau^2,\, -\ii \tau^1\big)$
and $\overline\sigma^\mu  =  \big( \mathbf{1},\, \ii \tau^3,\, -\ii \tau^2,\, \ii \tau^1\big)$.
The matrices $\big(\sigma^\mu\big)_{\alpha\dot\beta}$ and
$\big(\overline\sigma^\mu\big)^{\dot\alpha\beta}$ act on spinors of definite chirality $\psi_{\alpha}$ and $\psi^{\dot\alpha}$ as
\begin{equation}
\big(\sigma^\mu\big)_{\alpha\dot\beta} \,\psi^{\dot\beta}\quad\mbox{and}\quad
\big(\overline\sigma^\mu\big)^{\dot\alpha\beta} \,\psi_{\beta}~.
\end{equation}
After the rearrangement of indices, the charge conjugation matrix becomes
\begin{equation}
C_4= \tau^2 \otimes \tau^3~;
\label{c44}
\end{equation}
thus it is block diagonal with $\big(C_4\big)^{\alpha\beta}=- \ii \,\epsilon^{\alpha\beta}$ and
$\big(C_4\big)_{\dot\alpha\dot\beta}= \ii \, \epsilon_{\dot\alpha\dot\beta}$.

\paragraph{$\Gamma$-matrices in six dimensions}
The $d=6$ matrices
$\gamma^m$ which can be read from Eqs. (\ref{gm10}) and (\ref{split}), are
\begin{equation}
\begin{aligned}
&\gamma^4 = \tau^1 \otimes \mathbf{1} \otimes \mathbf{1} ~~,~~~
\gamma^5 = \tau^2 \otimes \mathbf{1} \otimes \mathbf{1}  ~~,~~~~
\gamma^6 = \tau^3 \otimes \tau^1 \otimes \mathbf{1}\\
&\gamma^7 = \tau^3 \otimes \tau^2 \otimes \mathbf{1} ~~,~~
\gamma^8 = \tau^3 \otimes \tau^3 \otimes \tau^1 ~~,~~
\gamma^9 = \tau^3 \otimes \tau^3 \otimes \tau^2
\end{aligned}
\label{gm2}
\end{equation}
while the corresponding chirality and charge conjugation matrices are
\begin{equation}
\label{gamma7}
\gamma_{(7)} = \ii\, \gamma^4 \gamma^5 \ldots \gamma^9\,=\,
\tau^3 \otimes \tau^3 \otimes \tau^3\quad\mbox{and}\quad
C_6= \tau^2 \otimes \tau^1 \otimes \tau^2~.
\end{equation}
In this case the eight spinor indices are ordered according to
\begin{equation}
\begin{tabular}{c|c}
\label{stati}
$2\vec{\epsilon}$ & chirality \\
\hline
$(+++)$ & $ + $ \\
$(-++)$ & $ - $ \\
$(+-+)$ & $ - $ \\
$(--+)$  & $ + $ \\
$(++-)$  & $ - $ \\
$(-+-)$  & $ + $ \\
$(+--)$  & $ + $ \\
$(---)$  & $ - $
\end{tabular}
\end{equation}
but again they can be rearranged in such a way to put first the chiral ones and then the anti-chiral
ones, and have the chirality matrix in the standard
form $\gamma_{(7)}= \mathbf{1} \otimes \mathbf{1} \otimes \tau^3$. In this basis the
matrices $\gamma^m\,C_6^{-1}$ may be written in the block diagonal form
\begin{equation}
\label{gamma6cdef}
\gamma^m \, C_6^{-1}=
\begin{pmatrix}
\Sigma^m & 0 \\
 0 & \overline\Sigma^{\,m}
\end{pmatrix}
\end{equation}
where $(\Sigma^m)^{AB}$ and $(\overline\Sigma^{\,m})_{AB}$ are $4\times 4$ anti-symmetric matrices.

If we order the four chiral indices as $\{(+++),(+--),(-+-),(--+)\}$ and the four anti-chiral indices
as $\{(---),(-++),(+-+),(++-)\}$ (see also \eq{spintransf} below), we have
\begin{equation}
\begin{aligned}
\Sigma^m &= \big(\eta^3, -\ii\overline\eta^3,\eta^2,
 -\ii\overline\eta^2,\eta^1,\ii\overline\eta^1\big)~,
\\
\overline\Sigma^{\,m} &= \big(\eta^3,\ii\overline\eta^3,-\eta^2,
-\ii\overline\eta^2,\eta^1,-\ii\overline\eta^1\big)~,
\end{aligned}
\label{Sigma}
\end{equation}
where $\eta^c$ and $\overline\eta^c$ are, respectively, the self-dual and anti-self-dual 't Hooft
symbols. Proceeding in a similar way for the antisimmetrized product of three matrices, we find
\begin{equation}
\label{g36cdef}
\gamma^{mnp} \, C_6^{-1}=
\begin{pmatrix}
\Sigma^{mnp} & 0 \\
0 &  \overline\Sigma^{\,mnp}
\end{pmatrix}~,
\end{equation}
where $(\Sigma^{mnp})^{AB}$ and $(\overline\Sigma^{\,mnp})_{AB}$ the $4\times 4$ symmetric matrices
that appear in Section \ref{sec:effects}. Using the properties of the chirality and charge
conjugation matrices, it is easy to show the following imaginary self-duality properties
\begin{equation}
*_6\Sigma^{mnp} = -\ii\,\Sigma^{mnp}\quad,\quad
*_6\overline\Sigma^{mnp} = +\ii\,\overline\Sigma^{mnp}~.
\label{sigmaSD1}
\end{equation}

\paragraph{Useful formulas}
The previous formulas allow us to obtain the explicit expressions for the fermion bilinears which have been discussed in Sections \ref{sec:CFT} and \ref{sec:effects}.
In this respect we point out that in writing a fermion bilinear, like the one appearing for instance
in Eq.(\ref{ampltot2}), we always understand the inverse charge conjugation matrix $C^{-1}$.
The precise expression of the bilinear is then
\begin{equation}
\label{th}
\Theta\Gamma^{mnp}\Theta \equiv
\Theta_{\cal A}\, (\Gamma^{mnp}C^{-1})^{{\cal A}{\cal B}} \,\Theta_{\cal B}~.
\end{equation}
Using the 4+6 decomposition discussed above, we obtain
\begin{equation}
\Gamma^{mnp}C^{-1} = \big(\gamma_{(5)} C_4^{-1} \big)\otimes \big(\gamma^{mnp}C_6^{-1}\big)
=
\begin{pmatrix}
\tau^2 &  0 \\
0 & \tau^2
 \end{pmatrix}
\otimes
\begin{pmatrix}
\Sigma^{mnp} & 0 \\
0 &  \overline\Sigma^{\,mnp}
\end{pmatrix}~,
\end{equation}
so that Eq. (\ref{th}) can be rewritten as
\begin{equation}
\label{th1}
\Theta_{\cal A}\, (\Gamma^{mnp}C^{-1})^{{\cal A}{\cal B}} \,\Theta_{\cal B} =
- \ii\,\Theta^{\alpha  A}\,\epsilon_{\alpha \beta} \Theta^{\beta B}
(\overline\Sigma^{\,mnp})_{AB} \,- \ii \,
\Theta_{\dot \alpha  A}\,\epsilon^{\dot \alpha \dot \beta} \Theta_{\dot \beta B}
(\Sigma^{mnp})^{AB}
\end{equation}
which coincides with Eq. (\ref{decomp}).

Finally, we observe that the natural ordering (\ref{stati}) of the spinor indices
for the tensor product representation (\ref{gm2}) is particularly convenient if one
uses the complex basis in the internal six-dimensional space. Indeed, computing the holomorphic
and anti-holomorphic products $\gamma^{123}$ and $\gamma^{\bar 1 \bar 2 \bar 3}$ and combining
them with the charge conjugation matrix we find
\begin{equation}
\begin{aligned}
\gamma^{123} \,C_6^{-1} &= -\,\begin{pmatrix}
0\phantom 0&  0 \\
0\phantom 0 & 1
 \end{pmatrix}
\otimes
\begin{pmatrix}
0\phantom 0&  0 \\
0\phantom 0 & 1
 \end{pmatrix}
\otimes\begin{pmatrix}
0\phantom 0&  0 \\
0\phantom 0 & 1
 \end{pmatrix}
\\
\gamma^{\bar 1\bar 2 \bar 3} \,C_6^{-1} &= +\,\begin{pmatrix}
1\phantom 0&  0\\
0\phantom 0 & 0
 \end{pmatrix}
\otimes
\begin{pmatrix}
1\phantom 0&  0 \\
0\phantom 0 & 0
 \end{pmatrix}
\otimes\begin{pmatrix}
1\phantom 0&  0 \\
0\phantom 0& 0
 \end{pmatrix}
\end{aligned}
\end{equation}
from which we immediately see that $\Sigma^{123}=\overline\Sigma^{\,\bar 1\bar 2\bar3}=0$ and that
the only non-vanishing entries of the matrices $\Sigma^{\bar 1\bar 2 \bar 3}$ and $\overline\Sigma^{\,123}$ are, respectively, the upper most left and the lower most right, that is
\begin{equation}
\big(\Sigma^{\bar 1\bar 2 \bar 3})^{+++,+++}=1\quad\mbox{and}\quad
\big(\overline\Sigma^{\,123}\big)_{---,---}=-1~.
\label{sigma00}
\end{equation}

\subsection{The orbifold $\mathcal{T}_6/(\mathbb{Z}_2\times\mathbb{Z}_2)$}
\label{subapp:T6orb} The orbifold group
$\mathbb{Z}_2\times\mathbb{Z}_2$ acting on the orthonormal complex
coordinates $Z^i$ of $\mathcal{T}_6$ as in Table \ref{tablez2z2} is
a discrete subgroup of $\mathrm{SO}(6)$ that contains 4 elements
$h_I$ ($I=0,1,2,3$), with $h^0\equiv e$ being the identity element,
and
\begin{equation}
\label{frac2bis} h^1 = \ee^{\ii\pi (J_3 - J_2)}~,~~~~ h^2 =
\ee^{\ii\pi (J_1 - J_3)}~,~~~~ h^3\equiv h_1 h_2 = \ee^{\ii\pi (J_1
- J_2)}
\end{equation}
where $J_{1,2,3}$ are the generators of rotations in the 4-5, 6-7
and 8-9 planes respectively. We may summarize the transformation
properties  for the conformal fields $\partial Z^i$ and $\Psi^i$
($i=1,2,3$) in the Neveu-Schwarz sector by means of the following
table:
\begin{equation}
\label{ZPsirep}
\begin{tabular}{c|c}
conf. field & irrep \\
\hline
$\Big.\partial Z^i$, $\Psi^i$ & $R_i$
\end{tabular}~,
\end{equation}
where $\{R_A\}=\{R_0,R_i\}$ are the irreducible representations of $\mathbb{Z}_2\times
\mathbb{Z}_2$, identified by writing the character table $\ch_A^I = \tr_{R_A}\big(h^I\big)$
of the group
\begin{equation}
\label{frac3}
\begin{tabular}{c|cccc}
 & $e$ & $h_1$ & $h_2$ & $h_3$ \\
\hline
$R_0$ & 1 & ~1 & ~1 & ~1 \\
$R_1$ & 1 & ~1 & $-1$ & $-1$ \\
$R_2$ & 1 & $-1$ & ~1 & $-1$ \\
$R_3$ & 1 & $-1$ & $-1$ & ~1
\end{tabular}
\end{equation}
The Clebsch-Gordan series for these representations is simply given by
\begin{equation}
\label{frac4}
R_0\otimes R_A = R_A~~~~,~~~~
R_i\otimes R_j = \delta_{ij}  R_0 + |\epsilon_{ijk}| R_k~~,
\end{equation}
and is crucial in determining the open string spectrum.

Recall that through the bosonization procedure \cite{Friedan:1985ge}
the chiral spin fields $S^A \sim \ee^{\ii
\vec{\epsilon}^{\,A}\cdot\vec\varphi}$ of $\mathrm{SO}(6)$
and the anti-chiral ones $S_{A} \sim \ee^{\ii
\vec{\epsilon}_{\,A}\cdot\vec\varphi}$ are associated respectively to the
$\mathrm{SO}(6)$ spinor weights
$\vec{\epsilon}^{\,A}= \frac 12(\pm,\pm,\pm)$ with the product of signs being positive,
and $\vec{\epsilon}_{\,A}= \frac 12(\pm,\pm,\pm)$ with the product of signs being negative.
Using this information, we easily deduce from (\ref{frac2bis}) the
transformation properties of the various spin fields, which are
summarized in the following table
\begin{equation}
\label{spintransf}
\begin{tabular}{c|c|c}
irrep $R_A$ & $S^A$ &  $S_A$  \\
\hline
$R_0$ &  $S^0 \equiv S^{+++}$
& $ \Big.S_{0} \equiv S_{---}$
\\
$R_1$ & $S^1 \equiv S^{+--}$
& $\Big.S_{1} \equiv S_{-++}$
\\
$R_2$ & $S^2\equiv S^{-+-}$
& $\Big. S_{2}\equiv S_{+-+}$
\\
$R_3$ &  $S^3\equiv S^{--+}$
& $\Big.S_{3}\equiv S_{++-}$
\end{tabular}
\end{equation}
In other words, we can order the internal spinor indices so that
$S^A$ and $S_{A}$ transform in the irrep $R_A$.

Closed strings on the orbifold have different sectors. The untwisted
sector simply contains  the closed string states defined on the
covering space $\mathcal{T}_6$ which are invariant under the
orbifold action. The twisted sectors are in correspondence with the
16 fixed planes ($a=1,\ldots,16$) of the action of a nontrivial
element $h^i$. The vertex operators in a twisted sector contain
left- and right-moving twist fields $\Delta^i_a(z)$ and
$\tilde\Delta^i_a(z)$, and must be invariant under the orbifold. If,
as explained in the main text, we assume that the orbifold group
does not act on the twist fields, it is not difficult to write down
all the massless vertices.

Let us also notice that the orbifold projection leaves only two bulk supercharges, whose Weyl components in the $-1/2$ picture are
\begin{equation}
 \label{sclm}
Q_{\alpha} = \oint \frac{dz}{2\pi\ii}\, S_\alpha S_0 \,\ee^{-\frac{\phi}{2}}(z)~,~~~
Q^{\dot\alpha} = \oint \frac{dz}{2\pi\ii} \,S^{\dot\alpha} S^{0}\, \ee^{-\frac{\phi}{2}}(z)
\end{equation}
for the left-moving ones, and
\begin{equation}
 \label{scrm}
{\tilde Q}_{\alpha} = \oint \frac{d\bar z}{2\pi\ii}\, {\tilde S}_\alpha {\tilde S}_0
\,\ee^{-\frac{\tilde\phi}{2}}(z)~,~~~
{\tilde Q}^{\dot\alpha} = \oint \frac{d\bar z}{2\pi\ii} \,{\tilde S}^{\dot\alpha} {\tilde S}^{0} \,\ee^{-\frac{\tilde\phi}{2}}(z)
\end{equation}
for the right-moving ones.

\subsection{Soft supersymmetry breaking on fractional D9 branes}
\label{subsec:fD9}
In the orbifold $\mathcal T_6/{\mathbb Z_2}\times {\mathbb Z_2}$, we can realize an $\mathcal{N}=1$ $d=4$ gauge theory using fractional D9 branes that completely wrap the internal compact space.
Such brane configuration preserves a different  $\mathcal{N}=1$ supersymmetry
with respect to the fractional D3 branes considered in Section \ref{sec:n1int}, and thus
the moduli fields organize into chiral multiplets with respect to this new supersymmetry.
The bulk Lagrangian (which is in fact the same since we have not changed the compactification manifold) can be rewritten in terms of these multiplets via a different
K\"ahler potential and superpotential. Again, this allows to relate the flux-induced gaugino mass term to the value of the auxiliary component of the gauge kinetic function.

For the reasons already explained in the case of D3 branes, we are interested mostly
in the untwisted couplings, which can be deduced by reducing to four
dimensions the usual DBI-WZ action of a D9 brane
on $\mathcal{T}_6/(\mathbb{Z}_2\times\mathbb{Z}_2)$ given by
\begin{equation}
 \label{D9BI}
-T_9 \int_{D3}\int_{\mathcal{T}_6/(\mathbb{Z}_2\times\mathbb{Z}_2)}
  \ee^{-\varphi}\sqrt{-{\rm det}(G_{(10)} + \mathcal{F} )} + T_9
\int_{D3}\int_{\mathcal{T}_6/(\mathbb{Z}_2\times\mathbb{Z}_2)}
\sum_{n=0}^5 C_{2n} \,\ee^{  \mathcal{F} }
\end{equation}
where $T_9 = (4\pi)^{-1}
(2\pi\alpha')^{-2}(2\pi\sqrt{\alpha'})^{-6}$. From this expression
it follows that the quadratic part in the gauge fields $F$, after
promoting the latter to the non-abelian case and switching to the
Einstein frame, is
\begin{equation}
 \label{YMD9}
-\frac{\ee^{\varphi/2}\mathcal{V}}{32\pi}\int_{D_3} d^4x \,{\mathrm{Tr}} \big(F_{\mu\nu} F^{\mu\nu}
\big) + \frac{\widetilde{C}}{32\pi}\int_{D_3} d^4x\,
{\mathrm{Tr}} \big(F_{\mu\nu}{}^*F^{\mu\nu}\big)~,
\end{equation}
where $\widetilde C$ is%
\footnote{This pseudoscalar modulus is in fact related, through the duality between $F_7$ and $F_3$, to
the two-form $C_2$ with indices in the space-time directions. Notice also that
\begin{equation*}
 \label{intg6}
\int_{\mathcal{T}_6/(\mathbb{Z}_2\times\mathbb{Z}_2)}\!\!\! d^6y \sqrt{-g_{(6)}} =
\frac 14\, (2\pi\sqrt{\alpha'})^6\, \ee^{3\varphi/2}\mathcal{V}
\end{equation*}
since it corresponds to the internal volume in the string frame, which is related by the factor of $\ee^{3\phi/2}$
to the Einstein frame volume $\mathcal{V}$. This explains the prefactor in \eq{YMD9}.}
\begin{equation}
 \label{defct}
\int_{\mathcal{T}_6/(\mathbb{Z}_2\times\mathbb{Z}_2)} C_6 =
\frac 14 (2\pi\sqrt{\alpha'})^{-6}\, \tilde{C}~.
\end{equation}
Comparing with \eq{biwzf}, we see that the untwisted part of the gauge kinetic function $f_A^{(9)}$
for any type $A$ of fractional D9 branes reads
\begin{equation}
 \label{gkf9}
f_A^{(9)}=\frac{s}{4}\quad\mbox{with}\quad
s = \tilde{C} + \ii \ee^{\varphi/2}\mathcal{V}~.
\end{equation}
In a similar way one can consider D5 branes wrapped on untwisted
cycles $e^i$, which preserve the same supersymmetry of the D9
branes. The way to combine the  untwisted moduli into chiral
multiplets is again suggested by the gauge kinetic functions
$f^{(5)}_i$. Extracting the quadratic terms in the gauge fields from
the wrapped D5-brane DBI-WZ action, it is straightforward to obtain
\begin{equation}
 \label{f5i}
f^{(5)}_i = \frac 14 r^i \quad\mbox{with}\quad r^i \equiv c^i + \ii
\ee^{-\varphi/2}\,v^i~.
\end{equation}
Notice that $c^i$ and $v^i$  are invariant under the O9 orientifold
projection appropriate to our situation.

By supersymmetry, the complex scalars  represents the lowest
component of a chiral superfield, and the supersymmetric
$\mathcal{N}=1$ Lagrangian (\ref{gkf}), contains the coupling of the
gaugino to the auxiliary components of this multiplet. For D9-branes
\begin{equation}
\label{gaugaux9}
-\frac{\ii}{32\pi} F^s\,\Tr \big(\Lambda^\alpha \Lambda_\alpha\big) + \cc
\end{equation}
This has to be compared to the gaugino mass term which follows from the D9 flux coupling
indicated in Table \ref{Dbranes}. To do so we have to adapt the steps
used to arrive at \eq{massD3} in the D3 case,
since now the normalization $c_F$ contains contains the topological factor $\mathcal C_{(10)}$
suitable for D9 disk amplitude, namely \cite{Billo:2002hm}
\begin{equation}
\mathcal C_{(10)} = \frac{\mathcal C_{(4)}}{(2\pi\sqrt{\alpha'})^6}~.
\label{c10}
\end{equation}
On the other hand to obtain the four dimensional couplings, we have to
dimensionally reduce on $\mathcal{T}_6/(\mathbb{Z}_2\times\mathbb{Z}_2)$, gaining a factor of
$ (2\pi\sqrt{\alpha'})^6\,\ee^{3\varphi/2}\,\mathcal{V}$, to obtain the 4d coupling. In the end, we find the term
\begin{equation}
 \label{gc9}
-\frac{\ii}{4\pi}\,\ee^{\varphi/2}\,\mathcal{V}\,(2\pi\alpha')^{-\frac12}\,\mathcal{N}_F\, F_{(3,0)}
=-\frac{\ii}{4\pi}\,\ee^{\varphi}\, F_{(3,0)}
\end{equation}
where in the second step we used the normalization of the flux vertex already fixed in \eq{NFvalue}
so that by comparison with \eq{gaugaux9} the auxiliary field must be given by
\begin{equation}
\label{Fs}
F^s = 8 \,\ee^{\varphi}\, F_{(3,0)}~.
\end{equation}

Analogously to what we did in the D3-brane case, we restrict to the
slice of moduli space spanned by $s$ and by the overall scale
\begin{equation}
 \label{runico}
r \equiv r^1 = r^2 = r^3~;
\end{equation}
this scale is related to the volume by $(\im r)^3 =
\ee^{-3\varphi/2}\,\mathcal{V}$, as it follows from \eq{vol}. In
this slice of the moduli space the bulk theory can be rewritten in
the standard $\mathcal{N}=1$ form employing the chiral fields
$s(\theta)$ and $r(\theta)$. The K\"ahler potential reads
\begin{equation}
 \label{Kahlersr}
K = - \log(\im s) - 3 \log (\im r)~,
\end{equation}
and it coincides with \eq{KP} re-expressed in the new set of variables. The superpotential is
given by
\begin{equation}
 \label{suppotrs}
W = \frac{1}{\kappa_{10}^2}\int F\wedge \Omega~,
\end{equation}
where, with respect to \eq{WGO},  the O9 projection eliminates the NS-NS flux.
It is straightforward to check that
\begin{equation}
 \label{FWs}
{\overline F}^{\,\bar s} = -\ii \kappa_4^2\,\ee^{K/2} K^{\bar{s} s} D_{s} W
= 8 \,\ee^{\varphi}\, F_{(0,3)}~,
\end{equation}
in agreement with \eq{Fs}.

\providecommand{\href}[2]{#2}\begingroup\raggedright\endgroup


\begin{thebibliography}{10}

\bibitem{Blumenhagen:2005mu}
R.~Blumenhagen, M.~Cvetic, P.~Langacker, and G.~Shiu, \emph{{Toward realistic
  intersecting D-brane models}}, Ann. Rev. Nucl. Part. Sci. {\bf 55} (2005)
  71--139,
\href{http://arxiv.org/abs/hep-th/0502005}{{\tt arXiv:hep-th/0502005}}.

\bibitem{Blumenhagen:2006ci}
R.~Blumenhagen, B.~Kors, D.~Lust, and S.~Stieberger, \emph{{Four-dimensional
  String Compactifications with D-Branes, Orientifolds and Fluxes}},
  \href{http://dx.doi.org/10.1016/j.physrep.2007.04.003}{Phys. Rept. {\bf 445}
  (2007)  1--193},
\href{http://arxiv.org/abs/hep-th/0610327}{{\tt arXiv:hep-th/0610327}}.

\bibitem{Marchesano:2007de}
F.~Marchesano, \emph{{Progress in D-brane model building}},
  \href{http://dx.doi.org/10.1002/prop.200610381}{Fortsch. Phys. {\bf 55}
  (2007)  491--518},
\href{http://arxiv.org/abs/hep-th/0702094}{{\tt arXiv:hep-th/0702094}}.

\bibitem{Grana:2005jc}
M.~Grana, \emph{{Flux compactifications in string theory: A comprehensive
  review}}, \href{http://dx.doi.org/10.1016/j.physrep.2005.10.008}{Phys. Rept.
  {\bf 423} (2006)  91--158},
\href{http://arxiv.org/abs/hep-th/0509003}{{\tt arXiv:hep-th/0509003}}.

\bibitem{Douglas:2006es}
M.~R. Douglas and S.~Kachru, \emph{{Flux compactification}},
  \href{http://dx.doi.org/10.1103/RevModPhys.79.733}{Rev. Mod. Phys. {\bf 79}
  (2007)  733--796},
\href{http://arxiv.org/abs/hep-th/0610102}{{\tt arXiv:hep-th/0610102}}.

\bibitem{Denef:2007pq}
F.~Denef, M.~R. Douglas, and S.~Kachru, \emph{{Physics of string flux
  compactifications}},
  \href{http://dx.doi.org/10.1146/annurev.nucl.57.090506.123042}{Ann. Rev.
  Nucl. Part. Sci. {\bf 57} (2007)  119--144},
\href{http://arxiv.org/abs/hep-th/0701050}{{\tt arXiv:hep-th/0701050}}.

\bibitem{Gukov:1999ya}
S.~Gukov, C.~Vafa, and E.~Witten, \emph{{CFT's from Calabi-Yau four-folds}},
  \href{http://dx.doi.org/10.1016/S0550-3213(00)00373-4}{Nucl. Phys. {\bf B584}
  (2000)  69--108},
\href{http://arxiv.org/abs/hep-th/9906070}{{\tt arXiv:hep-th/9906070}}.

\bibitem{Taylor:1999ii}
T.~R. Taylor and C.~Vafa, \emph{{RR flux on Calabi-Yau and partial
  supersymmetry breaking}},
  \href{http://dx.doi.org/10.1016/S0370-2693(00)00005-8}{Phys. Lett. {\bf B474}
  (2000)  130--137},
\href{http://arxiv.org/abs/hep-th/9912152}{{\tt arXiv:hep-th/9912152}}.

\bibitem{Grana:2002nq}
M.~Grana, \emph{{MSSM parameters from supergravity backgrounds}},
  \href{http://dx.doi.org/10.1103/PhysRevD.67.066006}{Phys. Rev. {\bf D67}
  (2003)  066006},
\href{http://arxiv.org/abs/hep-th/0209200}{{\tt arXiv:hep-th/0209200}}.

\bibitem{Camara:2003ku}
P.~G. Camara, L.~E. Ibanez, and A.~M. Uranga, \emph{{Flux-induced SUSY-breaking
  soft terms}}, \href{http://dx.doi.org/10.1016/j.nuclphysb.2004.04.013}{Nucl.
  Phys. {\bf B689} (2004)  195--242},
\href{http://arxiv.org/abs/hep-th/0311241}{{\tt arXiv:hep-th/0311241}}.

\bibitem{Grana:2003ek}
M.~Grana, T.~W. Grimm, H.~Jockers, and J.~Louis, \emph{{Soft supersymmetry
  breaking in Calabi-Yau orientifolds with D-branes and fluxes}},
  \href{http://dx.doi.org/10.1016/j.nuclphysb.2004.04.021}{Nucl. Phys. {\bf
  B690} (2004)  21--61},
\href{http://arxiv.org/abs/hep-th/0312232}{{\tt arXiv:hep-th/0312232}}.

\bibitem{Camara:2004jj}
P.~G. Camara, L.~E. Ibanez, and A.~M. Uranga, \emph{{Flux-induced SUSY-breaking
  soft terms on D7-D3 brane systems}},
  \href{http://dx.doi.org/10.1016/j.nuclphysb.2004.11.035}{Nucl. Phys. {\bf
  B708} (2005)  268--316},
\href{http://arxiv.org/abs/hep-th/0408036}{{\tt arXiv:hep-th/0408036}}.

\bibitem{Lust:2004fi}
D.~Lust, S.~Reffert, and S.~Stieberger, \emph{{Flux-induced soft supersymmetry
  breaking in chiral type IIb orientifolds with D3/D7-branes}},
  \href{http://dx.doi.org/10.1016/j.nuclphysb.2004.11.030}{Nucl. Phys. {\bf
  B706} (2005)  3--52},
\href{http://arxiv.org/abs/hep-th/0406092}{{\tt arXiv:hep-th/0406092}}.

\bibitem{Conlon:2005ki}
J.~P. Conlon, F.~Quevedo, and K.~Suruliz, \emph{{Large-volume flux
  compactifications: Moduli spectrum and D3/D7 soft supersymmetry breaking}},
  JHEP {\bf 08} (2005)  007,
\href{http://arxiv.org/abs/hep-th/0505076}{{\tt arXiv:hep-th/0505076}}.

\bibitem{Conlon:2006wz}
J.~P. Conlon, S.~S. Abdussalam, F.~Quevedo, and K.~Suruliz, \emph{{Soft SUSY
  breaking terms for chiral matter in IIB string compactifications}}, JHEP {\bf
  01} (2007)  032,
\href{http://arxiv.org/abs/hep-th/0610129}{{\tt arXiv:hep-th/0610129}}.

\bibitem{Berg:2007wt}
M.~Berg, M.~Haack, and E.~Pajer, \emph{{Jumping Through Loops: On Soft Terms
  from Large Volume Compactifications}},
  \href{http://dx.doi.org/10.1088/1126-6708/2007/09/031}{JHEP {\bf 09} (2007)
  031},
\href{http://arxiv.org/abs/0704.0737}{{\tt arXiv:0704.0737 [hep-th]}}.

\bibitem{Witten:1995gx}
E.~Witten, \emph{{Small Instantons in String Theory}},
  \href{http://dx.doi.org/10.1016/0550-3213(95)00625-7}{Nucl. Phys. {\bf B460}
  (1996)  541--559},
\href{http://arxiv.org/abs/hep-th/9511030}{{\tt arXiv:hep-th/9511030}}.

\bibitem{Douglas:1995bn}
M.~R. Douglas, \emph{{Branes within branes}},
\href{http://arxiv.org/abs/hep-th/9512077}{{\tt arXiv:hep-th/9512077}}.

\bibitem{Green:2000ke}
M.~B. Green and M.~Gutperle, \emph{{D-instanton induced interactions on a
  D3-brane}}, JHEP {\bf 02} (2000)  014,
\href{http://arxiv.org/abs/hep-th/0002011}{{\tt arXiv:hep-th/0002011}}.

\bibitem{Billo:2002hm}
M.~Billo, M.~Frau, I.~Pesando, F.~Fucito, A.~Lerda, and A.~Liccardo,
  \emph{{Classical gauge instantons from open strings}}, JHEP {\bf 02} (2003)
  045,
\href{http://arxiv.org/abs/hep-th/0211250}{{\tt arXiv:hep-th/0211250}}.

\bibitem{Billo:2006jm}
M.~Billo, M.~Frau, F.~Fucito, and A.~Lerda, \emph{{Instanton calculus in R-R
  background and the topological string}}, JHEP {\bf 11} (2006)  012,
\href{http://arxiv.org/abs/hep-th/0606013}{{\tt arXiv:hep-th/0606013}}.

\bibitem{Akerblom:2006hx}
N.~Akerblom, R.~Blumenhagen, D.~Lust, E.~Plauschinn, and M.~Schmidt-Sommerfeld,
  \emph{{Non-perturbative SQCD Superpotentials from String Instantons}}, JHEP
  {\bf 04} (2007)  076,
\href{http://arxiv.org/abs/hep-th/0612132}{{\tt arXiv:hep-th/0612132}}.

\bibitem{Billo:2007sw}
M.~Billo, M.~Frau, I.~Pesando, P.~Di~Vecchia, A.~Lerda, and R.~Marotta,
  \emph{{Instantons in N=2 magnetized D-brane worlds}},
  \href{http://dx.doi.org/10.1088/1126-6708/2007/10/091}{JHEP {\bf 10} (2007)
  091},
\href{http://arxiv.org/abs/0708.3806}{{\tt arXiv:0708.3806 [hep-th]}}.

\bibitem{Billo:2007py}
M.~Billo, M.~Frau, I.~Pesando, P.~Di~Vecchia, A.~Lerda, and R.~Marotta,
  \emph{{Instanton effects in N=1 brane models and the Kahler metric of twisted
  matter}}, \href{http://dx.doi.org/10.1088/1126-6708/2007/12/051}{JHEP {\bf
  12} (2007)  051},
\href{http://arxiv.org/abs/0709.0245}{{\tt arXiv:0709.0245 [hep-th]}}.

\bibitem{Dorey:2002ik}
N.~Dorey, T.~J. Hollowood, V.~V. Khoze, and M.~P. Mattis, \emph{{The calculus
  of many instantons}},
  \href{http://dx.doi.org/10.1016/S0370-1573(02)00301-0}{Phys. Rept. {\bf 371}
  (2002)  231--459},
\href{http://arxiv.org/abs/hep-th/0206063}{{\tt arXiv:hep-th/0206063}}.

\bibitem{Blumenhagen:2006xt}
R.~Blumenhagen, M.~Cvetic, and T.~Weigand, \emph{{Spacetime instanton
  corrections in 4D string vacua - the seesaw mechanism for D-brane models}},
  \href{http://dx.doi.org/10.1016/j.nuclphysb.2007.02.016}{Nucl. Phys. {\bf
  B771} (2007)  113--142},
\href{http://arxiv.org/abs/hep-th/0609191}{{\tt arXiv:hep-th/0609191}}.

\bibitem{Ibanez:2006da}
L.~E. Ibanez and A.~M. Uranga, \emph{{Neutrino Majorana masses from string
  theory instanton effects}}, JHEP {\bf 03} (2007)  052,
\href{http://arxiv.org/abs/hep-th/0609213}{{\tt arXiv:hep-th/0609213}}.

\bibitem{Florea:2006si}
B.~Florea, S.~Kachru, J.~McGreevy, and N.~Saulina, \emph{{Stringy instantons
  and quiver gauge theories}}, JHEP {\bf 05} (2007)  024,
\href{http://arxiv.org/abs/hep-th/0610003}{{\tt arXiv:hep-th/0610003}}.

\bibitem{Bianchi:2007fx}
M.~Bianchi and E.~Kiritsis, \emph{{Non-perturbative and Flux superpotentials
  for Type I strings on the orbifold}},
  \href{http://dx.doi.org/10.1016/j.nuclphysb.2007.05.006}{Nucl. Phys. {\bf
  B782} (2007)  26--50},
\href{http://arxiv.org/abs/hep-th/0702015}{{\tt arXiv:hep-th/0702015}}.

\bibitem{Argurio:2007vqa}
R.~Argurio, M.~Bertolini, G.~Ferretti, A.~Lerda, and C.~Petersson,
  \emph{{Stringy Instantons at Orbifold Singularities}}, JHEP {\bf 06} (2007)
  067,
\href{http://arxiv.org/abs/0704.0262}{{\tt arXiv:0704.0262 [hep-th]}}.

\bibitem{Bianchi:2007wy}
M.~Bianchi, F.~Fucito, and J.~F. Morales, \emph{{D-brane Instantons on the
  orientifold}}, \href{http://dx.doi.org/10.1088/1126-6708/2007/07/038}{JHEP
  {\bf 07} (2007)  038},
\href{http://arxiv.org/abs/0704.0784}{{\tt arXiv:0704.0784 [hep-th]}}.

\bibitem{Ibanez:2007rs}
L.~E. Ibanez, A.~N. Schellekens, and A.~M. Uranga, \emph{{Instanton Induced
  Neutrino Majorana Masses in CFT Orientifolds with MSSM-like spectra}}, JHEP
  {\bf 06} (2007)  011,
\href{http://arxiv.org/abs/0704.1079}{{\tt arXiv:0704.1079 [hep-th]}}.

\bibitem{Antusch:2007jd}
S.~Antusch, L.~E. Ibanez, and T.~Macri, \emph{{Neutrino Masses and Mixings from
  String Theory Instantons}},
  \href{http://dx.doi.org/10.1088/1126-6708/2007/09/087}{JHEP {\bf 09} (2007)
  087},
\href{http://arxiv.org/abs/0706.2132}{{\tt arXiv:0706.2132 [hep-ph]}}.

\bibitem{Blumenhagen:2007zk}
R.~Blumenhagen, M.~Cvetic, D.~Lust, R.~Richter, and T.~Weigand,
  \emph{{Non-perturbative Yukawa Couplings from String Instantons}},
  \href{http://dx.doi.org/10.1103/PhysRevLett.100.061602}{Phys. Rev. Lett. {\bf
  100} (2008)  061602},
\href{http://arxiv.org/abs/0707.1871}{{\tt arXiv:0707.1871 [hep-th]}}.

\bibitem{Aharony:2007pr}
O.~Aharony and S.~Kachru, \emph{{Stringy Instantons and Cascading Quivers}},
  \href{http://dx.doi.org/10.1088/1126-6708/2007/09/060}{JHEP {\bf 09} (2007)
  060},
\href{http://arxiv.org/abs/0707.3126}{{\tt arXiv:0707.3126 [hep-th]}}.

\bibitem{Blumenhagen:2007bn}
R.~Blumenhagen, M.~Cvetic, R.~Richter, and T.~Weigand, \emph{{Lifting
  D-Instanton Zero Modes by Recombination and Background Fluxes}},
  \href{http://dx.doi.org/10.1088/1126-6708/2007/10/098}{JHEP {\bf 10} (2007)
  098},
\href{http://arxiv.org/abs/0708.0403}{{\tt arXiv:0708.0403 [hep-th]}}.

\bibitem{Camara:2007dy}
P.~G. Camara, E.~Dudas, T.~Maillard, and G.~Pradisi, \emph{{String instantons,
  fluxes and moduli stabilization}},
  \href{http://dx.doi.org/10.1016/j.nuclphysb.2007.11.026}{Nucl. Phys. {\bf
  B795} (2008)  453--489},
\href{http://arxiv.org/abs/0710.3080}{{\tt arXiv:0710.3080 [hep-th]}}.

\bibitem{Ibanez:2007tu}
L.~E. Ibanez and A.~M. Uranga, \emph{{Instanton Induced Open String
  Superpotentials and Branes at Singularities}},
  \href{http://dx.doi.org/10.1088/1126-6708/2008/02/103}{JHEP {\bf 02} (2008)
  103},
\href{http://arxiv.org/abs/0711.1316}{{\tt arXiv:0711.1316 [hep-th]}}.

\bibitem{GarciaEtxebarria:2007zv}
I.~Garcia-Etxebarria and A.~M. Uranga, \emph{{Non-perturbative superpotentials
  across lines of marginal stability}},
  \href{http://dx.doi.org/10.1088/1126-6708/2008/01/033}{JHEP {\bf 01} (2008)
  033},
\href{http://arxiv.org/abs/0711.1430}{{\tt arXiv:0711.1430 [hep-th]}}.

\bibitem{Petersson:2007sc}
C.~Petersson, \emph{{Superpotentials From Stringy Instantons Without
  Orientifolds}},
\href{http://arxiv.org/abs/0711.1837}{{\tt arXiv:0711.1837 [hep-th]}}.

\bibitem{Bianchi:2007rb}
M.~Bianchi and J.~F. Morales, \emph{{Unoriented D-brane Instantons vs Heterotic
  worldsheet Instantons}},
  \href{http://dx.doi.org/10.1088/1126-6708/2008/02/073}{JHEP {\bf 02} (2008)
  073},
\href{http://arxiv.org/abs/0712.1895}{{\tt arXiv:0712.1895 [hep-th]}}.

\bibitem{Blumenhagen:2008ji}
R.~Blumenhagen and M.~Schmidt-Sommerfeld, \emph{{Power Towers of String
  Instantons for N=1 Vacua}},
\href{http://arxiv.org/abs/0803.1562}{{\tt arXiv:0803.1562 [hep-th]}}.

\bibitem{Argurio:2008jm}
R.~Argurio, G.~Ferretti, and C.~Petersson, \emph{{Instantons and Toric Quiver
  Gauge Theories}},
\href{http://arxiv.org/abs/0803.2041}{{\tt arXiv:0803.2041 [hep-th]}}.

\bibitem{Cvetic:2008ws}
M.~Cvetic, R.~Richter, and T.~Weigand, \emph{{(Non-)BPS bound states and
  D-brane instantons}},
\href{http://arxiv.org/abs/0803.2513}{{\tt arXiv:0803.2513 [hep-th]}}.

\bibitem{Kachru:2008wt}
S.~Kachru and D.~Simic, \emph{{Stringy Instantons in IIB Brane Systems}},
\href{http://arxiv.org/abs/0803.2514}{{\tt arXiv:0803.2514 [hep-th]}}.

\bibitem{GarciaEtxebarria:2008pi}
I.~Garcia-Etxebarria, F.~Marchesano, and A.~M. Uranga, \emph{{Non-perturbative
  F-terms across lines of BPS stability}},
\href{http://arxiv.org/abs/0805.0713}{{\tt arXiv:0805.0713 [hep-th]}}.

\bibitem{Buican:2008qe}
M.~Buican and S.~Franco, \emph{{SUSY breaking mediation by D-brane
  instantons}},
\href{http://arxiv.org/abs/0806.1964}{{\tt arXiv:0806.1964 [hep-th]}}.

\bibitem{Grana:2002tu}
M.~Grana, \emph{{D3-brane action in a supergravity background: The fermionic
  story}}, \href{http://dx.doi.org/10.1103/PhysRevD.66.045014}{Phys. Rev. {\bf
  D66} (2002)  045014},
\href{http://arxiv.org/abs/hep-th/0202118}{{\tt arXiv:hep-th/0202118}}.

\bibitem{Marolf:2003vf}
D.~Marolf, L.~Martucci, and P.~J. Silva, \emph{{Actions and fermionic
  symmetries for D-branes in bosonic backgrounds}}, JHEP {\bf 07} (2003)  019,
\href{http://arxiv.org/abs/hep-th/0306066}{{\tt arXiv:hep-th/0306066}}.

\bibitem{Marolf:2003ye}
D.~Marolf, L.~Martucci, and P.~J. Silva, \emph{{Fermions, T-duality and
  effective actions for D-branes in bosonic backgrounds}}, JHEP {\bf 04} (2003)
   051,
\href{http://arxiv.org/abs/hep-th/0303209}{{\tt arXiv:hep-th/0303209}}.

\bibitem{Tripathy:2005hv}
P.~K. Tripathy and S.~P. Trivedi, \emph{{D3 brane action and fermion zero modes
  in presence of background flux}}, JHEP {\bf 06} (2005)  066,
\href{http://arxiv.org/abs/hep-th/0503072}{{\tt arXiv:hep-th/0503072}}.

\bibitem{Martucci:2005rb}
L.~Martucci, J.~Rosseel, D.~Van~den Bleeken, and A.~Van~Proeyen, \emph{{Dirac
  actions for D-branes on backgrounds with fluxes}},
  \href{http://dx.doi.org/10.1088/0264-9381/22/13/014}{Class. Quant. Grav. {\bf
  22} (2005)  2745--2764},
\href{http://arxiv.org/abs/hep-th/0504041}{{\tt arXiv:hep-th/0504041}}.

\bibitem{Bergshoeff:2005yp}
E.~Bergshoeff, R.~Kallosh, A.-K. Kashani-Poor, D.~Sorokin, and A.~Tomasiello,
  \emph{{An index for the Dirac operator on D3 branes with background fluxes}},
  JHEP {\bf 10} (2005)  102,
\href{http://arxiv.org/abs/hep-th/0507069}{{\tt arXiv:hep-th/0507069}}.

\bibitem{Linch:2008rw}
W.~D.~.~Linch, J.~McOrist and B.~C.~Vallilo,
\emph{{Type IIB Flux Vacua from the String Worldsheet}},
\href{http://arxiv.org/abs/0804.0613}{{\tt arXiv:0804.0613 [hep-th]}}.

\bibitem{Billo:2004zq}
M.~Billo, M.~Frau, I.~Pesando, and A.~Lerda, \emph{{N = 1/2 gauge theory and
  its instanton moduli space from open strings in R-R background}}, JHEP {\bf
  05} (2004)  023,
\href{http://arxiv.org/abs/hep-th/0402160}{{\tt arXiv:hep-th/0402160}}.

\bibitem{Lust:2004cx}
D.~Lust, P.~Mayr, R.~Richter, and S.~Stieberger, \emph{{Scattering of gauge,
  matter, and moduli fields from intersecting branes}},
  \href{http://dx.doi.org/10.1016/j.nuclphysb.2004.06.052}{Nucl. Phys. {\bf
  B696} (2004)  205--250},
\href{http://arxiv.org/abs/hep-th/0404134}{{\tt arXiv:hep-th/0404134}}.

\bibitem{Billo:2005jw}
M.~Billo, M.~Frau, F.~Lonegro, and A.~Lerda, \emph{{N = 1/2 quiver gauge
  theories from open strings with R-R fluxes}}, JHEP {\bf 05} (2005)  047,
\href{http://arxiv.org/abs/hep-th/0502084}{{\tt arXiv:hep-th/0502084}}.

\bibitem{Bertolini:2005qh}
M.~Bertolini, M.~Billo, A.~Lerda, J.~F. Morales, and R.~Russo, \emph{{Brane
  world effective actions for D-branes with fluxes}},
  \href{http://dx.doi.org/10.1016/j.nuclphysb.2006.02.044}{Nucl. Phys. {\bf
  B743} (2006)  1--40},
\href{http://arxiv.org/abs/hep-th/0512067}{{\tt arXiv:hep-th/0512067}}.

\bibitem{DiVecchia:1996uq}
P.~Di~Vecchia, L.~Magnea, A.~Lerda, R.~Russo, and R.~Marotta, \emph{{String
  techniques for the calculation of renormalization constants in field
  theory}}, \href{http://dx.doi.org/10.1016/0550-3213(96)00141-1}{Nucl. Phys.
  {\bf B469} (1996)  235--286},
\href{http://arxiv.org/abs/hep-th/9601143}{{\tt arXiv:hep-th/9601143}}.

\bibitem{Friedan:1985ge}
D.~Friedan, E.~J. Martinec, and S.~H. Shenker, \emph{{Conformal Invariance,
  Supersymmetry and String Theory}},
Nucl. Phys. {\bf B271} (1986)  93.

\bibitem{Aspinwall:1996mn}
P.~S. Aspinwall, \emph{{K3 surfaces and string duality}},
\href{http://arxiv.org/abs/hep-th/9611137}{{\tt arXiv:hep-th/9611137}}.

\bibitem{Douglas:1996sw}
M.~R. Douglas and G.~W. Moore, \emph{{D-branes, Quivers, and ALE Instantons}},
\href{http://arxiv.org/abs/hep-th/9603167}{{\tt arXiv:hep-th/9603167}}.

\bibitem{DiVecchia:1999rh}
P.~Di~Vecchia and A.~Liccardo, \emph{{D branes in string theory. I}}, NATO Adv.
  Study Inst. Ser. C. Math. Phys. Sci. {\bf 556} (2000)  1--59,
\href{http://arxiv.org/abs/hep-th/9912161}{{\tt arXiv:hep-th/9912161}}.

\bibitem{DiVecchia:1999fx}
P.~Di~Vecchia and A.~Liccardo, \emph{{D-branes in string theory. II}},
\href{http://arxiv.org/abs/hep-th/9912275}{{\tt arXiv:hep-th/9912275}}.

\bibitem{Billo:2000yb}
M.~Billo, B.~Craps, and F.~Roose, \emph{{Orbifold boundary states from Cardy's
  condition}}, JHEP {\bf 01} (2001)  038,
\href{http://arxiv.org/abs/hep-th/0011060}{{\tt arXiv:hep-th/0011060}}.

\bibitem{Bertolini:2001gg}
M.~Bertolini, P.~Di~Vecchia, G.~Ferretti, and R.~Marotta, \emph{{Fractional
  branes and N = 1 gauge theories}},
  \href{http://dx.doi.org/10.1016/S0550-3213(02)00178-5}{Nucl. Phys. {\bf B630}
  (2002)  222--240},
\href{http://arxiv.org/abs/hep-th/0112187}{{\tt arXiv:hep-th/0112187}}.

\bibitem{DiVecchia:2005vm}
P.~Di~Vecchia, A.~Liccardo, R.~Marotta, and F.~Pezzella, \emph{{On the gauge /
  gravity correspondence and the open/closed string duality}},
  \href{http://dx.doi.org/10.1142/S0217751X05024900}{Int. J. Mod. Phys. {\bf
  A20} (2005)  4699--4796},
\href{http://arxiv.org/abs/hep-th/0503156}{{\tt arXiv:hep-th/0503156}}.

\bibitem{Grimm:2004uq}
T.~W. Grimm and J.~Louis, \emph{{The effective action of N = 1 Calabi-Yau
  orientifolds}},
  \href{http://dx.doi.org/10.1016/j.nuclphysb.2004.08.005}{Nucl. Phys. {\bf
  B699} (2004)  387--426},
\href{http://arxiv.org/abs/hep-th/0403067}{{\tt arXiv:hep-th/0403067}}.

\bibitem{Billo:2008pg}
  M.~Billo', L.~Ferro, M.~Frau, F.~Fucito, A.~Lerda and J.~F.~Morales,
\emph{{Non-perturbative effective interactions from fluxes}},
\href{http://arxiv.org/abs/0807.4098}{{\tt arXiv:0807.4098 [hep-th]}}.


\end{thebibliography}

\end{document}